\documentclass[aps, prd, reprint, nofootinbib, longbibliography]{revtex4-1}

\usepackage{graphics,graphicx,epsfig}
\usepackage{amsmath, amssymb, color}
\usepackage[dvipsnames]{xcolor}
\usepackage[colorlinks=true, linkcolor=blue, citecolor=magenta]{hyperref}
\usepackage{url}

\usepackage{float, placeins}
\usepackage{mwe}
\usepackage[caption=false]{subfig}
\usepackage{morefloats}
\usepackage{empheq}
\usepackage{bm}
\usepackage{booktabs} 
\usepackage{multirow}
\usepackage{makecell}
\usepackage{array}
\newcommand\Tstrut{\rule{0pt}{2.6ex}}         % = `top' strut
\newcommand\Bstrut{\rule[-0.9ex]{0pt}{0pt}}   % = `bottom' strut
\usepackage{colortbl}
\graphicspath{{./Figures/}}

\newcommand{\beq}{\begin{equation}}
\newcommand{\beqn}{\begin{align}}
\newcommand{\eeq}{\end{equation}}
\newcommand{\eeqn}{\end{align}}
\newcommand\Mpl{M_{\textrm{pl}}}

\definecolor{light-gray}{gray}{0.85}
\begin{document}

\title{Flows into de Sitter space from anisotropic initial conditions: \\ An effective field theory approach}

\author{Feraz Azhar}
\email[Email address: ]{fazhar@nd.edu}
\affiliation{Department of Philosophy, University of Notre Dame, South Bend, IN 46556 USA}

\author{David I.~Kaiser}
\email[Email address: ]{dikaiser@mit.edu}
\affiliation{Department of Physics, Massachusetts Institute of Technology, Cambridge, MA 02139 USA}

\date{\today}

\begin{abstract}
For decades, physicists have analyzed various versions of a ``cosmic no-hair" conjecture, to understand under what conditions a spacetime that is initially spatially anisotropic and/or inhomogeneous will flow into an isotropic and homogeneous state. Wald's theorem, in particular, established that homogeneous but anisotropic spacetimes, if filled with a positive cosmological constant plus additional matter sources that satisfy specific energy conditions, will necessarily flow toward an (isotropic) de Sitter state at late times. In this paper we study the flow of homogeneous but anisotropic spacetimes toward isotropic states under conditions more general than those to which Wald's theorem applies. We construct an effective field theory (EFT) treatment for generic ``single-clock" systems in anisotropic spacetimes---which are not limited to realizations compatible with scalar-field constructions---and identify fixed points in the resulting phase space. We identify regions of this phase space that flow to isotropic fixed points---including a de Sitter fixed point---even in the absence of a bare cosmological constant, and for matter sources that do not obey the energy conditions required for Wald's theorem. Such flows into de Sitter reveal the emergence of an effective cosmological constant.
\end{abstract}

\maketitle

%\tableofcontents

\section{Introduction}

Cosmic inflation provides the leading account of the dynamics of the very early universe. (See Refs.~\cite{guth+kaiser_05, bassett+al_06, Lyth:2009zz,martin+al_14,Guth:2013sya,Baumann:2014nda} for reviews.) Models of inflation describe a brief period of accelerated expansion of space and can account for the highly symmetric spacetime we observe today---namely, one that is homogeneous and isotropic to high accuracy on large scales, with relatively small differences in energy density from place to place.

A natural question that arises is: how inhomogeneous and anisotropic can the universe be {\it prior} to inflation such that (i) inflation initiates and (ii) assuming that it does initiate, the inhomogeneities and anisotropies dissipate? If highly symmetric initial states are required for (i) and (ii), then models of inflation would face an ``initial conditions problem," such that the conditions that inflation looks to explain would be {\it required} as starting assumptions. This potential problem was recognized early in the history of the development of cosmic inflation and has received renewed interest of late. For recent discussion, see Refs.~\cite{East:2015ggf,kleban+senatore_16,Clough:2016ymm,Clough:2017efm,brandenberger_17, linde_18,Bloomfield:2019rbs,Chowdhury:2019otk,Aurrekoetxea:2019fhr,Creminelli:2019pdh,Creminelli:2020zvc,azhar_20,Joana:2020rxm,Tenkanen:2020cvw,Wang:2021hzv,Corman:2022alv}. 

The question of whether initially inhomogeneous and anisotropic backgrounds can evolve via inflation into homogeneous and isotropic backgrounds has crystallized into a ``cosmic no-hair'' conjecture. There are various versions of this conjecture (see Ref.~\cite[\S 9.5]{schmidt_04}) but it primarily amounts to the claim that---given certain constraints placed on the matter content of the universe, such as the presence of a cosmological constant, and assuming a metric theory of gravity, such as general relativity---initially inhomogeneous and/or anisotropic degrees of freedom dissipate as the universe expands. The presence of a cosmological constant in certain formulations of the conjecture serves as a proxy for a broad class of inflationary scenarios that employ potential-energy domination as a means of driving accelerated expansion. 

Probing spacetime dynamics assuming generic inhomogeneities and anisotropies of space is a formidable challenge. A more tractable approach is to relax the assumption of isotropy of the background space while preserving homogeneity, and investigate any ensuing ``isotropization." This approach was influentially addressed by~\citet{wald_83}. In particular, Wald demonstrated that initially expanding homogeneous backgrounds (generically for all Bianchi spacetimes except for Bianchi type IX)---with a positive cosmological constant and matter degrees of freedom that satisfy the dominant and strong energy conditions---will isotropize. Moreover, such conditions generically yield de Sitter-like expansion at late times, in which the cosmological constant eventually dominates the energy density of the background, leading to accelerated expansion. (See Refs.~\cite{watanabe+al_09, maleknejad+sheikh-jabbari_12,Cannone:2015rra,East:2015ggf,kleban+senatore_16,andreasson+ringstrom_16,Clough:2016ymm,Clough:2017efm,brandenberger_17, linde_18,carroll+chatwin-davies_18,Bloomfield:2019rbs,Chowdhury:2019otk,Aurrekoetxea:2019fhr,Creminelli:2019pdh,Creminelli:2020zvc,azhar_20,Joana:2020rxm,Tenkanen:2020cvw,Wang:2021hzv,moss+sahni_86, kitada+maeda_93, rendall_04,kleban+senatore_16,schmidt_04,Corman:2022alv} for more recent, related work.)

Wald's powerful theorem \cite{wald_83} requires that the matter sources (other than the bare, positive cosmological constant) satisfy both the strong and the dominant energy conditions. Yet since Wald's original work, several groups have demonstrated that fairly simple models can yield well-behaved cosmological dynamics despite failing to satisfy {\it any} of the energy conditions on which Wald's theorem depends \cite{Kandrup:1992xw,Barcelo:1999hq,Visser:1999de,Barcelo:2000zf,Bellucci:2001cc,Barcelo:2002bv,Dubovsky:2005xd,Nicolis:2009qm,Rubakov:2014jja,Martin-Moruno:2017exc}. Such examples point to physically relevant scenarios that fall entirely outside the domain to which Wald's theorem applies. To assess the dynamics of anisotropic spacetimes in  such scenarios, one must develop distinct methods, complementary to those with which Wald’s theorem has been explored to date. 

We pursue such a new approach in this work. In our analysis, we do not assume an explicit cosmological constant, nor do we place {\it a priori} conditions (such as energy conditions) on the matter degrees of freedom. Rather than imposing various energy conditions---the limitations of which have been scrutinized in detail \cite{Kandrup:1992xw,Barcelo:1999hq,Visser:1999de,Barcelo:2000zf,Bellucci:2001cc,Barcelo:2002bv,Dubovsky:2005xd,Nicolis:2009qm,Rubakov:2014jja,Martin-Moruno:2017exc}---we adapt the powerful effective field theory (EFT) approach to inflationary dynamics \cite{cheung+al_08,Weinberg:2008hq}. The EFT approach enables us to analyze the evolution of matter sources, including anisotropic pressures and other features associated with imperfect fluids, in quantitative detail yet also in a model-independent way. In particular, within this framework, the accelerated expansion of space need not be driven by the potential energy $V (\phi^I)$ associated with one or more scalar fields $\phi^I$. More generally, our ``single-clock” formalism, akin to the one developed in Ref.~\cite{Azhar:2018nol}, can accommodate scenarios that are not limited to those compatible with canonical scalar-field constructions.

Within this general framework, we identify regions within the effective phase space that flow into (isotropic) de Sitter-like expansion at late times, {\it without} having included a bare cosmological constant, and for matter sources that need not satisfy {\it either} the strong or dominant energy conditions.\footnote{By ``late times," we mean times $t \gg \tau$, where $\tau$ is the relevant dynamical time-scale set by the inverse of the trace of the extrinsic curvature.} Such dynamics fall entirely outside the conditions under which Wald’s theorem~\cite{wald_83} holds. These examples do not contradict Wald’s theorem. Rather, they indicate a new physical phenomenon---the emergence of an {\it effective cosmological constant}---that could not be identified, let alone analyzed, in physical systems to which Wald’s theorem applies.

In more detail, we construct an EFT of anisotropic backgrounds that describes the simplest anisotropic spacetime, a Bianchi I cosmological model. Our formalism consists of an action with effective matter degrees of freedom that (i) reflect certain symmetries of the background space at early times and (ii) are minimally coupled to (Einstein) gravity. We construct a dynamical system from this EFT, without imposing energy conditions nor inserting an explicit cosmological constant. We focus on the flow of such systems within the ``principal phase space," the lowest-dimensional phase space of the class of dynamical models developed here.

We investigate the stability of initially isotropic backgrounds in the presence of anisotropic matter sources, identifying conditions under which the space flows back to isotropy. (We do not treat more general questions about the stability of de Sitter spacetime, related to possible infrared divergences associated with light scalar fields \cite{burgess+al_16,gorbenko+senatore_19, baumgart+sundrum_20}.) We also investigate the evolution of initially anisotropic backgrounds, identifying conditions under which such spacetimes isotropize over time. For each of these scenarios, we find nontrivial regions of parameter space that flow into isotropic, de Sitter-like expansion at late times, despite the presence of significant anisotropic matter sources at early times.

Our plan for this paper is as follows. In Sec.~\ref{SEC:Formalism} we develop our EFT formalism for treating anisotropic backgrounds and translate this EFT into a dynamical system. We describe fixed points and stability properties for these fixed points for the principal phase space. In Sec.~\ref{SEC:Emerge} we describe our results where, in Sec.~\ref{SEC:ISOstability}, we probe stability properties of isotropic backgrounds, and in Sec.~\ref{SEC:anISOics} we demonstrate cases that isotropize at late times, 
with space undergoing accelerated expansion due to the emergence of an effective (rather than a bare) cosmological constant. Concluding remarks follow in Sec.~\ref{SEC:Discussion}, and we collect some relevant results in Appendixes~\ref{APP:MultipleScalars}--\ref{APP:Kasner}.

\section{  EFT formalism }\label{SEC:Formalism}

We begin by introducing, for the class of models in which we are interested: the relevant effective degrees of freedom, a characterization of energy conditions (to facilitate a comparison with previous work), and dynamical equations. We build upon EFT techniques first introduced in Refs.~\cite{cheung+al_08,Weinberg:2008hq}; techniques for handling dynamical systems introduced in Ref.~\cite{frusciante+al_14} (further developed in Ref.~\cite{Azhar:2018nol}); and a parameterization for anisotropic backgrounds introduced in Ref.~\cite{vandenhoogen+coley_95}. We work in units where $\hbar=c= 1$, so that the reduced Planck mass may be written as $\Mpl=(8\pi G)^{-1/2}\simeq 2.4\times 10^{18}\,\textrm{GeV}$. We restrict attention to four spacetime dimensions and adopt the metric signature $(-,+,+,+)$. Lower-case Greek letters $\alpha,\beta,\dots = 0, 1, 2, 3$, label spacetime indices and lower-case Latin letters $i,j,\dots = 1, 2, 3$, label spatial indices.

\subsection{An EFT of anisotropic backgrounds}\label{SEC:EFTdev}

Wald derived his influential cosmic no-hair theorem \cite{wald_83} for the case of a universe filled with a bare, positive cosmological constant $\Lambda$ plus additional matter sources described by an energy-momentum tensor $\tilde{T}_{\mu\nu}$, such that the total energy-momentum tensor for the system took the form
%%%%
\beq
T_{\mu\nu} = - \Lambda g_{\mu\nu} + \tilde{T}_{\mu\nu} .
\label{WaldTotalT}
\eeq
Wald's proof relied upon several postulates. First, that the dynamics of the system were governed by Einstein's field equations of ordinary general relativity, with $T_{\mu\nu}$ of the form in Eq.~(\ref{WaldTotalT}). Second, that the trace of the extrinsic curvature of the spacetime was positive at an initial time, corresponding to an initial volume-averaged expansion of space. Third---and crucially---that $\tilde{T}_{\mu\nu}$ satisfied both the dominant energy condition,
%%%%
\beq
\tilde{T}_{\mu\nu} u^\mu u^\nu \geq 0 ,
\label{WaldDEC}
\eeq
and the strong energy condition,
%%%%%
\beq
\left( \tilde{T}_{\mu\nu} - \frac{1}{2} g_{\mu\nu} \tilde{T} \right) u^\mu u^\nu \geq 0 ,
\label{WaldSEC}
\eeq
where $u^\mu$ is a future-directed time-like unit vector orthogonal to the spatial hypersurfaces. Under those conditions, Wald elegantly demonstrated that the square of the shear tensor $\sigma^2\equiv\sigma_{\alpha \beta}\sigma^{\alpha \beta}/2$ must vanish exponentially quickly over time, on a time-scale set by $M_{\rm pl} / \Lambda^{1/2}$. The shear tensor is defined as \cite{Hawking:1973uf,wald_83text,visser_95}
%%%%
\beq
\sigma_{\alpha \beta} \equiv \frac{1}{2} \left( \nabla_\nu u_\mu + \nabla_\mu u_\nu \right) h^\mu_{\>\> \alpha} h^\nu_{\>\> \beta} - \frac{1}{3} \nabla_\mu u^\mu \, h_{\alpha \beta} ,
\label{sheardef}
\eeq
where
%%%
\beq
h_{\mu\nu} \equiv g_{\mu\nu} + u_\mu u_\nu
\label{hmn}
\eeq
is the metric on spatial hypersurfaces. The quantity $\sigma^2$ vanishes identically for spacetimes that are homogeneous and isotropic, but in general is nonvanishing for anisotropic spacetimes. Therefore the condition $\sigma^2 \rightarrow 0$ indicates isotropization. 

Critical to Wald's theorem are {\it both} the presence of a bare, positive cosmological constant, $\Lambda > 0$, {\it and} the constraint that the other sources of matter obey both the dominant and strong energy conditions of Eqs.~(\ref{WaldDEC})--(\ref{WaldSEC}). Given the discussion in Refs.~\cite{Barcelo:1999hq,Visser:1999de,Barcelo:2000zf,Barcelo:2002bv,Bellucci:2001cc,Dubovsky:2005xd,Nicolis:2009qm,Rubakov:2014jja,Martin-Moruno:2017exc} of how readily these (and related) energy conditions can be violated by simple, well-behaved systems, on the other hand, we study the dynamics of effective systems in which we set $\Lambda = 0$, and we do not require that $\tilde{T}_{\mu\nu}$ obey either the dominant or strong energy conditions. 

In order to study whether anisotropic spacetimes can isotropize over time under conditions more general than those to which Wald's theorem applies, we must therefore develop a complementary approach to the one deployed so powerfully in Ref.~\cite{wald_83}. We develop the key ingredients of our alternative approach in this section.

We study dynamical flows for a homogeneous spacetime that is generally anisotropic, with a line element corresponding to a Bianchi I background given by
\beq\label{EQN:Bprelim}
ds^2 = - dt^2 + a^{2}_{i}(t) (dx^{i})^2.
\eeq
The three functions $a_{i}(t)$ (for $i=1,2,3$) encode (generally different) scale factors for expansion in each of three independent spatial directions. One may define an average scale factor, $a(t)$, as the geometric mean of these three scale factors: 
\beq\label{EQN:geomean}
a(t)\equiv\left[a_{1}(t)a_{2}(t)a_{3}(t)\right]^{1/3}.
\eeq
One can then rewrite Eq.~(\ref{EQN:Bprelim}) as
\beq
ds^2 = - dt^2 + a^2 (t) \, \delta_{ij} \, e^{2 \beta_i (t) } \, dx^i \, dx^j \, ,
\label{ds}
\eeq
so that the functions $\beta_i(t)$ correspond to the anisotropic parts of the scale factors $a_i(t)$. Equating coefficients in Eqs.~(\ref{EQN:Bprelim}) and~(\ref{ds}) and then using Eq.~(\ref{EQN:geomean}) yields a consistency condition:
\beq
\beta_1 (t) + \beta_2 (t) + \beta_3 (t) = 0 \, .
\label{betasum}
\eeq
Although the line element in Eq.~(\ref{ds}) represents a particularly simple form of a homogeneous yet anisotropic spacetime, for our purposes this ansatz is actually quite general. As demonstrated in Ref.~\cite{Cropp:2010yj}, the immediate vicinity of {\it any} timelike geodesic in an {\it arbitrary} spacetime can be described by a line element corresponding to a Bianchi I spacetime in the so-called ``ultra-local limit." Eq.~(\ref{ds}) therefore provides a fairly general basis for constructing an effective field theory treatment. 

To lowest order in our effective description, we consider fluid quantities that are spatially homogeneous (hence they depend only on cosmic time, $t$), but that are not necessarily spatially isotropic (hence they need not remain spherically symmetric at any given time $t$). In addition to the average scale factor $a(t)$ and its change over time, we also need to consider a nonzero shear tensor $\sigma_{\alpha \beta}$. For the line element of Eq.~(\ref{ds}), the nonvanishing components of $\sigma_{\alpha \beta}$ are given by
%%%%%%
\beq
\sigma_{ij} = a^2 \dot{\beta}_i e^{2 \beta_i} \, \delta_{ij} \quad {\rm (no \> sum \> on \>} i) \, ,
\label{sigmaij}
\eeq
where overdots denote a derivative with respect to cosmic time, $t$, and hence
%%%%%%%
\beq
\sigma^2 \equiv \frac{1}{2} \sigma_{\alpha \beta} \, \sigma^{\alpha \beta} = \frac{1}{2} \left( \dot{\beta}_1^2 + \dot{\beta}_2^2 + \dot{\beta}_3^2 \right) \, .
\label{sigma2}
\eeq
For a line element of the form in Eq.~(\ref{ds}), an isotropic spacetime corresponds to the case $\dot{\beta}_i \rightarrow 0$ for each $i$, and hence $\sigma^2 \rightarrow 0$.

Much as in Ref.~\cite{Azhar:2018nol}, we are interested in dynamics associated with an action that is linear in perturbations about the background (and, as usual, is described in unitary gauge). We are interested in dynamics of the background, so we ignore terms that are quadratic (and higher) in perturbations and that contain terms suppressed in the EFT expansion by an assumed large energy scale. (See Ref.~\cite{pereira+al_07} for a gauge-invariant treatment of spatially inhomogeneous perturbations about a Bianchi I spacetime.) We thus consider an effective action for matter degrees of freedom of the form:
\begin{widetext}
\begin{align}
S^{({\cal M})} & = \int d^{4}x \sqrt{-g}\left[-L(t)-c(t)g^{00}-X(t)\left(g^{11}-g^{22}\right)-Y(t)\left(g^{11}-g^{33}\right)\right]\label{SM1}\\
&\equiv \int d^4 x \sqrt{-g} \left[ - L (t) - c (t) g^{00}  - X_{11} (t) g^{11} - X_{22} (t) g^{22} - X_{33} (t) g^{33}  \right] \, ,
\label{SM2}
\end{align}
\end{widetext}
where `${\cal M}$' stands for `matter.' In Eq.~(\ref{SM1}), we have introduced four free functions of time, $L(t), c(t), X(t)$, and $Y(t)$. Whereas only two of these functions, $L (t)$ and $c(t)$, are required (at background order) to account for the dynamics of a homogeneous and isotropic Friedmann-Lema\^{i}tre-Robertson-Walker (FLRW) spacetime \cite{cheung+al_08,Weinberg:2008hq,Azhar:2018nol}, we require two more functions, $X (t)$ and $Y(t)$, for the anisotropic case, given the metric functions $\beta_i$ of Eq.~(\ref{ds}) subject to the constraint of Eq.~(\ref{betasum}). In Eq.~(\ref{SM2}), to more easily keep track of indices, we have introduced the identifications: $X_{11}(t)\equiv X(t)+Y(t)$, $X_{22}(t)\equiv -X(t)$, $X_{33}(t)\equiv -Y(t)$. As a result of these identifications, we have
\begin{align}\label{EQN:Xsum}
X_{11}(t)+X_{22}(t)+X_{33}(t)=0.
\end{align}

The authors of Ref.~\cite{Cannone:2015rra}, in considering backgrounds that perturb away from a 
FLRW background, include comparable terms proportional to the inverse metric in the effective action, and then work to lowest order in the terms that multiply the inverse metric. In our case, we need not consider the anisotropic source terms $X_{ii}$ to be perturbatively small. On the other hand, the authors of Ref.~\cite{Cannone:2015rra} include a term in their effective action proportional to $g^{0i}$ that would be sourced by a nonvanishing momentum flux $T^{0i}$, but, as we will describe shortly, such a term must vanish in general in a spatially homogeneous spacetime, so we neglect it here.

Varying $S^{( { \cal M} )}$ with respect to $g^{\mu\nu}$ yields an effective energy-momentum tensor:
%%%%%%%
\begin{widetext}
\beq
\begin{split}
T_{\mu\nu} &= g_{\mu\nu} \left[ - L(t) - c (t) g^{00} - \tilde{X} (t) \right] - 2 \frac{ \delta}{\delta g^{\mu\nu} } \left[ - L(t) - c (t) g^{00} - \tilde{X} (t) \right] \\
&= g_{\mu\nu} \left[ c (t) - L (t) - \tilde{X} (t) \right] + 2 c (t) \, \delta^0_{\>\> \mu} \delta^0_{\>\> \nu} + 2 \sum_{i} X_{ii} (t) \, \delta^i_{\>\> \mu} \delta^i_{\>\> \nu} \, . \label{Tmn2}
\end{split}
\eeq
\end{widetext}
For convenience, we have introduced the notation
%%%%%%
\beq
\tilde{X}(t) \equiv X_{11} (t) g^{11} + X_{22} (t) g^{22} + X_{33} (t) g^{33} \, .
\label{tildeXdef}
\eeq
We may write Eq.~(\ref{Tmn2}) in the canonical form,
%%%%%%
\beq
T_{\mu\nu} = \rho \, u_\mu u_\nu + P (g_{\mu\nu} + u_\mu u_\nu ) + \pi_{\mu\nu} \, ,
\label{Tmn1}
\eeq
where $\rho$ is the isotropic energy density, $P$ is the isotropic pressure, $\pi_{\mu\nu}$ is the anisotropic pressure, and $u^\mu$ is the effective fluid 4-velocity, which (to background order) takes the form $u^\mu = \delta^\mu_{\>\> 0}$. In this reference frame, the anisotropic pressure is orthogonal to the fluid velocity, $u^\mu \pi_{\mu\nu} = 0$, and hence $\pi_{0 \nu} = 0$ for all $\nu$ (again, to background order, that is, in the absence of spatial inhomogeneities). Likewise, $\pi_{\mu\nu}$ is symmetrical and traceless: $\pi_{\mu\nu} = \pi_{\nu\mu}$ and $\pi^\mu_{\>\> \mu} = 0$, respectively.

Equivalently, we may define the fluid components in terms of the mixed-index energy-momentum tensor (see, for example, Refs.~\cite{bassett+al_06,Lyth:2009zz}):
%%%%%%
\beq
\begin{split}
T^0_{\>\> 0} &= - \rho \, , \\
T^0_{\>\> i} &= \partial_i \delta q \, , \\
T^i_{\>\> j} &=  \delta^i_{\>\> j} P + \pi^i_{\>\> j} \, .
\end{split}
\label{Tmncomponents1}
\eeq
In a homogeneous universe, the spatial gradient of the momentum flux $\partial_i \delta q$ vanishes. This form also makes manifest that the anisotropic pressure $\pi_{ij}$ cannot be proportional to $\delta_{ij}$; if it were, then it would simply contribute to the ``ordinary" (isotropic) pressure $P$. Given that the Einstein tensor associated with the line element in Eq.~(\ref{ds}) has vanishing off-diagonal terms, $G_{ij} = 0$ for $i \neq j$, we expect the off-diagonal components within $\pi_{ij}$ to vanish as well. We therefore expect the anisotropic pressure for the case of interest to include up to three nonvanishing components, $\pi_{11}$, $\pi_{22}$, and $\pi_{33}$, which are not all equal to each other.

In Appendix~\ref{APP:MultipleScalars}, we demonstrate that if the universe were filled {\it only} with scalar fields, each with a canonical coupling to gravity {\it and} with gravity described by the usual Einstein-Hilbert action, then the anisotropic pressure would vanish (at background order, neglecting spatial inhomogeneities). In that case, $\pi_{ij} = 0$ for all $i, j$, and then $T_{ij} \propto g_{ij}$, even for multifield models with nontrivial field-space manifolds. On the other hand, if we consider generalizations beyond canonical general relativity, such as the broader class of Horndeski actions, then there {\it do} arise terms in the effective action that behave like an anisotropic pressure, even for spatially homogeneous spacetimes filled only with scalar fields. (See also Ref.~\cite{Rubakov:2014jja}.) More generally, anisotropic pressures can arise even with canonical general relativity if the universe contains sources of matter with nonzero spin, such as vector fields \cite{Cropp:2010yj}.

Dropping explicit dependences on time, we may write Eq.~(\ref{Tmn2}) as
%%%%%%%
\begin{widetext}
\beq
T_{\mu\nu} = \left( c + L + \tilde{X} \right) u_\mu u_\nu + \left( c - L - \frac{1}{3} \tilde{X} \right) \left( g_{\mu\nu} + u_\mu u_\nu \right) + \pi_{\mu\nu} \, ,
\label{Tmn3}
\eeq
\end{widetext}
with
%%%%%
\beq
\pi_{\mu\nu} \equiv 2 \sum_{i=1}^{3}X_{ii} \left( - \frac{1}{3} h_{\mu\nu} \, g^{ii} + \delta^i_{\> \mu} \, \delta^i_{\> \nu} \right) 
\label{pidef}
\eeq
and $h_{\mu\nu}$ defined in Eq.~(\ref{hmn}).
%%%%%%%
Comparing Eqs.~(\ref{Tmn1}) and (\ref{Tmn3}), we may define 
\beq
\rho \equiv  c + L + \tilde{X}  \,  , \> \> P \equiv c - L - \frac{1}{3} \tilde{X} \, ,
\label{Tmncomponents3}
\eeq
and then, as expected, 
%%%%%%
\beq
\pi_{\mu\nu} = \pi_{\nu\mu} \, , \>\> u^\mu \, \pi_{\mu\nu} = 0 \, , \>\> g^{\mu\nu} \, \pi_{\mu\nu} = 0 \, .
\label{piproperties}
\eeq
We may also introduce the three nonvanishing components of the anisotropic pressure that source the components of the shear tensor:
%%%%%
\beq
p_i \equiv (g_{ii} )^{-1} \, \pi_{ii} = - \frac{2}{3} \tilde{X} + 2 ( g_{ii} )^{-1} \, X_{ii} \>\> \>\> ( {\rm no \>\> sum \>\> on \>\>} i) \, .
\label{pidef2}
\eeq
The tracelessness of $\pi_{\mu\nu}$ translates to the constraint that
%%%%
\beq
p_1 + p_2 + p_3 = 0 \, .
\label{psum}
\eeq
Note that in the isotropic limit, with $X_{ii} = 0$ (for each $i$), $c$ contributes to $\rho$ and $P$ like the kinetic energy of a canonically normalized scalar field, and $L$ contributes like the scalar field's potential $V$ \cite{Azhar:2018nol}.

In order to compare properties of our system with previous investigations of cosmic no-hair theorems, including Wald's work \cite{wald_83}, we next identify salient pointwise energy conditions for the effective matter degrees of freedom (see Refs.~\cite{Hawking:1973uf, wald_83text, visser_95,Visser:1999de,Barcelo:2000zf,Barcelo:2002bv,Rubakov:2014jja,Martin-Moruno:2017exc}). The null energy condition (NEC) may be written (where there is no sum on $i$ in any term)
\beq
\textrm{NEC: } \> \rho + P + p_i \geqslant 0 \equiv c+g^{ii}X_{ii}\geqslant 0. 
\label{EQNnec}
\eeq
The weak energy condition (WEC) becomes
\begin{align}
\textrm{WEC: } \> &\rho\geqslant \;0\;\;\&\;\; \rho + P + p_i \geq 0 \nonumber \\ &\equiv c+L+\tilde{X} \geqslant 0 \;\;\&\;\; c+g^{ii}X_{ii}\geqslant 0. 
\label{EQNwec}
\end{align}
The dominant energy condition (DEC) may be written 
\begin{align}
&\textrm{DEC: } \>  \rho\geqslant  0\;\;\&\;\; \lvert P + p_i \rvert \leqslant \rho  \nonumber \\ & \;\equiv c+L+\tilde{X} \geqslant 0 \;\;\&\;\; 0\leqslant c+g^{ii}X_{ii}\leqslant c+L+\tilde{X}.
\label{EQNdec}
\end{align}
And finally, the strong energy condition (SEC) becomes
\begin{align}
\textrm{SEC: } \> &\rho +3P \geqslant \;0\;\;\&\;\; \rho + P + p_i \geq 0 \nonumber \\ &\equiv 2c-L\geqslant 0 \;\;\&\;\; c+g^{ii}X_{ii}\geqslant 0. 
\label{EQNsec}
\end{align}
As noted in Refs.~\cite{Barcelo:1999hq,Visser:1999de,Barcelo:2000zf,Barcelo:2002bv,Bellucci:2001cc,Dubovsky:2005xd,Nicolis:2009qm,Rubakov:2014jja,Martin-Moruno:2017exc},
%~\cite{Azhar:2018nol}, 
although many powerful theorems in general relativity rely on the assumption of various energy conditions, the pointwise energy conditions in Eqs.~(\ref{EQNnec})--(\ref{EQNsec}) can be violated by scalar fields in curved spacetimes (even at the classical level), without necessarily inducing unphysical instabilities. Therefore we do not impose constraints such as the DEC or SEC on our system {\it a priori}, in contrast to previous studies of isotropization. 

The Einstein field equations yield coupled equations of motion for various dynamical quantities. Given the form of $T_{\mu\nu}$ in Eq.~(\ref{Tmn3}), the generalized Friedmann equations for $H \equiv \dot{a} / a$ take the form
%%%%%%%%
\beq
H^2 = \frac{1}{3 M_{\rm pl}^2} \left( c+ L + \tilde{X} \right) + \frac{1}{3} \sigma^2 \, , 
\label{Friedmann}
\eeq
\beq
\dot{H} + H^2 = - \frac{1}{3 M_{\rm pl}^2} \left( 2c -  L \right) - \frac{2}{3} \sigma^2 \, .
\label{Friedmann2}
\eeq
Each anisotropic scale factor evolves as
%%%%%%%
\beq
\ddot{\beta}_i + 3 H \dot{\beta}_i = \frac{ 1}{ M_{\rm pl}^2} p_i \, ,
\label{betaeom}
\eeq
from which one may evaluate the evolution of the shear scalar
%%%%%%%
\beq
\frac{ d}{dt} \sigma^2 + 6 H \sigma^2 = \frac{1}{ M_{\rm pl}^2} \left( \dot{\beta}_1 p_1 + \dot{\beta}_2 p_2 + \dot{\beta}_3 p_3 \right) \, .
\label{sigmaeom}
\eeq
Finally, the continuity equation $\nabla_\nu \, T^{\nu}_{\>\> \mu} = 0$ yields
%%%%%%%
\beq
\dot{\rho} + 3 H (\rho + P) + \sigma_{\mu\nu} \, \pi^{\mu\nu} = 0 \, .
\label{continuity1}
\eeq
Upon using Eqs.~(\ref{pidef}) and (\ref{Tmncomponents3}), this expression may be written as
%%%%%%
\beq
\dot{c} + \dot{L} + \dot{\tilde{X}} + 3 H \left( 2c + \frac{2}{3} \tilde{X} \right) + \dot{\beta}_1 p_1 + \dot{\beta}_2 p_2 + \dot{\beta}_3 p_3 = 0 \, .
\label{continuity}
\eeq
One may readily check that the two Friedmann equations in Eqs.~(\ref{Friedmann}) and (\ref{Friedmann2}) are mutually consistent upon using Eq.~(\ref{continuity}). Upon subtracting Eq.~(\ref{Friedmann}) from Eq.~(\ref{Friedmann2}), we can solve for $\dot{H}$, and from that expression we can solve for the Hubble slow-roll parameter $\epsilon$:
%%%%
\beq
\epsilon \equiv - \frac{ \dot{H} }{H^2} = \frac{1}{ 3 M_{\rm pl}^2 H^2} \left( 3 c + \tilde{X} \right) + \frac{ \sigma^2}{ H^2} \, .
\label{eps1}
\eeq
As expected, this expression reduces to the usual case, in which $\epsilon \propto c$ \cite{Azhar:2018nol}, when anisotropic degrees of freedom vanish: $\tilde{X} , \sigma^2 \rightarrow 0$.

Much as in Wald's analysis \cite{wald_83}, we restrict attention to scenarios in which the initial extrinsic curvature is positive, corresponding to volume-averaged expansion of space. For the line element of Eq.~(\ref{ds}), the lapse function $N (x^\mu) = 1$ and, since $g_{0i} = 0$, the shift vector vanishes, so the extrinsic curvature takes the simple form \cite{wald_83text,Cropp:2010yj}
%%%%%
\beq
K_{ij} = \frac{1}{2} \partial_t h_{ij} = (H + \dot{\beta}_i) h_{ij} ,
\label{Kij}
\eeq
where $h_{ij} = a^2 (t) \delta_{ij} e^{2 \beta_i (t)}$ is the metric on equal-time spatial slices. Then we find
%%%%
\beq
K = h^{ij} K_{ij} = 3  H + \sum_i \dot{\beta}_i  = 3 H ,
\label{Ktrace}
\eeq
where the final expression follows upon using the time derivative of the constraint in Eq.~(\ref{betasum}). Hence we see that, much as for an FLRW spacetime, $K > 0$ corresponds to the case of (volume-averaged) spatial expansion, $H > 0$, upon which we will focus in our analysis.

\subsection{New variables}

We now introduce a change of variables that will allow us to rewrite the Einstein equations expressed in Eqs.~(\ref{Friedmann}),~(\ref{Friedmann2}), and~(\ref{betaeom}), as well as the continuity equation in Eq.~(\ref{continuity}), in a form that makes them amenable to a useful dynamical analysis. (See also Ref.~\cite{vandenhoogen+coley_95} for a complementary analysis.) 

First note that the shear tensor, $\sigma_{\alpha\beta}$, is traceless: $\sigma^{\mu}_{\;\;\mu} =0$. This reduces the three non-zero (purely spatial) components of $\sigma_{\alpha\beta}$ to two independent components. We can represent these two independent components with the following new variables:
\begin{align}
\sigma_{1} &\equiv \sigma^{1}_{\;\;1}-\sigma^{2}_{\;\;2},\\
\sigma_{2} &\equiv \sigma^{1}_{\;\;1}-\sigma^{3}_{\;\;3}.
\end{align}
For the Bianchi I background in Eq.~(\ref{ds}), these are given by
\begin{align}
\sigma_{1} &= \dot{\beta}_{1}-\dot{\beta}_{2},\\
\sigma_{2} &= \dot{\beta}_{1}-\dot{\beta}_{3}.
\end{align}
One may now show---using the (derivative of the) constraint on the sum of the $\beta_i$'s in Eq.~(\ref{betasum})---that the quantity we had previously used to signal the degree of anisotropy of the background space, namely $\sigma^2$, can be written as
\begin{align}\label{EQN:sigSqnew}
\sigma^2 = \frac{1}{3}( \sigma_1+\sigma_2 )^2-\sigma_1 \sigma_2.
\end{align}

A similar procedure can be implemented for the anisotropic pressure $\pi_{\mu\nu}$, since it is also traceless, symmetric, and purely spatial---which means it has just two independent components. We thus define
\begin{align}
\pi_{1} &\equiv \pi^{1}_{\;\;1}-\pi^{2}_{\;\;2},\\
\pi_{2} &\equiv \pi^{1}_{\;\;1}-\pi^{3}_{\;\;3}.
\end{align}
One finds, using Eqs.~(\ref{pidef}) and~(\ref{pidef2}), that $\pi^{i}_{\;\;i}=p_i$ (no sum on $i$). These new variables thus each encode differences in the anisotropic pressures: 
\begin{align}
\pi_{1} &= p_1-p_2,\label{pi1}\\
\pi_{2} &= p_1-p_3. \label{pi2}
\end{align}

In describing the Einstein field equations with the variables $\{\sigma_1, \sigma_2, \pi_1, \pi_2\}$, one may replace the evolution equations for the $\beta_i$'s in Eq.~(\ref{betaeom}) with equations of evolution for the $\sigma_A$'s, with $A = 1, 2$. Indeed, employing  Eq.~(\ref{betaeom}), one finds
\begin{align}
\dot{\sigma}_A & = -3H\sigma_A +\frac{1}{\Mpl^2}\pi_A.\label{EQN:dsig1}
\end{align}
Finally, the continuity equation of Eq.~(\ref{continuity}) can be rewritten in terms of these new variables by using the constraint on the sum of the $p_i$'s in Eq.~(\ref{psum}), and the derivative of the constraint on the sum of the $\beta_i$'s in Eq.~(\ref{betasum}). We find %
\begin{align}\label{EQN:betapinew}
\dot{\beta}_1 p_1+\dot{\beta}_2 p_2+\dot{\beta}_3 p_3 = \frac{1}{3}\left[\sigma_1\left(2\pi_1-\pi_2\right)+\sigma_2\left(2\pi_2-\pi_1\right)\right]. 
\end{align}
Using Eqs.~(\ref{EQN:sigSqnew}),~(\ref{EQN:dsig1}), and~(\ref{EQN:betapinew}), we may then rewrite the coupled dynamical equations as
\begin{widetext}
\begin{align}\label{EQN:FR1vc}
H^2 &= \frac{1}{3\Mpl^2}\left(c+L+\tilde{X}\right)+\frac{1}{9}(\sigma_1+\sigma_2)^2-\frac{1}{3}\sigma_1\sigma_2,\\  
\dot{H}+H^2 &=-\frac{1}{3\Mpl^2}(2c-L)-\frac{2}{9}(\sigma_1+\sigma_2)^2
+\frac{2}{3}\sigma_1\sigma_2,\label{EQN:FR2vc}\\
\dot{\sigma}_1 & = -3H\sigma_1 +\frac{1}{\Mpl^2}\pi_1,\label{EQN:dsig1vc}\\
\dot{\sigma}_2 & = -3H\sigma_2 +\frac{1}{\Mpl^2}\pi_2,\label{EQN:dsig2vc}\\
0&=\dot{c}+\dot{L}+\dot{\tilde{X}}+2 H \left(3c+\tilde{X}\right)+ \frac{1}{3}\left[\sigma_1\left(2\pi_1-\pi_2\right)+\sigma_2\left(2\pi_2-\pi_1\right)\right].
\label{EQN:consvc}
\end{align}
\end{widetext}

\subsection{A dynamical analysis}\label{SEC:DynAnalysis}

Equations (\ref{EQN:FR1vc})--(\ref{EQN:consvc}) suggest expansion-normalized, dimensionless dynamical variables, which yield a dynamical system with some useful features for our analysis of isotropization of homogeneous spaces. In particular, one may extract a closed dynamical system in which one can identify (i) de Sitter evolution of the background space and (ii) necessary and sufficient conditions for the satisfaction of each energy condition, purely in terms of the dynamical variables. This latter feature helps to facilitate a comparison with previous work, including Wald's theorem \cite{wald_83}.

Following Ref.~\cite{Azhar:2018nol} (see also Ref.~\cite{frusciante+al_14}) we define the dimensionless variables
\begingroup
\allowdisplaybreaks
\begin{align}
x &\equiv \frac{c+L+\tilde{X}}{ 3 \Mpl^2  H^2 },\label{EQN:x}\\
y &\equiv \frac{3c+\tilde{X}}{ 3 \Mpl^2  H^2 }\, , \qquad \lambda_l \equiv - \frac{ (3c+\tilde{X})^{(l + 1)} }{ H (3c+\tilde{X})^{(l)} }\, , \label{EQN:ylambda} \\
\omega_1 &\equiv \frac{\pi_1}{ 3 \Mpl^2  H^2 }\, , \qquad \mu_m \equiv - \frac{\pi_1^{(m + 1)} }{ H \pi_1^{(m)} }\, \label{EQN:ommu},\\
\omega_2 &\equiv \frac{\pi_2}{ 3 \Mpl^2  H^2 }\, , \qquad \nu_n \equiv - \frac{\pi_2^{(n + 1)} }{ H \pi_2^{(n)} }\, ,\\
z_1&\equiv\frac{\sigma_1}{3 H},\\
z_2&\equiv\frac{\sigma_2}{3 H}.\label{EQN:z2}
\end{align}
\endgroup
Here $(m)$ denotes the $m$th derivative with respect to cosmic time $t$, and $l,m,n \geq 0$. The expressions for $\lambda_l$, $\mu_m$, and $\nu_n$ introduce infinite towers of dimensionless variables that encode implicit choices for the functional dependences of $3c+\tilde{X}$, $\pi_1$, and $\pi_2$ (respectively) on $t$---though in practice one need only consider a finite number of such variables for a given phase-space analysis. Note further that in light of Eq.~(\ref{Tmncomponents3}), we may write 
\begin{align}
x &= \frac{\rho}{ 3 \Mpl^2  H^2 },\label{EQN:xInterp}\\
y &= \frac{\rho+P}{ 2 \Mpl^2  H^2 }.\label{EQN:yInterp}
\end{align}
Thus we can interpret $x$ as an expansion-normalized isotropic (effective) energy density, and $y$ as a normalized sum of the isotropic energy density and the isotropic pressure. As such, in a purely isotropic spacetime, the vanishing of $y$ corresponds to a background with a constant energy density---that is, effectively, to a background with a cosmological constant. 

One may also derive a set of coupled, ordinary differential equations with respect to which we may perform a dynamical-systems analysis. We find
\begingroup
\allowdisplaybreaks
\begin{align}
\frac{d x}{d\ln a} = 2 \epsilon x - 2 y & - z_1 (2 \omega_1-\omega_2)-z_2 (2 \omega_2-\omega_1),\label{EQN:DEsN1}\\
\frac{d y}{d\ln a} &= 2\epsilon y - \lambda_0 y,\label{EQN:DEsN2}\\
\frac{d \omega_1}{d\ln a} &= 2\epsilon \omega_1 - \mu_0 \omega_1, \label{EQN:domega1}\\
\frac{d \omega_2}{d\ln a} &= 2\epsilon \omega_2 - \nu_0 \omega_2, \label{EQN:domega2}\\
\frac{d z_1}{d\ln a} &= z_1(\epsilon-3)+\omega_1, \label{EQN:dz1}\\
\frac{d z_2}{d\ln a} &= z_2(\epsilon-3)+\omega_2, \label{EQN:dz2}\\
\frac{d \lambda_l}{d\ln a} &= \lambda_l\left(-\lambda_{l+1}+\lambda_{l}+\epsilon\right), \label{EQN:lamT}\\
\frac{d \mu_m}{d\ln a} &= \mu_m\left(-\mu_{m+1}+\mu_{m}+\epsilon\right), \label{EQN:muT}\\
\frac{d \nu_n}{d\ln a} &= \nu_n\left(-\nu_{n+1}+\nu_{n}+\epsilon\right),\label{EQN:DEsNend}
\end{align}
\endgroup
for $l,m,n \geq 0$ and where the Hubble slow-roll parameter, $\epsilon$, described in Eq.~(\ref{eps1}), is given by
\begin{align}
\epsilon \equiv - \frac{\dot{H}}{H^2} = y + 3 (z_1+ z_2)^2 - 9 z_1 z_2.\label{EQN:newep}
\end{align}
Furthermore, there is  a constraint, derived from the first Friedmann equation, Eq.~(\ref{EQN:FR1vc}):
\begin{align}\label{EQN:Co}
1=x+(z_1+z_2)^2-3 z_1 z_2.
\end{align}

Equations~(\ref{EQN:DEsN1})--(\ref{EQN:Co}) do not form a closed system due to the three infinite towers in Eqs.~(\ref{EQN:lamT})--(\ref{EQN:DEsNend}), but we can construct a closed system by fixing $\lambda_L, \mu_M, \nu_N$ to be constants for some fixed $L, M, N\geq 0$. In such a case, the effective phase space will naturally be coordinatized by the variables $x,y,\omega_1, \omega_2, z_1, z_2$, together with  $ \lambda_0, \lambda_1, \dots, \lambda_{L-1}, \mu_0, \mu_1, \dots, \mu_{M-1}, \nu_0, \nu_1, \dots, \nu_{N-1}$ (when $L, M, N > 0$). Taking into account the constraint in Eq.~(\ref{EQN:Co}), the dynamics will thus be $(6+L+M+N-1)$-dimensional. 

The dynamics can be further simplified because there exist invariant manifolds for any choice of $(L, M, N)$. For example, it is straightforward to show that the constraint surface is an invariant manifold because $d\left[x+(z_1+z_2)^2-3 z_1 z_2-1\right]/d\ln a = 2 \epsilon \left[x+(z_1+z_2)^2-3 z_1 z_2-1\right] = 0$, given Eq.~(\ref{EQN:Co}). Hence the system does not move off the constraint surface if it begins on it. Similarly, the following surfaces are also invariant manifolds: $y=0$; $\omega_A=0$, for $A=1,2$; $\lambda_l=0$, for $l = 0, 1, \dots, L-1$; $\mu_m=0$, for $m = 0, 1, \dots, M-1$; and $\nu_n=0$, for $n = 0, 1, \dots, N-1$.

For any particular phase space---defined by fixing, for some $(L, M, N)$, $\lambda_L, \mu_M$, and $\nu_N$ to be constants---the dynamics in that phase space are consistent with
\begin{align}
{(3c+\tilde{X})}^{(L)}(t) & = {(3c+\tilde{X})}^{(L)}(t_i)\left[\frac{a(t_i)}{a(t)}\right]^{\lambda_L},\label{EQN:3cXred}\\
\pi_1^{(M)}(t) & = \pi_1^{(M)}(t_i)\left[\frac{a(t_i)}{a(t)}\right]^{\mu_M},\label{EQN:pi1red}\\
\pi_2^{(N)}(t) & = \pi_2^{(N)}(t_i)\left[\frac{a(t_i)}{a(t)}\right]^{\nu_N}.\label{EQN:pi2red}
\end{align}
Here $t_i$ is some fixed initial time. Thus, for example, setting $\mu_M$ to be a constant corresponds to assuming that the $M$th time derivative of $\pi_1(t)$ scales as $[a (t) ]^{- \mu_M}$.

Note also that for any phase space one can uniquely identify when the background is isotropic. In particular (as shown in Appendix~\ref{APP:uniqueISO}) assuming a finite Hubble expansion rate $H$:
\beq\label{EQN:isoCOND}
\textrm{Background is isotropic}\iff z_1=0=z_2,
\eeq
since the variables $z_A$ (for $A=1,2$) are proportional to the independent components of the shear tensor $\sigma_{\alpha \beta}$.
Furthermore, we can identify when such a background undergoes de Sitter expansion: for $z_A=0$, the Hubble slow-roll parameter becomes $\epsilon = y$, which vanishes whenever $y=0$. Thus the background undergoes de Sitter evolution whenever $y=z_1=z_2=0$.  

One may also now express the energy conditions described in Sec.~\ref{SEC:EFTdev} purely in terms of the dynamical variables introduced in Eqs.~(\ref{EQN:x})--(\ref{EQN:z2}). The NEC amounts to the condition that, for each $i$,
\begin{align}
c+g^{ii}X_{ii}\geqslant 0 \iff \frac{c+g^{ii}X_{ii}}{3\Mpl^2 H^2}\geqslant 0.
\label{EQN:NEC2}
\end{align}
The three conditions on the right-hand side of Eq.~(\ref{EQN:NEC2}) can be expressed in terms of the dynamical variables described above, so that the NEC is satisfied iff each of the following three conditions is satisfied:
\begin{equation}
\textrm{NEC}
    \begin{cases}
	2y+\omega_1+\omega_2 &\geqslant 0,\\
	2y-2 \omega_1 + \omega_2&\geqslant 0,\\
	2y+\omega_1 -2\omega_2&\geqslant 0.
    \end{cases}       
\end{equation}
Similarly, the WEC is satisfied iff each condition for the NEC is satisfied, together with one extra condition:
\begin{equation}
\textrm{WEC: NEC \&}\;x \geqslant 0.
\end{equation}
The DEC is satisfied iff each of the following four conditions is satisfied:
\begin{equation}
\textrm{DEC}
    \begin{cases}
	6x&\geqslant 2y+\omega_1+\omega_2  \geqslant 0,\\
	6x &\geqslant2y-2 \omega_1 + \omega_2 \geqslant 0,\\
	6x &\geqslant2y+\omega_1 -2\omega_2  \geqslant 0,\\
	x &\geqslant 0.    
    \end{cases}       
\end{equation}
Finally, the SEC is satisfied iff each condition for the NEC is satisfied, together with one extra condition:
\begin{equation}
\textrm{SEC: NEC \&}\;y \geqslant x.
\end{equation}

\subsection{The principal phase space}\label{SEC:PPS}

The dynamical system described in the previous subsection contains a countably infinite number of possible phase spaces, depending on choices for the non-negative integers $(L, M, N)$. In this paper we will focus on the simplest (lowest-dimensional) phase space, which we will refer to as the {\it principal phase space}. This phase space is found by choosing  $(L, M, N)=(0, 0, 0)$ and thereby arises by fixing $\lambda_0, \mu_0$, and $\nu_0$ to be constants. Of course, if one chooses different nonnegative values for $(L, M, N)$, one will obtain phase spaces of higher dimensionality, and we expect that such phase spaces will contain richer dynamical flows. There is a trade-off, however, in that nonzero values for $(L, M, N)$ lead to larger phase spaces and hence expanded parameter spaces to try to characterize. Moreover, we found in our previous analysis \cite{Azhar:2018nol} (for the case of homogeneous and isotropic backgrounds) that each of the fixed points within the principal phase space also appeared as fixed points in the higher-dimensional generalizations, even as the total number of fixed points grew with the dimensionality of the phase space. Hence we focus on an analysis of the fixed points within the principal phase space in the present work and defer analysis of higher-dimensional generalizations to future work.

In addition to assuming $H > 0$, we also restrict attention in what follows to fixed points at which $\epsilon > 0$. (In FLRW spacetimes, $\epsilon < 0$ can yield a so-called ``big rip" singularity \cite{starobinsky_00, caldwell_02,caldwell+al_03}.) As we will see below, once we fix $(L, M, N) = (0, 0,0)$ to construct our principal phase space, the slow-roll parameter $\epsilon$ when evaluated at various fixed points is proportional to $\lambda_0, \mu_0$, or $\nu_0$. Furthermore, setting $\lambda_0, \mu_0$, or $\nu_0$ equal to zero would collapse various fixed points to the (isotropic) de Sitter fixed point, which would artificially favor flows to de Sitter at late times. Hence we consider $\lambda_0, \mu_0$, and $\nu_0$ to be positive.

For the principal phase space, the values of $\lambda_0, \mu_0$, and $\nu_0$ determine the rate at which various combinations of parameters decay with time, in accord with Eqs.~(\ref{EQN:3cXred}),~(\ref{EQN:pi1red}), and~(\ref{EQN:pi2red}). In particular $|3c+\tilde{X}|\sim a(t)^{-\lambda_0}$, $|\pi_1|\sim a(t)^{-\mu_0}$, and $|\pi_2|\sim a(t)^{-\nu_0}$, respectively. The latter two equations dictate the rate at which anisotropic pressures decay as the background expands; cf.~Eqs.~(\ref{pi1}) and~(\ref{pi2}) with Eq.~(\ref{psum}). If, in addition to such a decay of anisotropic pressures, one has that $|3c+\tilde{X}|\sim a(t)^{-\lambda_0}$, we see from Eq.~(\ref{EQN:consvc}) that
$\dot{\rho} = \dot{c}+\dot{L}+\dot{\tilde X} \to 0$. That is, the system is driven towards an energy-density dominated state, where this energy density is constant in time. Such a scenario corresponds to the emergence of an effective cosmological constant. So we can think of $\mu_0$ and $\nu_0$ as determining the rate of decay of relevant anisotropic degrees of freedom; a sufficiently large (positive) value of $\lambda_0$, in turn, can create conditions for an effective cosmological constant. 

Equations of evolution for trajectories in this phase space are given by
\begingroup\label{EQN:DEsPPS}
\allowdisplaybreaks
\begin{align}
\frac{d x}{d\ln a} = 2 \epsilon x - 2 y &- z_1 (2 \omega_1-\omega_2)-z_2 (2 \omega_2-\omega_1),\label{EQN:DEsN1PPS}\\
\frac{d y}{d\ln a} &= 2\epsilon y - \lambda_0 y,\label{EQN:DEsN2PPS}\\
\frac{d \omega_1}{d\ln a} &= 2\epsilon \omega_1 - \mu_0 \omega_1,\label{EQN:DEsN3PPS}\\
\frac{d \omega_2}{d\ln a} &= 2\epsilon \omega_2 - \nu_0 \omega_2,\label{EQN:DEsN4PPS}\\
\frac{d z_1}{d\ln a} &= z_1(\epsilon-3)+\omega_1,\label{EQN:DEsN5PPS}\\
\frac{d z_2}{d\ln a} &= z_2(\epsilon-3)+\omega_2 \label{EQN:DEsN6PPS}
\end{align}
\endgroup
where the equation for the Hubble slow-roll parameter, Eq.~(\ref{EQN:newep}), continues to hold, as does the constraint  in Eq.~(\ref{EQN:Co}).

\begin{figure*}[htb]
            \includegraphics[width=0.475\textwidth]{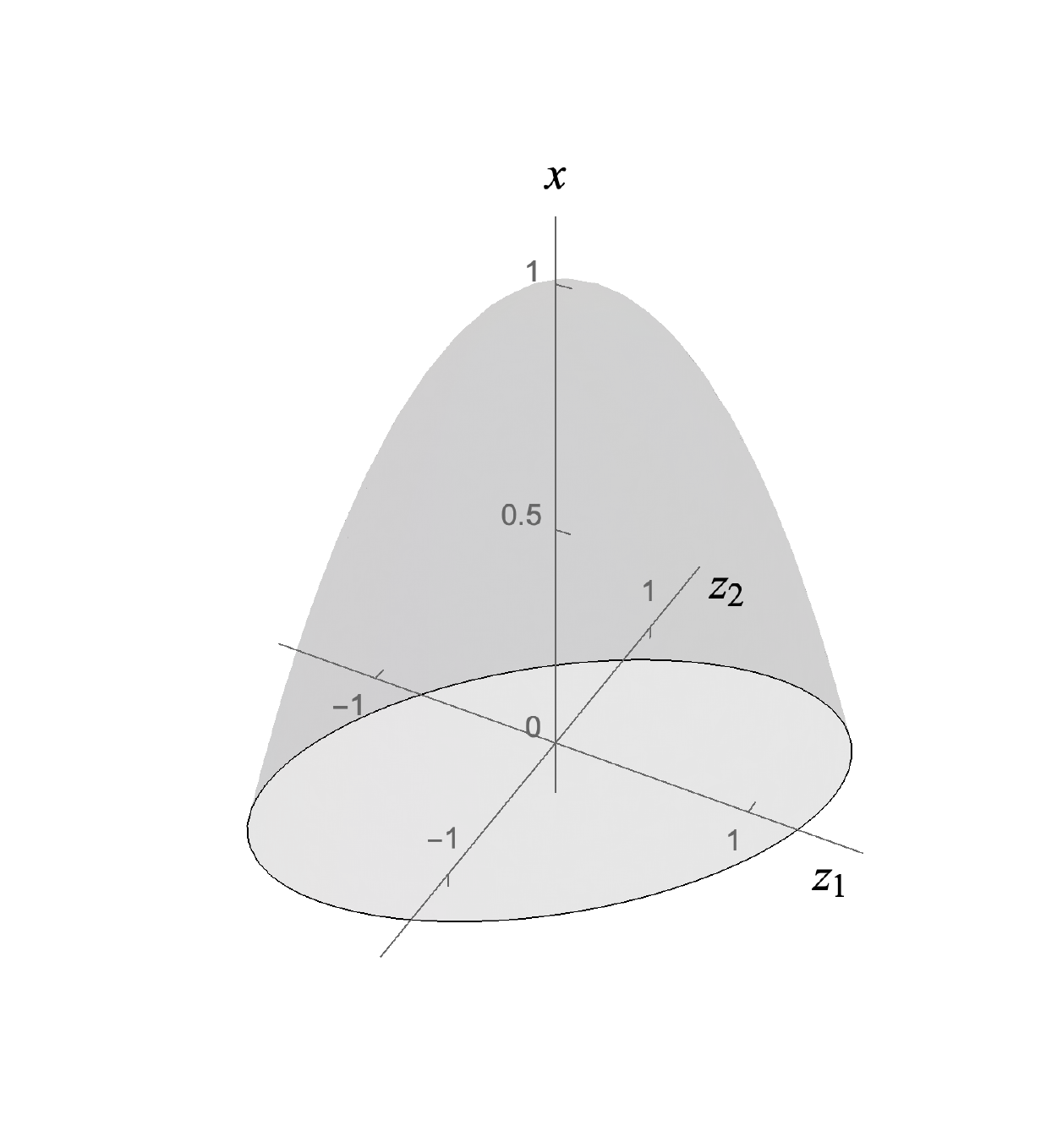}
        \hfill
            \includegraphics[width=0.47\textwidth]{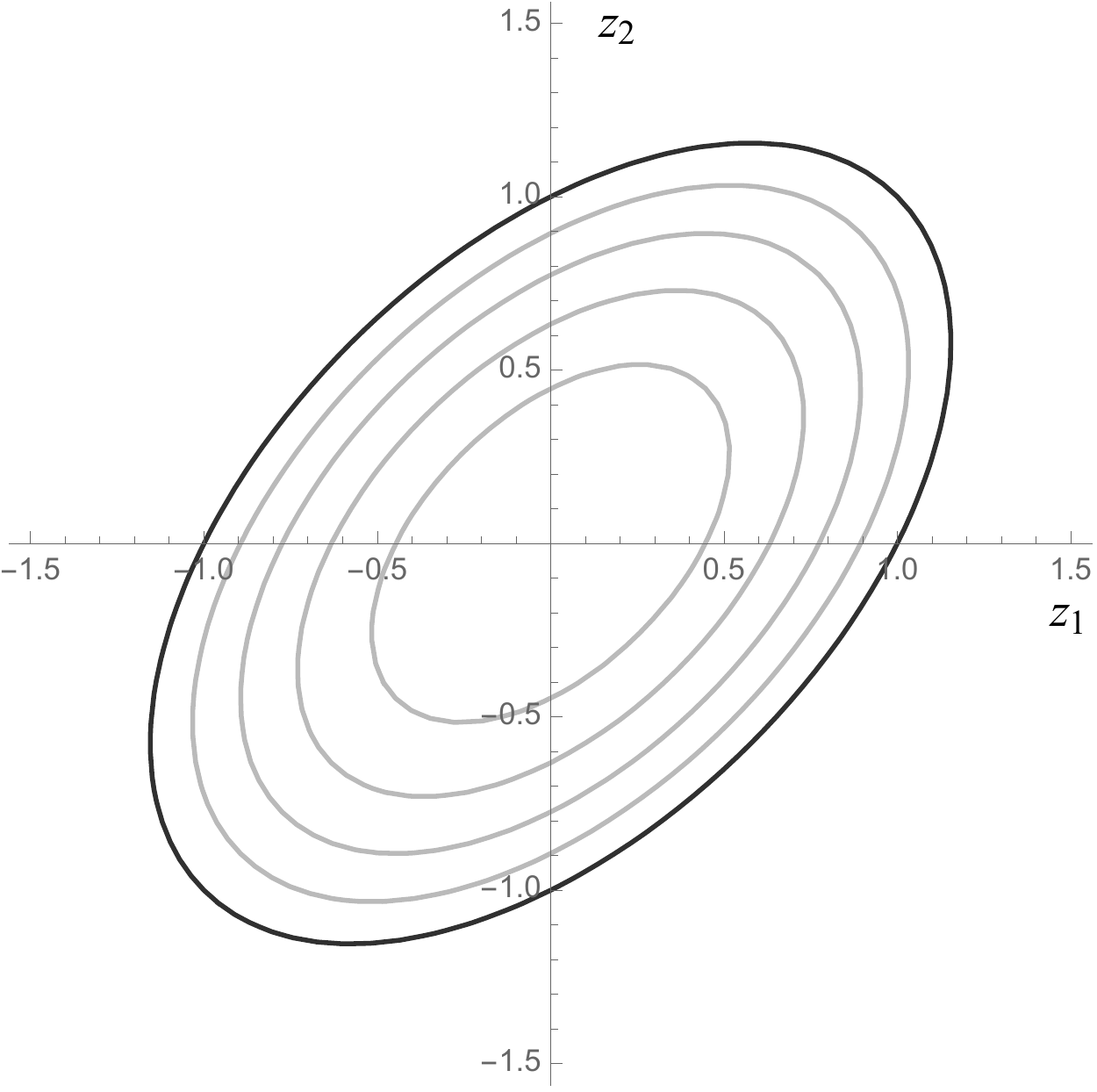}
        \caption[]
        {\small (Left) The constraint surface (upside down elliptic paraboloid in gray) for $0 \leq x \leq 1$. (Right) The projection of the constraint surface onto the $z_1$-$z_2$ plane. The outermost ellipse (in black) corresponds to the contour where $x=0$. The contours in gray correspond to $x=0.2, 0.4, 0.6$, and $0.8$ (moving inwards, towards the origin, from the black ellipse).} 
        \label{FIG:ZOPSpace}
    \end{figure*}

One may picture the five-dimensional principal phase space---formed from the six directions $\{x,y, \omega_1, \omega_2, z_1, z_2\}$ with the constraint of Eq.~(\ref{EQN:Co})---as a product of two spaces (essentially $\mathbb{R}^2\times\mathbb{R}^3$): one (constrained) space coordinatized by $(z_1,z_2, x)$ with the other space coordinatized by $(\omega_1, \omega_2, y)$. Note that the constraint surface is described purely in terms of $(z_1,z_2, x)$ whereas the energy conditions largely reference $(\omega_1, \omega_2, y)$. We will refer to the space coordinatized by $(\omega_1, \omega_2, y)$ as the ``internal space." The constraint surface is depicted in Fig.~\ref{FIG:ZOPSpace} (left) and the projection of this surface onto the $z_1$-$z_2$ plane is depicted in Fig.~\ref{FIG:ZOPSpace} (right). At each point of the constraint surface, one may envision three further (``internal") directions coordinatized by $(\omega_1, \omega_2, y)$.

\subsubsection{Fixed points}

Dynamical flows in the principal phase space are naturally studied  by considering fixed points in this space. One finds these fixed points by setting the left-hand sides of Eqs.~(\ref{EQN:DEsN1PPS})--(\ref{EQN:DEsN6PPS}) to zero, subject to the constraint of Eq.~(\ref{EQN:Co}).

One can more efficiently calculate the fixed points by considering two distinct cases, namely (i) $x=0$ and (ii) $x\neq 0$. The former case leads to a one-parameter set of fixed points that is consistent with the {\it Kasner circle}, a one-parameter set of {\it anisotropic} vacuum solutions of the Einstein field equations first described by~\citet{kasner_21}.\footnote{These are solutions for which the energy density vanishes and yet the background has anisotropies that are ``frozen in." We do not believe these anisotropies can be removed by a coordinate transformation, though $\sigma^2$ does decrease quadratically with time, as we show in Appendix~\ref{APP:Kasner}.} (See also Ref.~\cite[\S 30.2]{misner+al_73}.) In what follows we will compute the fixed points explicitly for the case $x=0$, and then simply quote results for the case with $x \neq 0$. 

When $x=0$, the constraint of Eq.~(\ref{EQN:Co}) becomes
\begin{align}\label{EQN:CoNew}
1=(z_1+z_2)^2-3 z_1 z_2,
\end{align}
and this implies that the Hubble slow-roll parameter, $\epsilon$, is given by 
\begin{align}\label{EQN:epNew}
\epsilon=y+3(z_1+z_2)^2-9 z_1 z_2 = y+3.
\end{align}
To find the fixed points we need to solve:
\begingroup
\allowdisplaybreaks
\begin{align}
0 &= - 2 y - z_1 (2 \omega_1-\omega_2)-z_2 (2 \omega_2-\omega_1),\label{EQN:FPx}\\
0 &= 2\epsilon y - \lambda_0 y,\label{EQN:FPy}\\
0 &= 2\epsilon \omega_1 - \mu_0 \omega_1,\label{EQN:FPw1}\\
0 &= 2\epsilon \omega_2 - \nu_0 \omega_2,\label{EQN:FPw2}\\
0 &= z_1(\epsilon-3)+\omega_1,\label{EQN:FPz1}\\
0 &= z_2(\epsilon-3)+\omega_2.\label{EQN:FPz2}
\end{align}
\endgroup
Equation~(\ref{EQN:FPy}) together with Eq.~(\ref{EQN:epNew}) yields
\begin{align}
0 & = 2(y+3)y-\lambda_0 y = y(2y+6-\lambda_0)\nonumber\\ & \implies y=0\;\;{\rm or}\;\;y = \frac{\lambda_0-6}{2}.
\end{align}
One can show numerically that there is no solution to the system of equations of Eqs.~(\ref{EQN:FPx})--(\ref{EQN:FPz2}) when $y\neq 0$, so we proceed under the assumption that $y=0$. When $y=0$ we have, from Eq.~(\ref{EQN:epNew}), that $\epsilon =3$. This means that Eqs.~(\ref{EQN:FPz1}) and~(\ref{EQN:FPz2}) are satisfied (indeed, for any value of $z_1$ and $z_2$) iff $\omega_1=0=\omega_2$. Equations~(\ref{EQN:FPx}),~(\ref{EQN:FPw1}), and~(\ref{EQN:FPw2}) are then automatically satisfied. Thus, when $x=0$ we find that the fixed points (which we will refer to as {\bf FPK} for ``fixed point Kasner") are given by
\begin{align}
%\label{EQN:FPs1}
{\bf FPK:}\;\;x&=y=\omega_1=\omega_2=0 \;\;\&\;\; \nonumber\\ 1 &= (z_1+z_2)^2-3 z_1 z_2.
\end{align}
In Appendix~\ref{APP:Kasner} we derive the metric that results from assuming these values for the dynamical variables, showing that they are consistent with Einstein field equations whose solution is given by the {\it Kasner metric}. 

Below, we will separate the fixed points {\bf FPK} into two distinct sets of fixed points, {\bf FPK}$^-$ and {\bf FPK}$^+$. The former set corresponds to the lower portion of the ellipse traced out in the $z_1$-$z_2$ plane by $1 = (z_1+z_2)^2-3 z_1 z_2$, while the latter corresponds to the upper portion of the ellipse. We consider each such set to be given via a function, $z_2(z_1)$, where $z_1$ takes values in a prescribed domain. We collect all fixed points (including those for which $x\neq 0$) in Table~\ref{TAB:FPs}.
\begin{table*}
\begin{center}
 \begin{tabular}{ l | c | c | c | c | c | c || c }
 {\it Label}  & $z_1$ & $z_2$ & $x$  & $\omega_1$ & $\omega_2$ & $y$ & $\epsilon$\\ [1ex] 
 \Xhline{2pt}
{\bf dS}  & $0$ & $0$ & $1$  & $0$ & $0$ & $0$ & $0$\Tstrut \\ [0.5ex] 
{\bf FPb}  & $0$ & $0$ & $1$  & $0$ & $0$ & $\frac{\lambda_0}{2}$ & $\frac{\lambda_0}{2}$\\ [0.5ex] 
{\bf FPc} & $0$ & $-\sqrt{\frac{\nu_0}{6}}$ & $1 - \frac{\nu_0}{6}$  & $0$ & $\frac{1}{2}\sqrt{\frac{\nu_0}{6}}(\nu_0-6)$ & $0$ & $\frac{\nu_0}{2}$ \\ [0.5ex] 
{\bf FPd}  & $0$ & $\sqrt{\frac{\nu_0}{6}}$ & $1 - \frac{\nu_0}{6}$  & $0$ & $-\frac{1}{2}\sqrt{\frac{\nu_0}{6}}(\nu_0-6)$ & $0$ & $\frac{\nu_0}{2}$ \\ [0.5ex] 
{\bf FPe} & $-\sqrt{\frac{\mu_0}{6}}$ & $0$ & $1 - \frac{\mu_0}{6}$   & $\frac{1}{2}\sqrt{\frac{\mu_0}{6}}(\mu_0-6)$ & $0$ & $0$ &  $\frac{\mu_0}{2}$ \\ [0.5ex] 
{\bf FPf}  & $\sqrt{\frac{\mu_0}{6}}$ & $0$ & $1 - \frac{\mu_0}{6}$   & $-\frac{1}{2}\sqrt{\frac{\mu_0}{6}}(\mu_0-6)$ & $0$ & $0$ & $\frac{\mu_0}{2}$ \\ [0.5ex]  \hline
{\bf FPK}$^-$  & $[-\frac{2}{\sqrt{3}},\frac{2}{\sqrt{3}})$ & $\frac{1}{2}(z_1 - \sqrt{4 - 3 z_1^2})$ & $0$  & $0$ & $0$ & $0$ & $3$\Tstrut\\
{\bf FPK}$^+$  & $(-\frac{2}{\sqrt{3}},\frac{2}{\sqrt{3}}]$ & $\frac{1}{2}(z_1 + \sqrt{4 - 3 z_1^2})$ & $0$  & $0$ & $0$ & $0$ & $3$\Tstrut
 \end{tabular}
 \caption{Fixed points in the principal phase space. There is a single (isotropic) de Sitter fixed point, {\bf dS}, as well as another isotropic fixed point {\bf FPb}. The fixed points corresponding to the Kasner circle, {\bf FPK}, have been separated into two one-parameter families of fixed points, labeled {\bf FPK}$^-$ and {\bf FPK}$^+$.}
 \label{TAB:FPs}
 \end{center}
 \end{table*}

Note that there is a single fixed point that corresponds to de Sitter expansion of the background, which we label {\bf dS}. There is a second fixed point that corresponds to isotropic expansion, {\bf FPb}, which (for $\lambda_0 \neq 0$) does not correspond to de Sitter expansion of the background. The isotropic fixed point {\bf FPb} has the same properties as those identified as fixed point {\bf FP0b} in Ref.~\cite{Azhar:2018nol}, for the case of dynamical flows within homogeneous and isotropic background spacetimes. In particular, for constant $\lambda_0 > 0$, the (volume-averaged) scale factor $a(t)$ grows as $a(t) = a (t_i) (t / t_i)^p$ with $p = 2/\lambda_0$, which yields accelerated expansion for $0 < \lambda_0 < 2$. Such dynamics can be realized in canonical single-scalar-field models with an effective potential of the form \cite{Azhar:2018nol}
%%%
\beq
V (\phi) = V_0 \exp \left[ - \alpha \left(\frac{ \phi}{M_{\rm pl} } \right)^\beta \right]
\label{Veff}
\eeq
with $\alpha = \sqrt{\lambda_0}$ and $\beta = 1$.

The remaining fixed points (for arbitrary positive values of $\mu_0$ and $\nu_0$) each correspond to anisotropic expansion. In Appendix \ref{APP:dsAnisotropic} we consider the spacetime line elements associated with each of these anisotropic fixed points.

\subsubsection{Stability properties of fixed points}

One can probe the stability properties of the fixed points by considering the entire (unconstrained) phase space, or by focusing only on the region of the phase space in which the constraint of Eq.~(\ref{EQN:Co}) is satisfied. In what follows, we will focus on the latter case, as the constraint surface is an invariant manifold and all trajectories we will study will satisfy the constraint for all time. In this case, there is a straightforward argument (for any dynamical system derived from our framework) to the effect that we can simply ignore the dynamical variable $x$ and focus on the other variables. More specifically, on the constraint surface, we can rearrange the constraint of Eq.~(\ref{EQN:Co}) such that
\begin{align}\label{EQN:CoARR}
x=1-\left[(z_1+z_2)^2-3 z_1 z_2\right].
\end{align}
The dynamical equations for the remaining variables $\{y, \omega_1, \omega_2, z_1, z_2\}$ are indeed {\it closed}:
\begingroup\label{EQN:DEsN1constraint}
\allowdisplaybreaks
\begin{align}
\frac{d y}{d\ln a} &= 2\epsilon y - \lambda_0 y,\label{EQN:DEsN2constraint}\\
\frac{d \omega_1}{d\ln a} &= 2\epsilon \omega_1 - \mu_0 \omega_1,\\
\frac{d \omega_2}{d\ln a} &= 2\epsilon \omega_2 - \nu_0 \omega_2,\\
\frac{d z_1}{d\ln a} &= z_1(\epsilon-3)+\omega_1,\\
\frac{d z_2}{d\ln a} &= z_2(\epsilon-3)+\omega_2,
\label{EQN:DEsNendconstraint}
\end{align}
\endgroup
where the Hubble slow-roll parameter, $\epsilon$, is given by Eq.~(\ref{EQN:newep}).

Results for the case in which we focus on stability properties for the constrained phase space are shown in Table~\ref{TAB:Stability}, for select values of $\lambda_0, \mu_0$, and $\nu_0$.
\begin{table*}
\begin{center}
 \begin{tabular}{ l | c | c | c } 
 {\it Label} & {\it Eigenvalues of Jacobian} & {\it Sample condition} & {\it Stability}\\ [1ex] 
 \Xhline{2pt}
 
 {\bf dS}& $\{-3, -3, -\lambda_0, -\mu_0, -\nu_0\}$ & $\lambda_0>0$, $\mu_0>0$, $\nu_0 >0$ & Attractor\Tstrut\Bstrut\\
 
{\bf FPb}& $\left\{\frac{1}{2}(-6 + \lambda_0), 
 \frac{1}{2}(-6 + \lambda_0), \lambda_0, \lambda_0 - \mu_0, \
\lambda_0 - \nu_0\right\}$ &$\lambda_0=2$, $\mu_0=1/3$, $\nu_0=1$
 & Saddle\Tstrut\Bstrut\\
 
{\bf FPc}& $\left\{-\lambda_0 + \nu_0, -\mu_0 + \nu_0, \frac{1}{2}(-6 + \nu_0), c^-, c^+\right\}$ &$\lambda_0=2$, $\mu_0=1/3$, $\nu_0=1$ & Saddle\Tstrut\Bstrut\\
 
 {\bf FPd}& $\left\{-\lambda_0 + \nu_0, -\mu_0 + \nu_0, \frac{1}{2}(-6 + \nu_0), c^-, c^+\right\}$ &$\lambda_0=2$, $\mu_0=1/3$, $\nu_0=1$
 & Saddle\Tstrut\Bstrut\\
 
 {\bf FPe}& $\left\{-\lambda_0 + \mu_0, \frac{1}{2}(-6 + \mu_0), d^-, d^+, \mu_0-\nu_0\right\}$ &$\lambda_0=2$, $\mu_0=1/3$, $\nu_0=1$
 & Saddle\Tstrut\Bstrut\\

 {\bf FPf}& $\left\{-\lambda_0 + \mu_0, \frac{1}{2}(-6 + \mu_0), d^-, d^+, \mu_0-\nu_0\right\}$ &$\lambda_0=2$, $\mu_0=1/3$, $\nu_0=1$
 & Saddle\Tstrut\Bstrut\\ [0.5ex]
 
 \hline 
 
 {\bf FPK}$^-$ & $\left\{6,0,6-\lambda_0 ,6-\mu_0 ,6-\nu_0 \right\}$  & $0 < \lambda_0, \mu_0, \nu_0 < 6$ & Unstable$^*$\Tstrut\Bstrut\\
  
{\bf FPK}$^+$ & $\left\{6,0,6-\lambda_0 ,6-\mu_0 ,6-\nu_0 \right\}$  & $0 < \lambda_0, \mu_0, \nu_0 < 6$ & Unstable$^*$\Tstrut\Bstrut

 \end{tabular}
 \caption{Stability properties of fixed points for the principal phase space. Sample conditions for {\bf FPb}--{\bf FPf} yield the stability properties shown in the rightmost column (in each case, the fixed point is a saddle point for the sample condition). The parameters $c^\pm$ and $d^\pm$ are defined in Eqs.~(\ref{EQN:cpm}) and~(\ref{EQN:dpm}). The superscript $^{*}$ denotes that the fixed points are {\it normally hyperbolic}. That is, one has a set of non-isolated fixed points for which the only eigenvalues with zero real part are those whose eigenvectors are tangent to the set of points. In such a case, one may read off the stability properties of the fixed points from the remaining eigenvalues. (See Ref.~\cite{coley_99}.)} 
 \label{TAB:Stability}
\end{center}
\end{table*}
In Table~\ref{TAB:Stability}, the quantities $c^\pm$ and $d^\pm$ are given by
\begin{align}
c^\pm &= \frac{1}{4} \left(3 \nu_0 -6\pm\sqrt{-7 \nu_0 ^2+60 \nu_0 +36}\right),\label{EQN:cpm} \\
d^\pm &= \frac{1}{4} \left(3 \mu_0 -6\pm\sqrt{-7 \mu_0 ^2+60 \mu_0 +36}\right).\label{EQN:dpm}
\end{align}
Generally, the de Sitter fixed point is an attractor, the remaining non-Kasner fixed points are saddle points, and the Kasner fixed points are unstable.

\section{An emergent cosmological constant}\label{SEC:Emerge}

We now analyze flows through the principal phase space. We begin, in Sec.~\ref{SEC:ISOstability}, with an analysis that probes the stability of isotropic backgrounds. In particular, we start dynamical trajectories at $z_1=0=z_2$ (and therefore at $x=1$)---so that the background is initially isotropic---and identify regions of the  internal space coordinatized by $(\omega_1, \omega_2,y)$ that flow to {\bf dS} (the de Sitter fixed point) or to {\bf FPb} (the non-de Sitter isotropic fixed point), even in the presence of nonvanishing anisotropic pressure. In Sec.~\ref{SEC:anISOics} we analyze  states that begin with an anisotropic background characterized by nonvanishing shear, which then subsequently flow to {\bf dS} or {\bf FPb}. 

For such analyses there are various choices for parameters that need to be made. First, recall that the principal phase space is defined by fixing $\lambda_0, \mu_0$, and $\nu_0$ to be constants, which we take to be real-valued and positive. In order to probe a regime in which anisotropic degrees of freedom do not dissipate arbitrarily quickly, we select relatively small values for the parameters $\mu_0$ and $\nu_0$, such as $\{ \mu_0, \nu_0 \} \in \{ 1/10, 2/3 \}$. (We also analyze a scenario in which the average of $\mu_0$ and $\nu_0$ is $2/3$, with $\mu_0=1/3$ and $\nu_0=1$.) Furthermore, to explore various possibilities for the emergence of a quasi-constant energy density for the background, we consider a broader range of values for $\lambda_0$, with $\lambda_0 \in \{ 1/10, 1, 2 \}$. Recall from Eqs.~(\ref{EQN:ylambda}) and (\ref{EQN:yInterp}) that $y \sim \rho + P$ and $\lambda_0 = - \partial_t (\rho + P) / [H (\rho + P)]$. De Sitter expansion corresponds to $y = 0$ together with each $X_{ii} = 0$. Hence a large value of $\lambda_0$ would seem to favor the emergence of an effective cosmological constant (especially if combined with large values of $\mu_0$ and $\nu_0$, which drive rapid dissipation of the anisotropic pressure contributions), whereas, prima facie, smaller values of $\lambda_0$ would not favor such asymptotic behavior. (See also the discussion near the beginning of Sec.~\ref{SEC:PPS}.)

Second, our EFT framework (like any EFT framework) has a limited range of validity. The dimensionless variable $x \sim \rho$ is constrained via Eq.~(\ref{EQN:Co}) to be $x \leq 1$. We therefore analyze initial conditions such that the magnitude of each of our dimensionless dynamical variables is less than or equal to 1. We consider our EFT no longer valid if, during subsequent evolution of the system, the magnitude of at least one dimensionless dynamical variable grows to be greater than 10. 

Third, we distinguish between cases in which the initial value $y (t_i) < 0$ from those in which $y (t_i) \geq 0$. Note from Eq.~(\ref{EQN:yInterp}) that $y < 0$ corresponds to $P < - \rho$. Scenarios with $y (t_i) < 0$ cannot be realized with single scalar field models for which the scalar field has a minimal coupling to Einstein gravity and a canonical kinetic term. Nonetheless, such scenarios can be studied within our more general EFT framework. In an isotropic spacetime, an equation of state corresponding to $y < 0$ could yield a ``big rip" singularity \cite{starobinsky_00,caldwell_02,caldwell+al_03}. As we discuss further at the end of Sec.~\ref{SEC:ISOstability}, however, scenarios exist within our effective system for which $y < 0$ does not trigger such singular behavior.

\subsection{Flows back to isotropy}\label{SEC:ISOstability}

We first study the stability of isotropic backgrounds in the presence of nonvanishing anisotropic matter sources. In particular, we consider spacetimes that are initially isotropic, with $z_1 = 0 = z_2$ and hence $x = 1$, and establish the range of initial conditions in the internal space 
$\{ \omega_1, \omega_2, y \}$ that yield an isotropic background at late times. Recall from Eqs.~(\ref{EQN:ommu})--(\ref{EQN:z2}) that $\omega_A \sim \pi_A$ and $z_A \sim \sigma_A$ (for $A=1,2$), and from Eqs.~(\ref{EQN:dz1})--(\ref{EQN:dz2}) that nonzero anisotropic pressures $\omega_A$ will induce nonvanishing components of the shear $z_A$.

\begin{figure*}
\begin{minipage}{.3\linewidth}
\centering
\subfloat[]{\includegraphics[scale=0.425]{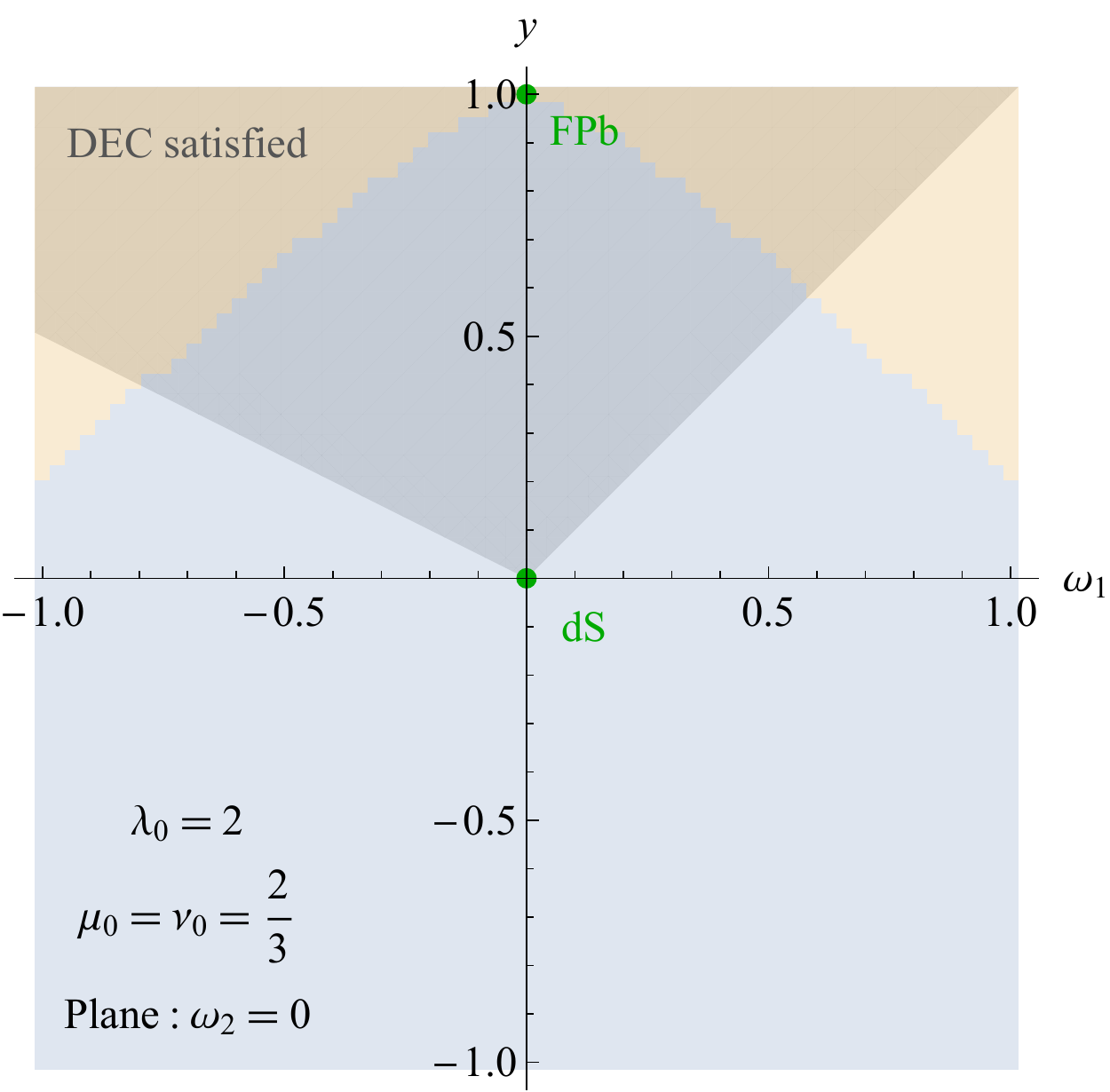}}
\end{minipage}
\begin{minipage}{.3\linewidth}
\centering
\subfloat[]{\includegraphics[scale=0.425]{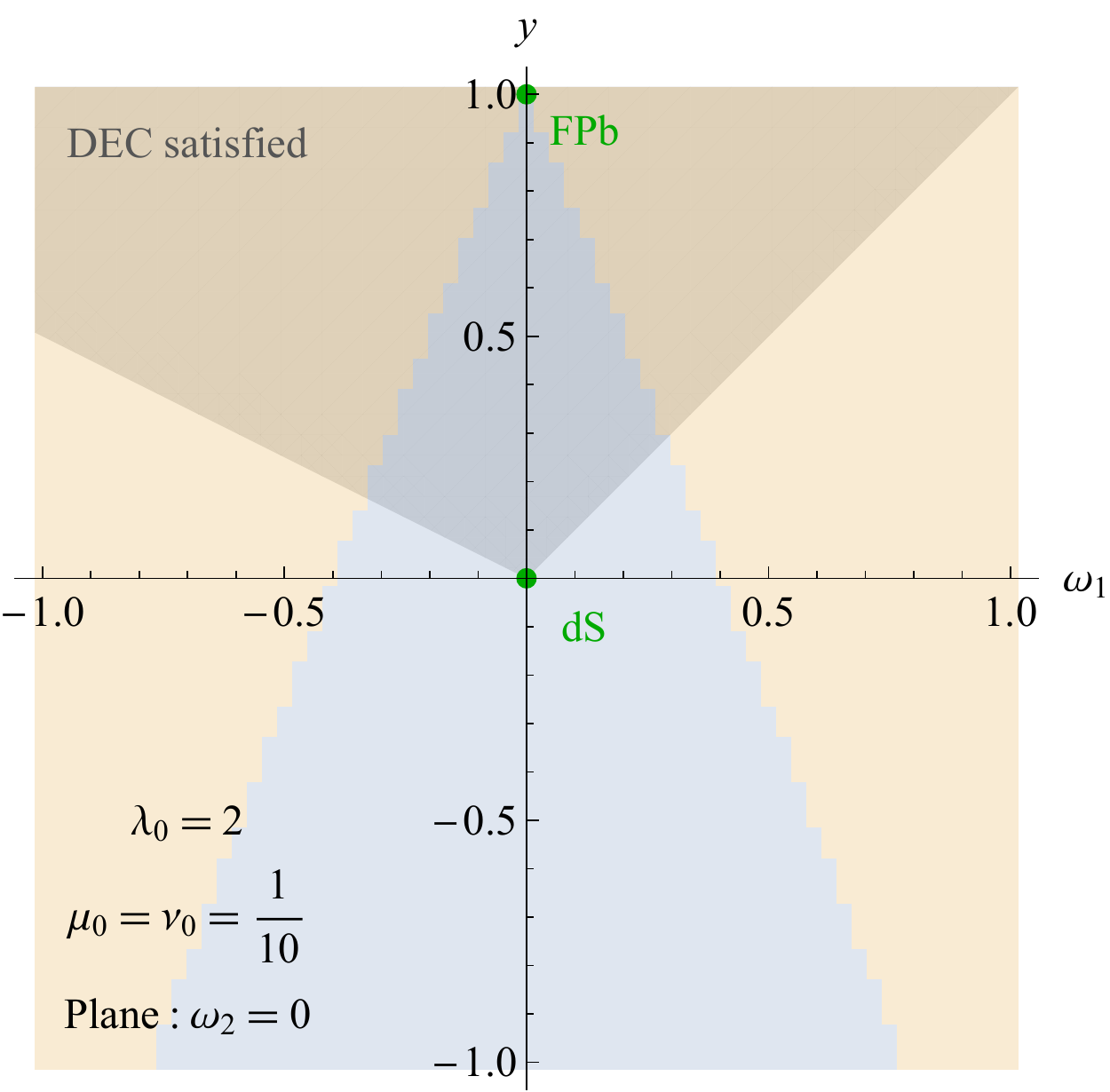}}
\end{minipage}
\begin{minipage}{.3\linewidth}
\centering
\subfloat[]{\includegraphics[scale=0.425]{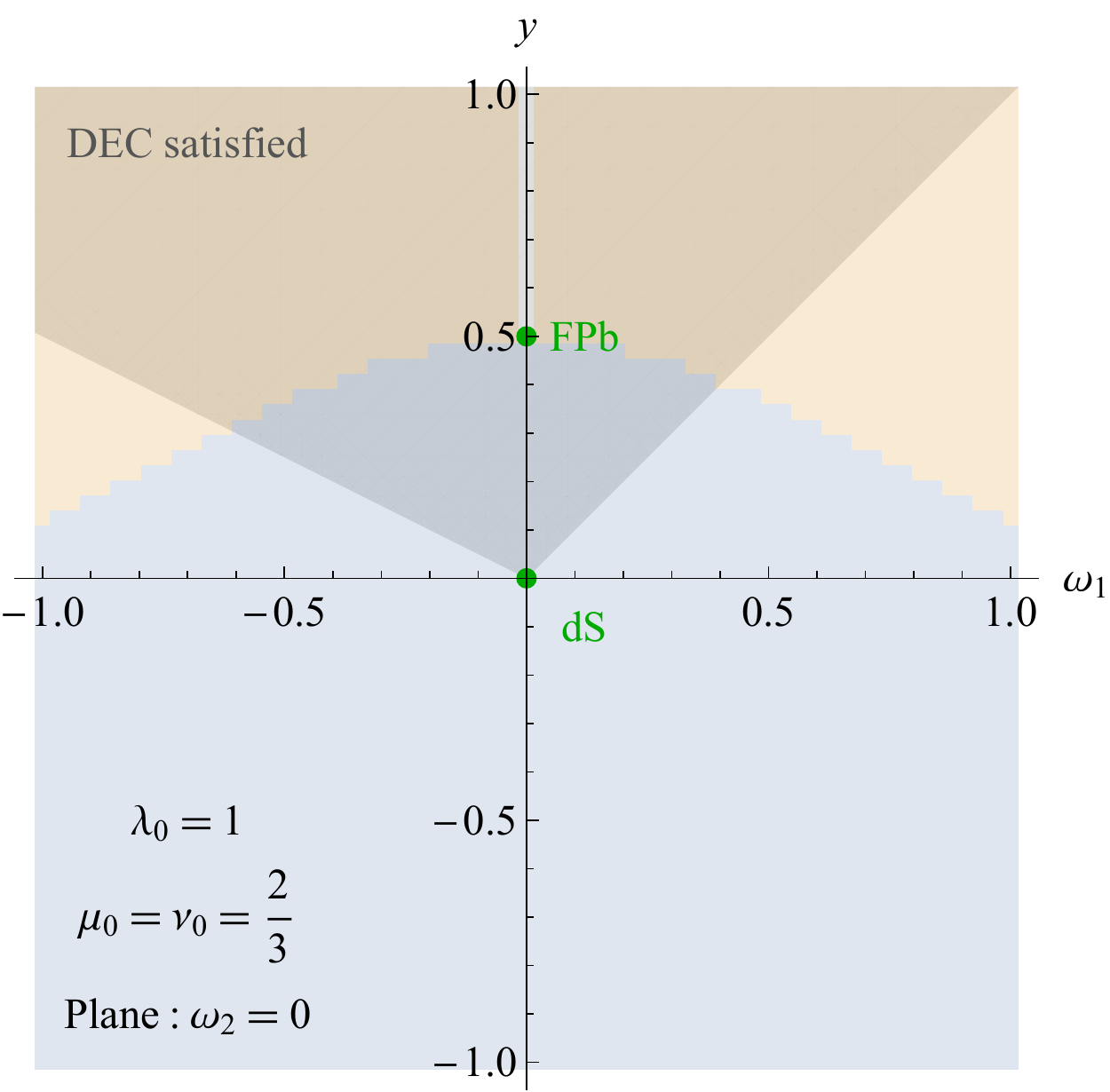}}
\end{minipage}
\begin{minipage}{.3\linewidth}
\centering
\subfloat[]{\includegraphics[scale=0.425]{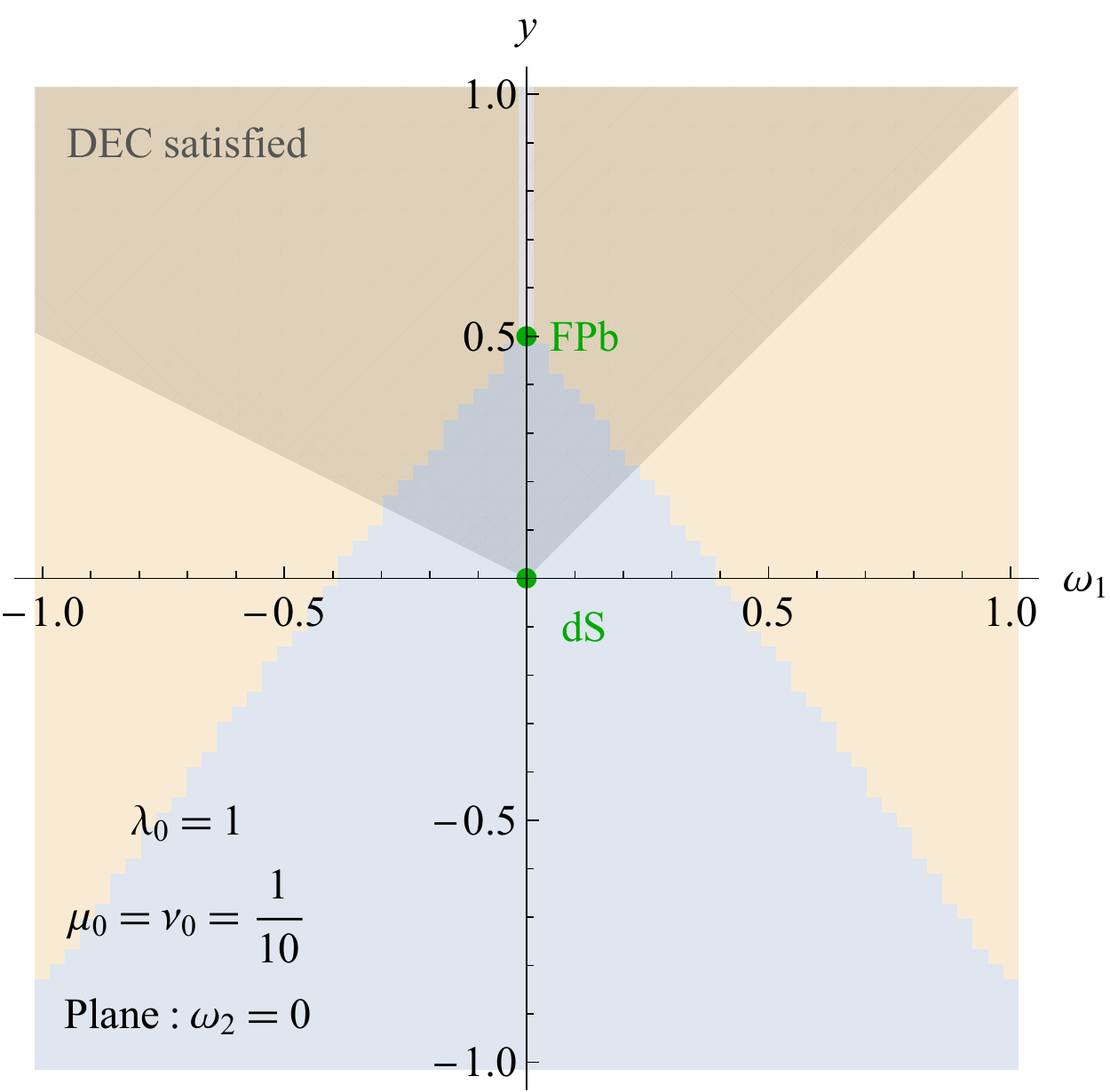}}
\end{minipage}
\begin{minipage}{.3\linewidth}
\centering
\subfloat[]{\includegraphics[scale=0.425]{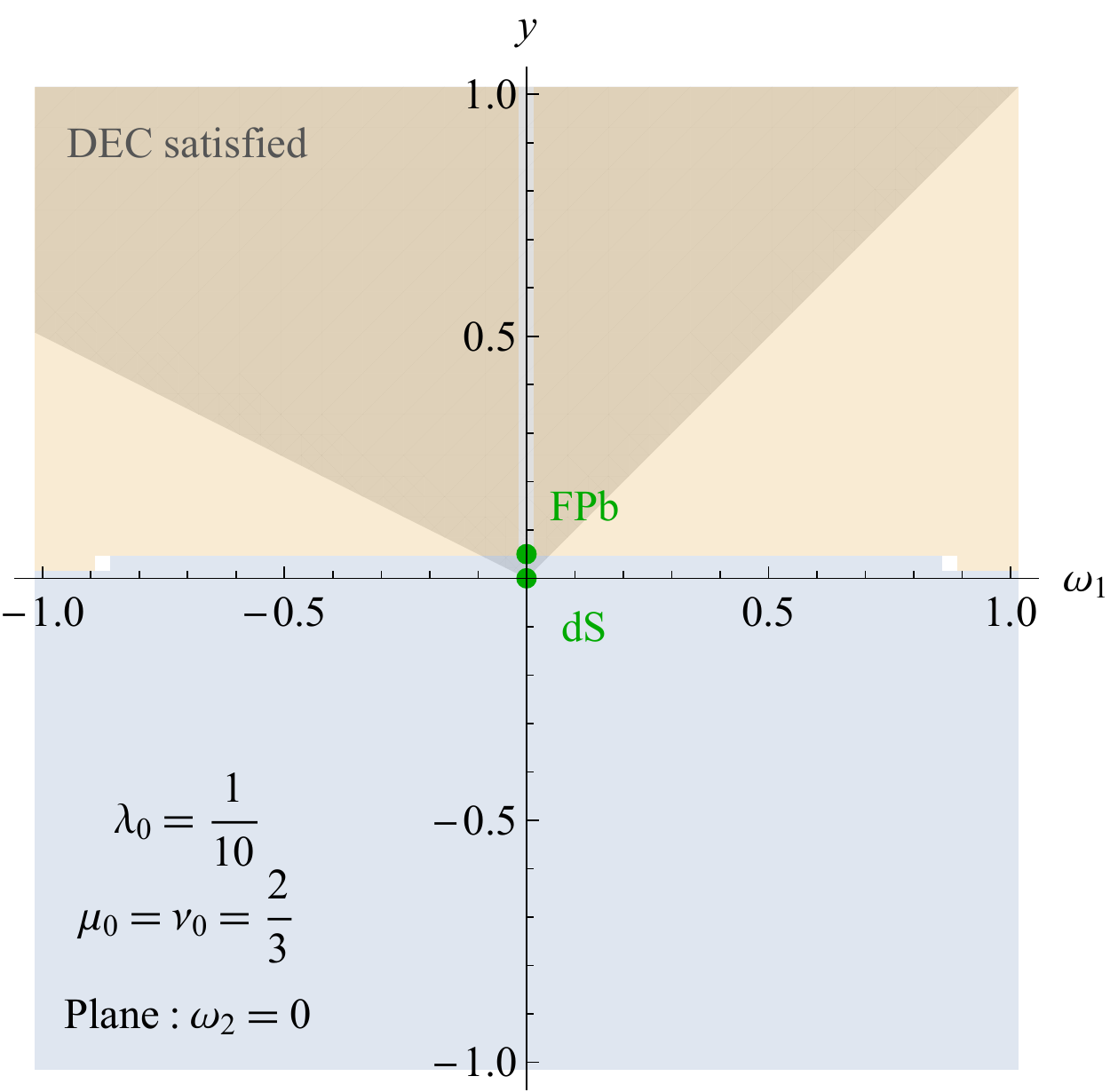}}
\end{minipage}
\begin{minipage}{.3\linewidth}
\centering
\subfloat[]{\includegraphics[scale=0.425]{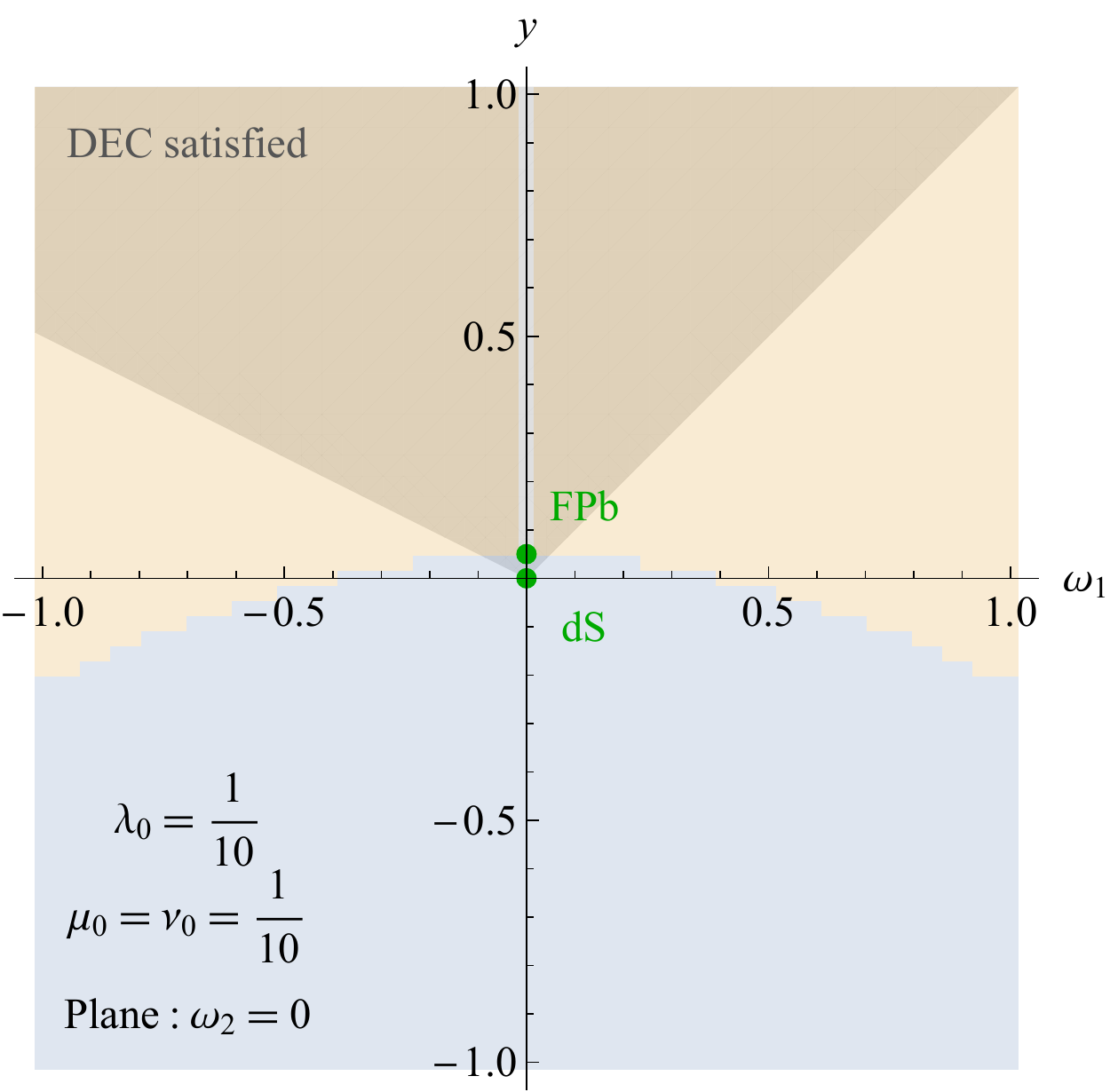}}
\end{minipage}
\caption{Slices (corresponding to $\omega_2=0$) through the internal space for six dynamical systems. The light blue region denotes initial conditions that flow to the dS fixed point, {\bf dS}. The orange region denotes initial conditions that exit the EFT region with a background that is anisotropic. Fractions of the internal space that flow to different states are displayed in Table~\ref{TAB:Stability_1}. The dominant energy condition (DEC) is satisfied for initial conditions that lie in the shaded region. 
}
\label{FIG:Stability_1}
\end{figure*}

In Fig.~\ref{FIG:Stability_1}, we present plots of a slice through initial conditions in the internal space $\{ \omega_1, \omega_2,y \}$.
The six plots correspond to six different choices for $\lambda_0$ and $\mu_0=\nu_0$. For each plot, the light blue region describes initial conditions that flow to {\bf dS} (the isotropic dS fixed point) while the orange region describes initial conditions that exit the regime of validity of our EFT dynamical system with a background that is anisotropic. In other words, points within the orange region represent evolution such that the magnitude of at least one of the dynamical variables grows to be greater than 10 at some finite time, and at that time at least one of the $z_A$ remains nonzero. There are also some cases (that appear in white) in which initial conditions yield flows that exit the regime of validity of our EFT dynamical system with a background that is isotropic, even though they did not reach either isotropic fixed point, {\bf dS} or {\bf FPb}. We set the threshold to consider a late-time state sufficiently isotropic to be $\vert z_A \vert \leq 10^{-5}$ for $A = 1, 2$, comparable to the observed anisotropy in our own universe on long length-scales as measured in the cosmic microwave background radiation.

These plots exhibit evolution from initial conditions beyond the domain to which Wald's theorem \cite{wald_83} applies. As noted around Eq.~(\ref{WaldTotalT}), Wald derived his cosmic no-hair theorem for systems described by an energy-momentum tensor $T_{\mu\nu}$ that included a bare, positive cosmological constant $\Lambda$ (which violates the SEC), as well as additional matter sources $\tilde{T}_{\mu\nu}$ that obey both the SEC and the DEC. Hence Wald's theorem applies to systems in which $T_{\mu\nu}$ always satisfies the DEC and may or may not satisfy the SEC (depending on the magnitude of $\Lambda$ compared to the components of $\tilde{T}_{\mu\nu}$). The plots in Fig.~\ref{FIG:Stability_1}, on the other hand, include scenarios
(outside of the shaded regions) in which the initial conditions violate the DEC, but for which the system nonetheless flows to the de Sitter fixed point. Indeed, the fraction of points within the (full) internal space that exhibit such behavior can be significant, as suggested by the slice shown in Fig.~\ref{FIG:Stability_1}a.

For each case shown in Fig.~\ref{FIG:Stability_1}, we initialize the system with $-1 \leq \{ \omega_1 (t_i), \omega_2 (t_i), y (t_i) \} \leq 1$. (This range includes sets of initial conditions that lie exactly along the $y$-axis, for which the initial anisotropic pressures vanish identically.) The fraction of the internal space $\{ \omega_1, \omega_2, y\}$ that subsequently flows to various late-time states is shown in Table~\ref{TAB:Stability_1}. Within the table, the row that yields the highest fraction of the space that flows to the fixed point {\bf dS}, for $y (t_i) \geq 0$, is highlighted in gray. For $y (t_i) \geq 0$, we find that decreasing $\lambda_0$ (while keeping $\mu_0=\nu_0$ fixed) leads to a decrease in the fraction of points that flow to {\bf dS}; whereas decreasing $\mu_0=\nu_0$  (while keeping $\lambda_0$ fixed) leads to a pronounced decrease in the fraction of points that flow to {\bf dS}. This pattern makes sense, given that smaller values of $\mu_0$ and $\nu_0$ correspond to scenarios in which the components of the anisotropic pressure dissipate more slowly over time, whereas larger values of $\lambda_0$ more effectively drive $(\rho + P) \rightarrow 0$. On the other hand, if we include cases where $y (t_i) < 0$ we find an interesting pattern: decreasing $\lambda_0$ while holding $\mu_0 = \nu_0 = 1/10$ fixed yields a {\it larger} fraction of points that flow to {\bf dS}. Trajectories with $y (t_i) < 0$ begin with $P < - \rho$ (see Eq.~[\ref{EQN:yInterp}]), which yields super-accelerated expansion at early times. Such early growth quickly dilutes the anisotropic pressures, driving the system toward the {\bf dS} fixed point.

\begin{table*}
\begin{center}
 \begin{tabular}{ c | c | c | c | c | c | c | c} 
  		&Initial $y$-values&&& \multicolumn{4}{c}{Fraction of the full 3D internal space that flows to \dots} \\ [1ex]
 {\it Fig.} & considered & $\lambda_0$ & $\mu_0=\nu_0$ & {\bf dS} & {\bf FPb} & {\it isotropic EXIT} & {\it anisotropic EXIT}\\ [1ex] 
 \Xhline{2pt}
\ref{FIG:Stability_1}a& $|y(t_i)| \leq 1$ & 2& 2/3&0.7030& -- &0&0.2970\Tstrut\Bstrut\\
\rowcolor{light-gray}\ref{FIG:Stability_1}a& $0\leq y(t_i) \leq 1$ & 2& 2/3&0.4336& -- &0&0.5664\Tstrut\Bstrut\\
 \Xhline{0.25pt}
\ref{FIG:Stability_1}b& $|y(t_i)| \leq 1$ & 2& 1/10&0.1841& -- &0&0.8159\Tstrut\Bstrut\\
\ref{FIG:Stability_1}b& $0\leq y(t_i) \leq 1$ & 2& 1/10&0.0487& -- &0&0.9513\Tstrut\Bstrut\\
 \Xhline{0.25pt}
\ref{FIG:Stability_1}c& $|y(t_i)| \leq 1$ & 1& 2/3&0.6122& -- &0.0001&0.3877\Tstrut\Bstrut\\
\ref{FIG:Stability_1}c& $0\leq y(t_i) \leq 1$ & 1& 2/3&0.2469& -- &0.0001&0.7530\Tstrut\Bstrut\\
 \Xhline{0.25pt}
\ref{FIG:Stability_1}d& $|y(t_i)| \leq 1$ & 1& 1/10&0.2710& -- &0.0001&0.7289\Tstrut\Bstrut\\
\ref{FIG:Stability_1}d& $0\leq y(t_i) \leq 1$ & 1& 1/10&0.0261& -- &0.0001&0.9738\Tstrut\Bstrut\\
 \Xhline{0.25pt}
 \ref{FIG:Stability_1}e& $|y(t_i)| \leq 1$ & 1/10& 2/3&0.5138&0&0.0009&0.4853\Tstrut\Bstrut\\
\ref{FIG:Stability_1}e& $0\leq y(t_i) \leq 1$ & 1/10& 2/3&0.0465&0&0.0017&0.9518\Tstrut\Bstrut\\
 \Xhline{0.25pt}
 \ref{FIG:Stability_1}f& $|y(t_i)| \leq 1$ & 1/10& 1/10&0.4350&0&0.0001&0.5649\Tstrut\Bstrut\\
\ref{FIG:Stability_1}f& $0\leq y(t_i) \leq 1$ & 1/10& 1/10&0.0058&0&0.0002&0.9940\Tstrut\Bstrut
 \end{tabular}
 \caption{For certain values of $\lambda_0, \mu_0$, and $\nu_0$, we display the fraction of the full internal space---coordinatized by $\{\omega_1, \omega_2, y\}$---that either flows to the isotropic de Sitter fixed point ({\bf dS}) or the isotropic (non-de Sitter) fixed point ({\bf FPb}). Trajectories that reach neither fixed point and that leave the regime of validity of the EFT in an (an)isotropic state undergo `{\it (an)isotropic EXIT}\,'. A dash `--' indicates a value $<0.0001$. 
 }
 \label{TAB:Stability_1}
\end{center}
\end{table*}

We next consider three trajectories of interest that begin with isotropic initial conditions, $( z_1, z_2 , x ) = ( 0, 0, 1 )$, and for which the anisotropic pressures vanish, $\omega_A = 0$ for $A = 1, 2$. For finite $H$, the case $\omega_A = 0$ corresponds to $\pi_A = 0$, which in turn means that $p_1 = p_2 = p_3$. The components $p_i$ also satisfy $p_1 + p_2 + p_3 = 0$ from Eq.~(\ref{psum}), so we find
%%%%%
\beq
p_i = 0 \>\>\> {\rm for} \>\> i = 1, 2, 3 \quad ({\rm for} \>\> \omega_A = 0).
\label{pvanish}
\eeq
Given that both $\omega_A = 0$ (for $A = 1, 2$) are invariant manifolds, $p_i = 0$ therefore persists under the system's evolution.

To explore such scenarios, we choose $(\lambda_0, \mu_0, \nu_0) = (1, 1/10, 1/10)$ so that {\bf FPb}, with coordinates $( \omega_1, \omega_2, y) = (0, 0, 1/2)$, lies within the range of initial conditions under consideration. We then vary the initial value of $y$ and consider three cases.

{\it Case 1}: $y (t_i) = 0.49$. The trajectory flows straight down, along the $y$-axis, to the de Sitter fixed point {\bf dS}, while maintaining $x = 1$, $z_A = 0$. In this scenario the background is driven to the de Sitter fixed point and not to {\bf FPb}. This is consistent with {\bf dS} being an attractor for $(\lambda_0, \mu_0, \nu_0)=(1, 1/10, 1/10)$.

{\it Case 2}: $y (t_i) = 0.51$. The trajectory flows straight up, along the $y$-axis, while maintaining $x = 1$, $z_A = 0$. In this case, initially inflating initial conditions (as diagnosed by initial conditions for which $\epsilon = y  < 1$) are blocked. Some insight can be gained into the nature of this trajectory by noting that if $x=1$ then $3\Mpl^2 H^2 = c+L+\tilde{X}$, in which case
\begin{align}
y\equiv\frac{3c+\tilde{X}}{3\Mpl^2 H^2}=\frac{3c+\tilde{X}}{c+L+\tilde{X}}.
\end{align}
In this case, a growing $y$ means that $3c$ grows relative to $c+L$ (or $2c$ grows relative to $L$)---and this corresponds, as described in Ref.~\cite{Azhar:2018nol}, to a kinetic-energy-density dominated regime, in which one would expect inflation to come to an end.

This observation about evolution under conditions dominated by kinetic energy can be made more precise by considering the behavior of $\tilde{X}$. As noted in Eq.~(\ref{pvanish}), $p_i = 0$ (for $i = 1, 2, 3$) when $\omega_A = 0$ (for $A = 1,2$). But $p_i\equiv-\frac{2}{3}\tilde{X}+2g^{ii}X_{ii}$ (no sum on $i$). So $p_i=0 \iff \tilde{X} = 3g^{ii}X_{ii}$ (no sum on $i$). Thus we have that 
\begin{align}
g^{11}X_{11}&=g^{22}X_{22}=g^{33}X_{33}\nonumber \\ & \implies \frac{X_{11}}{a^2e^{2\beta_1}} = \frac{X_{22}}{a^2e^{2\beta_2}} = \frac{X_{33}}{a^2e^{2\beta_3}}.
\end{align}
These equations yield 
\begin{align}
X_{22} &= e^{2 (\beta_2 - \beta_1 )} X_{11} \label{EQN:deriv1}, \\
X_{33} &= e^{2 (\beta_3 - \beta_1 )} X_{11}. \label{EQN:deriv2}
\end{align}
Note further that the $z_A$'s remain zero along the trajectory, in which case, for finite $H$, we have that $\sigma_1=0\iff \dot{\beta}_1=\dot{\beta}_2$ and $\sigma_2=0\iff \dot{\beta}_1=\dot{\beta}_3$, and hence $\dot{\beta}_1=\dot{\beta}_2=\dot{\beta}_3$. From the derivative of the constraint on the $\beta_i$'s in Eq.~(\ref{betasum}), we therefore find that $\dot{\beta}_i=0$ for $i=1,2,3$. 
Upon substituting Eqs.~(\ref{EQN:deriv1}) and (\ref{EQN:deriv2}) into the constraint on the components $X_{ii}$ in Eq.~(\ref{EQN:Xsum}), we find
\begin{align}
X_{11} \left[ 1 + e^{2 (\beta_2 - \beta_1 )} + e^{ 2 (\beta_3 - \beta_1 )} \right] = 0 .
\end{align}
The term in brackets is a positive constant. Therefore we find that $X_{11} = 0$ and, from Eqs.~(\ref{EQN:deriv1}) and~(\ref{EQN:deriv2}), $X_{22}=0=X_{33}$. Therefore $\tilde{X}=0$, and our system reduces to precisely the FLRW case studied in Ref.~\cite{Azhar:2018nol}. As analyzed there, trajectories in which $2c$ grows relative to $L$ do indeed correspond to the case of kinetic-energy density growing relative to the potential-energy density, and hence a flow away from inflating conditions.

{\it Case 3}: $y (t_i) = - 10$. Given $\omega_A = 0$ for $A = 1, 2$, this trajectory begins with large and negative $y$ and flows straight up the $y$-axis, maintaining $x = 1, z_A = 0, \omega_A = 0$, to {\bf dS}, the de Sitter fixed point. Just as in Case 2, this trajectory satisfies $\tilde{X} = 0$.

We may define an equation of state for this trajectory, $w\equiv {P}/{\rho}$. Using Eq.~(\ref{Tmncomponents3}) with $\tilde{X} = 0$, we find
\begin{align}
w=\frac{c-L}{c+L}.
\end{align}
Given that
\begin{align}
c-L &= 3\Mpl^2 H^2 \left(-x+\frac{2}{3}y\right),\\
c+L &= 3\Mpl^2 H^2 x,
\end{align}
we find
\begin{align}
w=-1+\frac{2y}{3x}=-1+\frac{2y}{3},
\end{align}
since $x=1$ along this trajectory. So $y<0\iff w<-1$ and  
\begin{align}
y\to 0^{-} \iff w\to -1^{-}.
\end{align}
That is, the trajectory of Case 3 evolves with an equation of state that varies with time and approaches that of a cosmological constant, $w = -1$, from below. Such trajectories represent EFT realizations of the ``pseudo-rip" scenarios identified in Ref.~\cite{Frampton:2011aa}, in which $H$ flows to a finite constant at late times, unlike the more familiar ``big rip" singularities usually associated with $w < -1$ \cite{starobinsky_00,caldwell_02,caldwell+al_03}. As in the scenarios described in Refs.~\cite{Stefancic:2004kb,Nojiri:2005sx,Nojiri:2005sr,Frampton:2011sp,Frampton:2011rh,Frampton:2011aa}, our Case 3 presumably avoids the big-rip singularity by violating various energy conditions.

\subsection{Flows to isotropy starting from anisotropic backgrounds}\label{SEC:anISOics}

We now consider scenarios in which the background is initially anisotropic. In these cases, the $z_A$'s (proportional to components of the shear) are nonzero at early times, and hence, from the constraint of Eq.~(\ref{EQN:Co}), $x\neq 1$.

\begin{figure}
\begin{minipage}{\linewidth}
\centering
\subfloat[]{\includegraphics[scale=0.5]{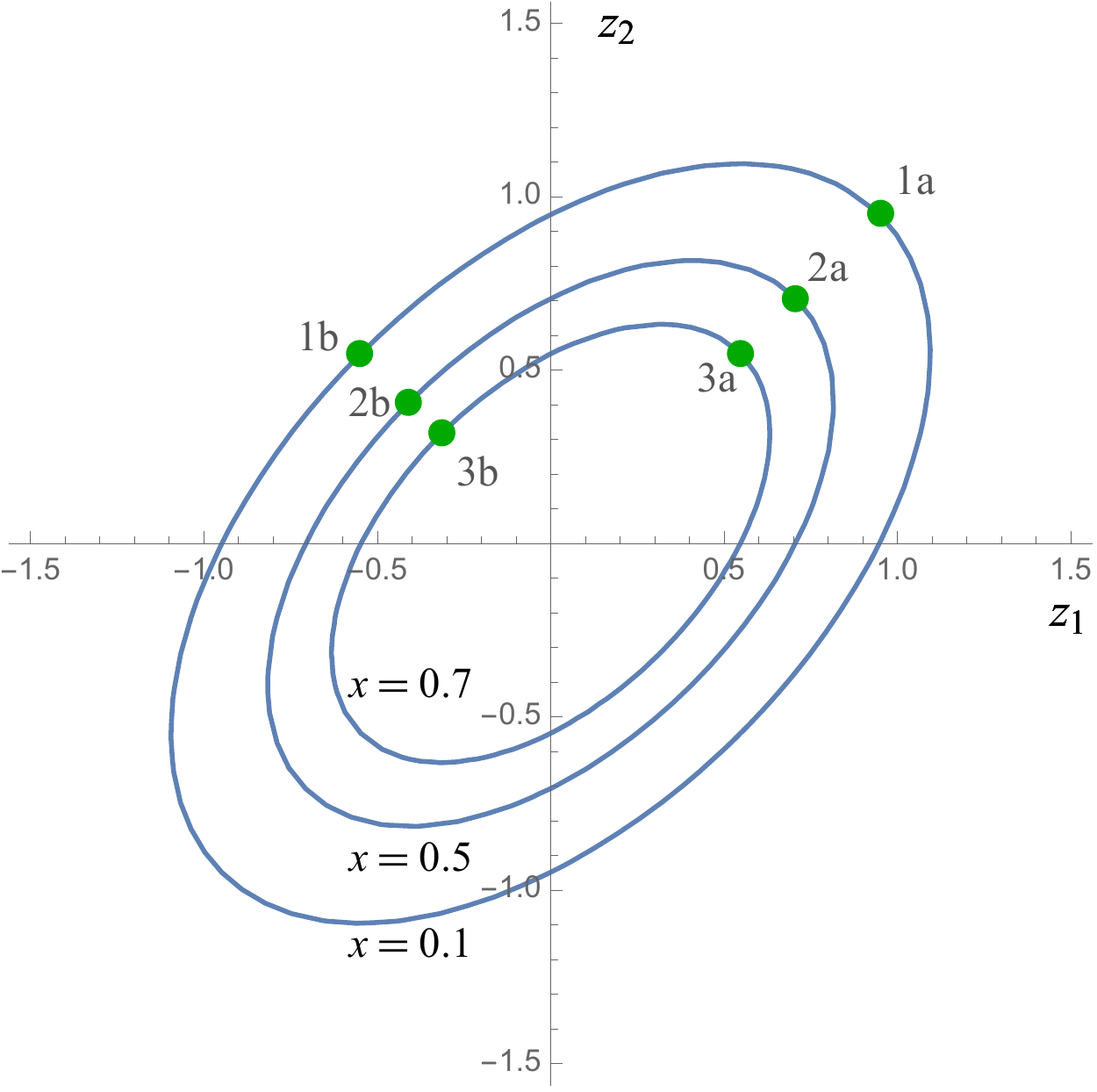}}
\end{minipage}
\caption{Constraint surface projected onto the $z_1$-$z_2$ plane. Starting points of trajectories analyzed in this section are presented in green. For each chosen set of values of $\{ \lambda_0, \mu_0, \nu_0 \}$, we consider six different starting points, labeled 1a, 1b, 2a, 2b, 3a, and 3b.}
\label{FIG:ProjectedPoints}
\end{figure}

To study such scenarios, we first select values for $\lambda_0, \mu_0$, and $\nu_0$. For each chosen value of $\lambda_0, \mu_0$, and $\nu_0$, we consider six different starting points on the constraint surface described by Eq.~(\ref{EQN:Co}) and depicted in Fig.~\ref{FIG:ProjectedPoints}. 
Those starting points correspond to three different values of $x$:  $x=0.1, 0.5, 0.7$. Increasing initial values of $x$ correspond to decreasing degrees of initial anisotropy. (Recall that $x=1$ corresponds to an isotropic state.) For each value of $x$, we select two different points, denoted by the letters `a' and `b' in Fig.~\ref{FIG:ProjectedPoints}. Points marked with an `a' satisfy $z_1=z_2=\sqrt{1-x}$, whereas points marked with a `b' satisfy $z_1=-z_2=-\sqrt{(1-x)/{3}}$. We first describe two representative cases, then consider more general behavior of such systems.

\begin{figure*}
\begin{minipage}{.3\linewidth}
\centering
\subfloat[]{\includegraphics[scale=0.425]{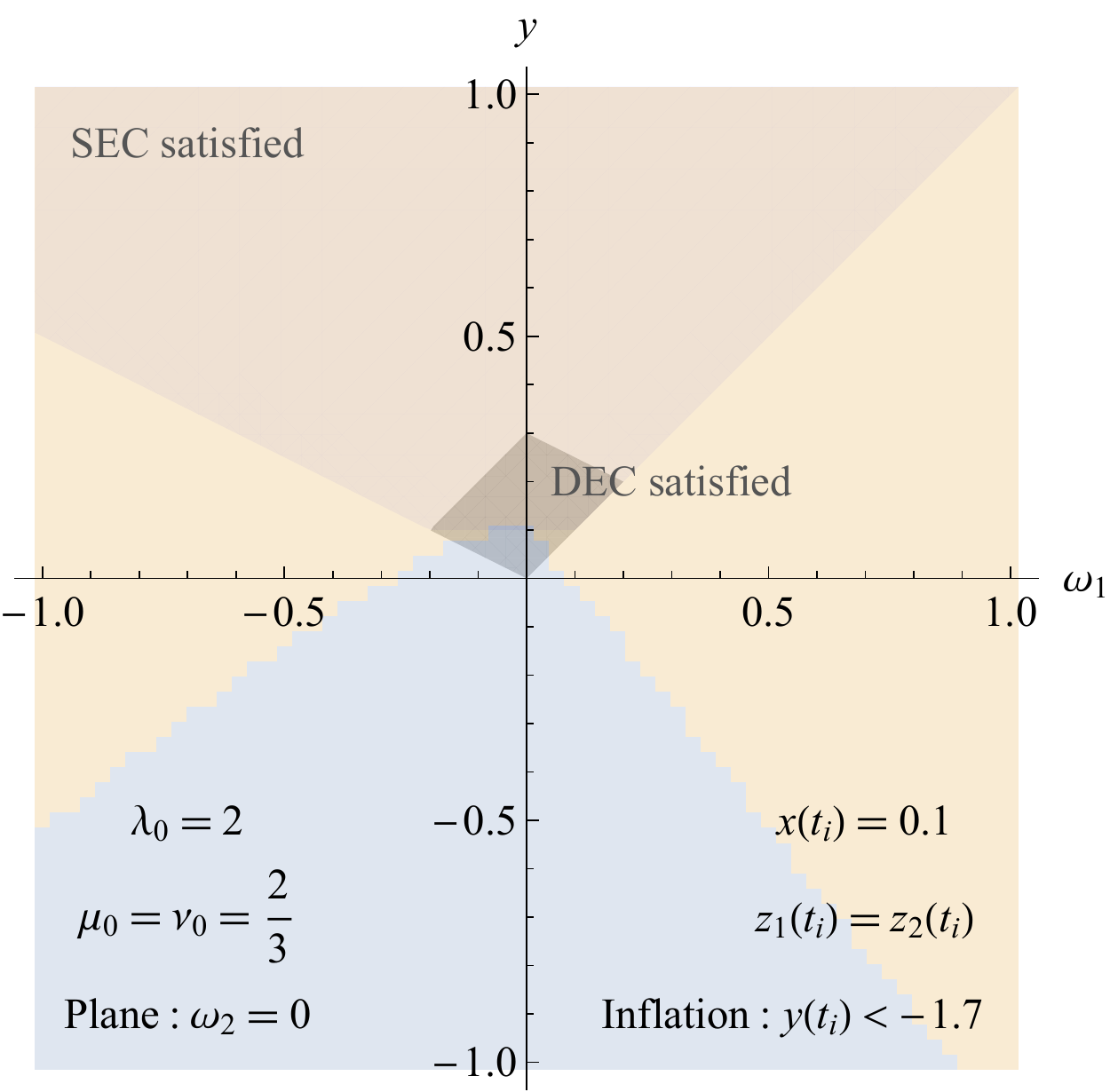}}
\end{minipage}
\begin{minipage}{.3\linewidth}
\centering
\subfloat[]{\includegraphics[scale=0.425]{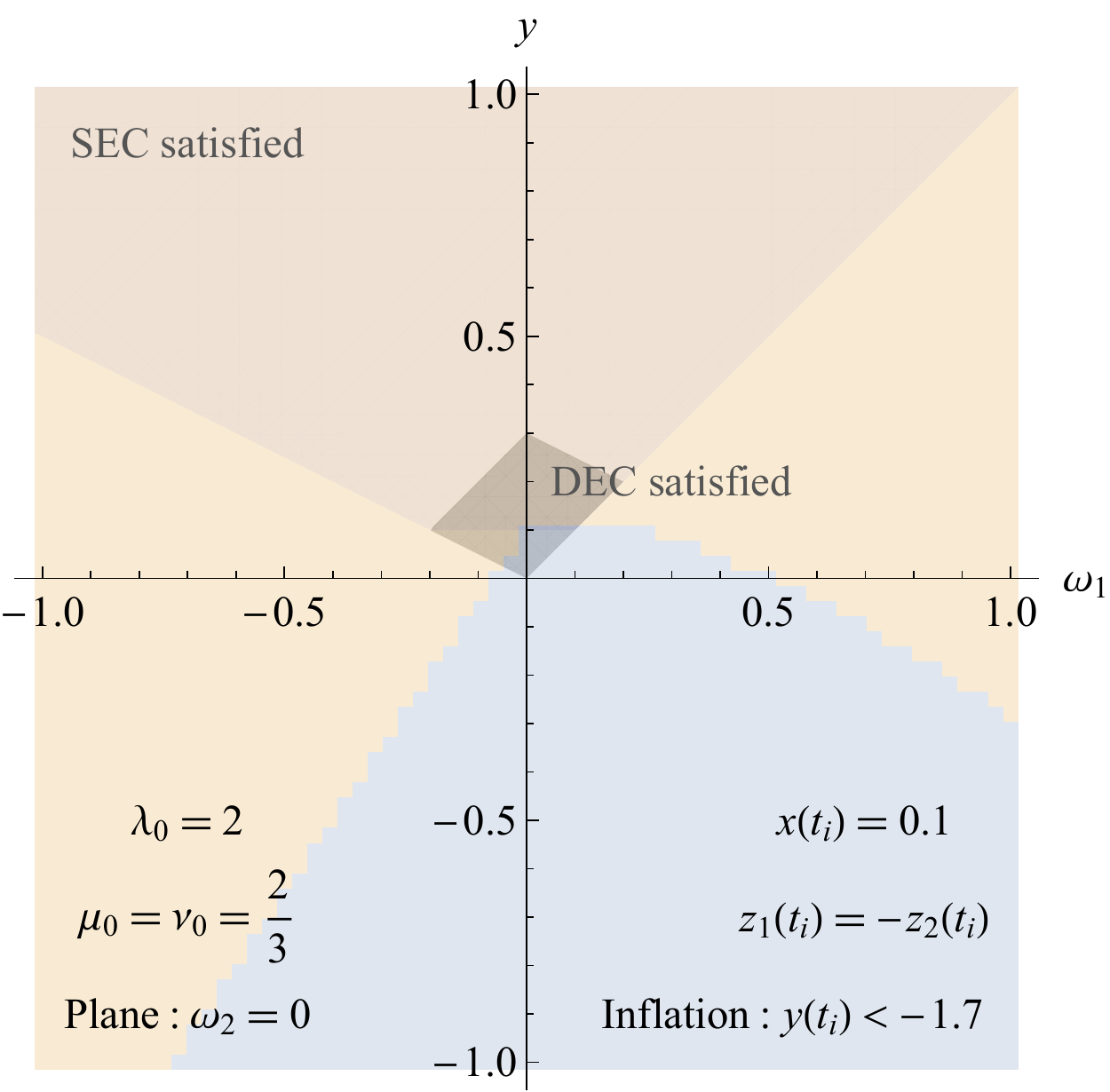}}
\end{minipage}
\begin{minipage}{.3\linewidth}
\centering
\subfloat[]{\includegraphics[scale=0.425]{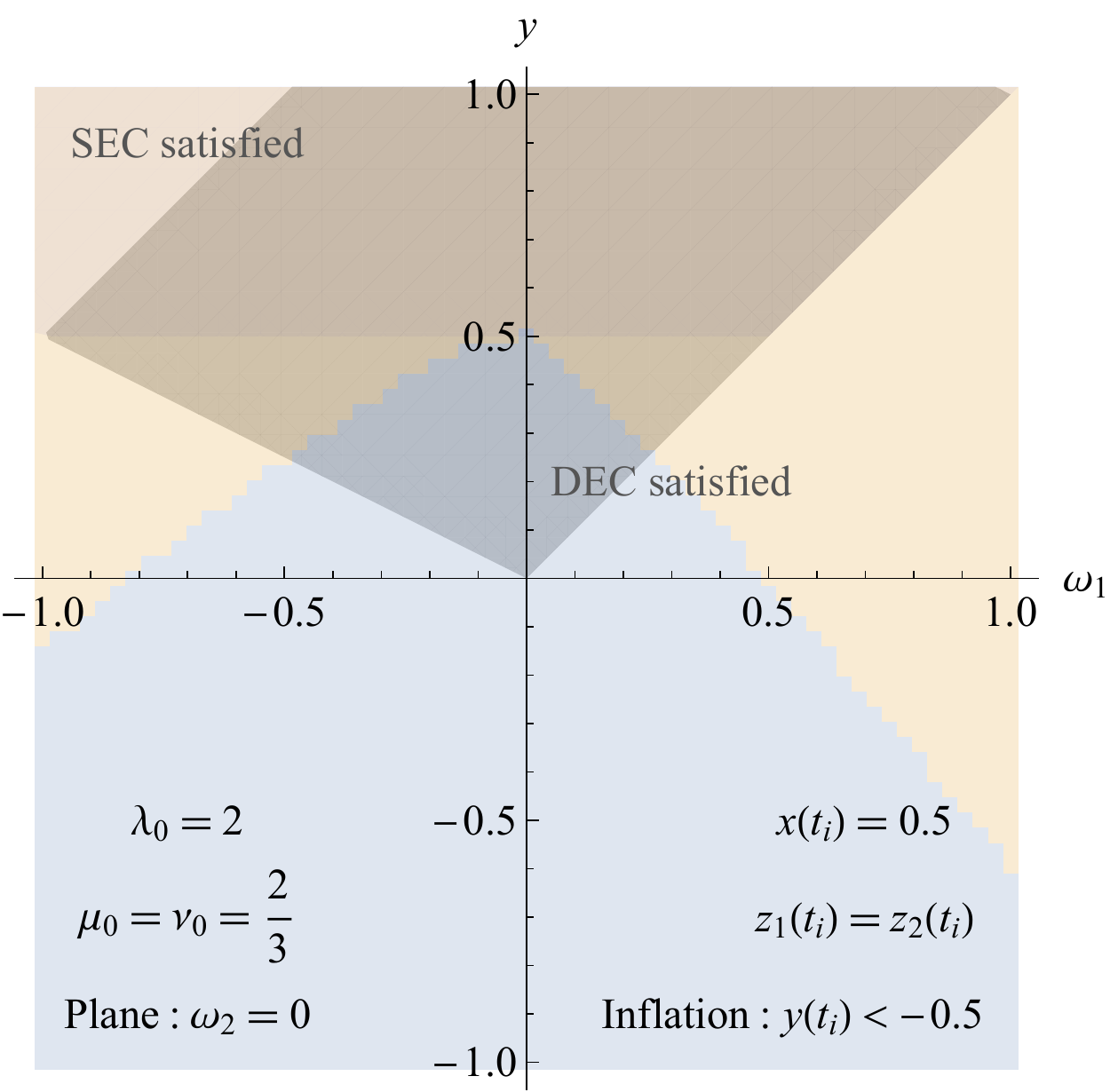}}
\end{minipage}
\begin{minipage}{.3\linewidth}
\centering
\subfloat[]{\includegraphics[scale=0.425]{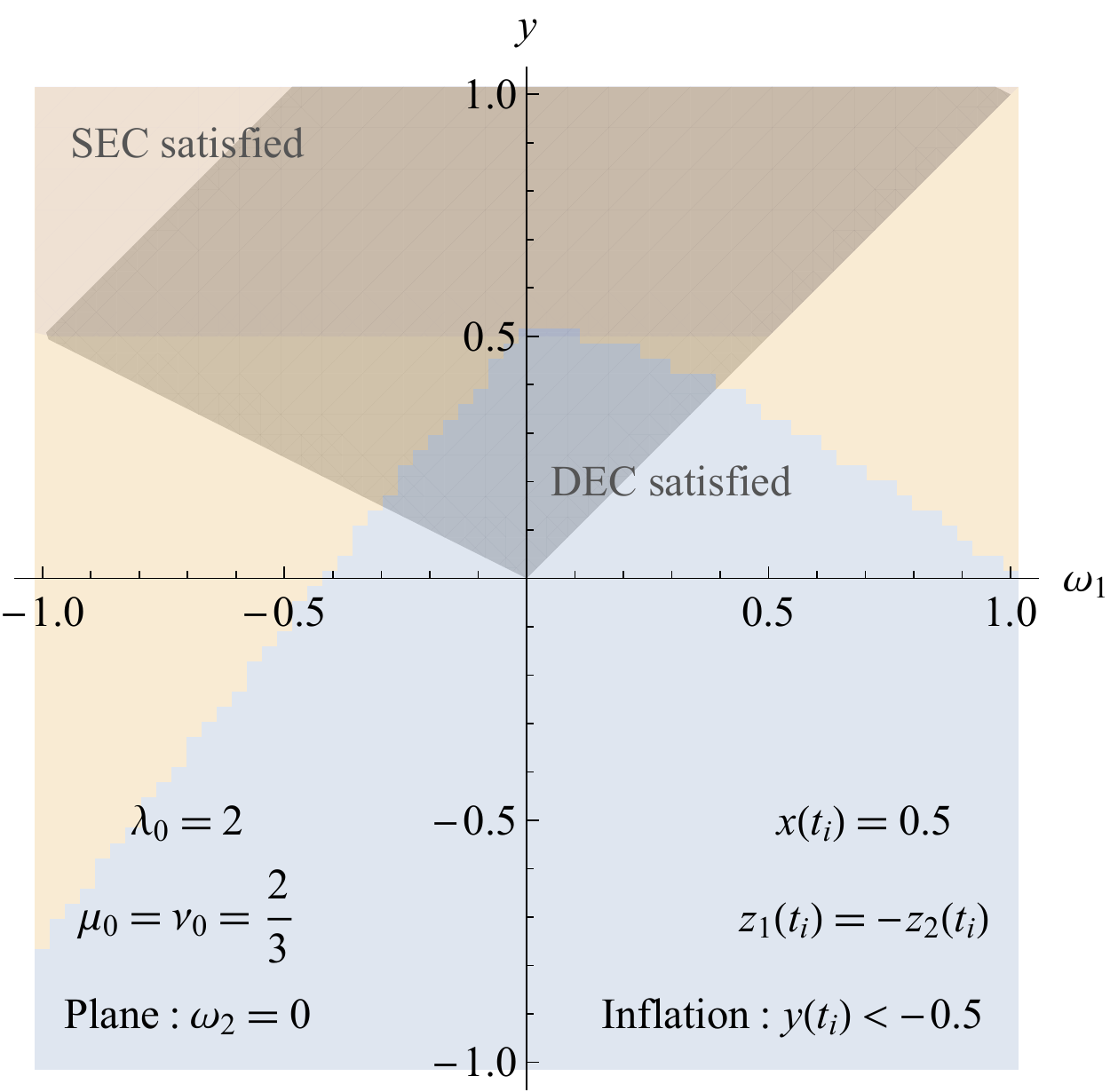}}
\end{minipage}
\begin{minipage}{.3\linewidth}
\centering
\subfloat[]{\includegraphics[scale=0.425]{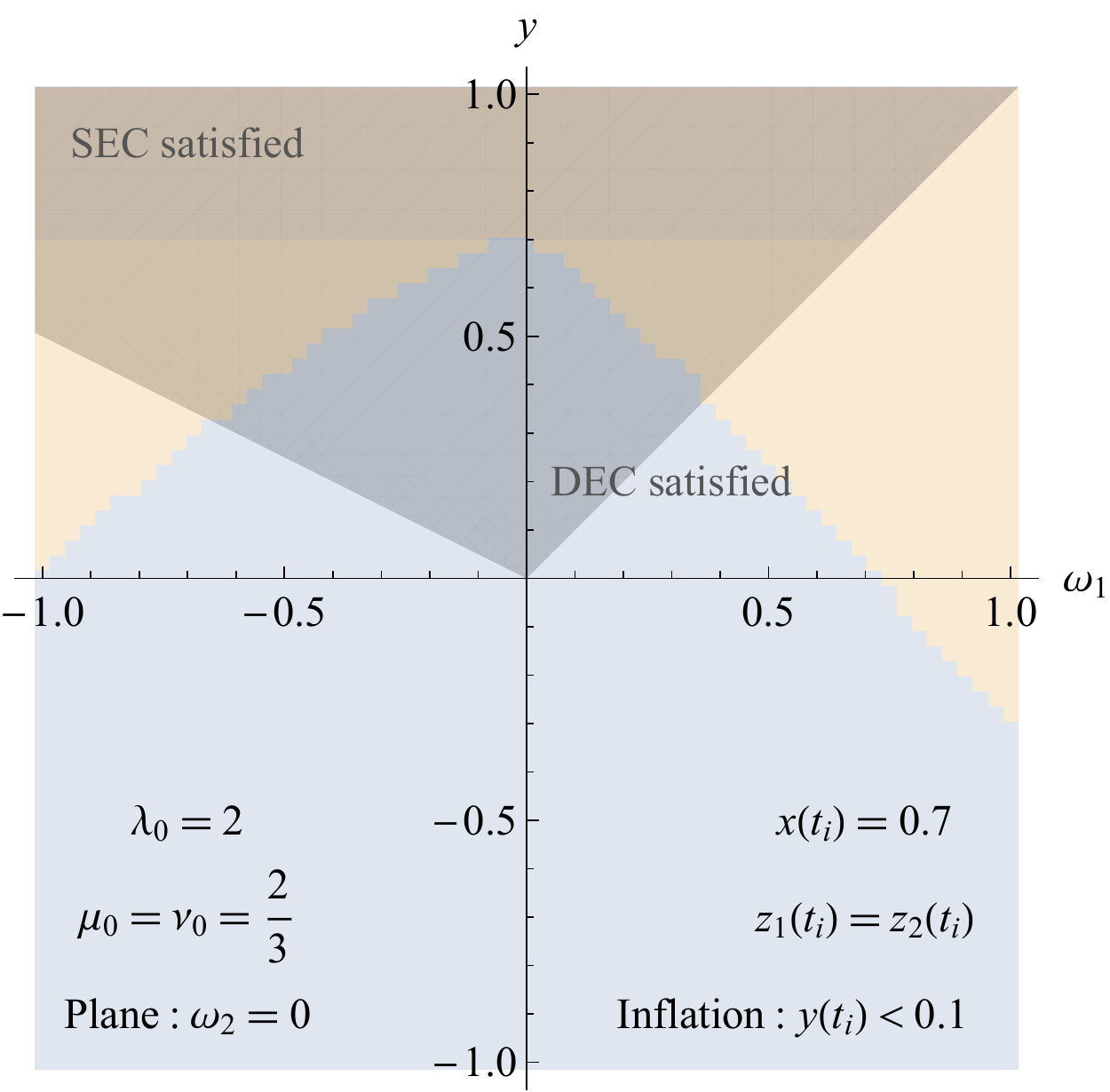}}
\end{minipage}
\begin{minipage}{.3\linewidth}
\centering
\subfloat[]{\includegraphics[scale=0.425]{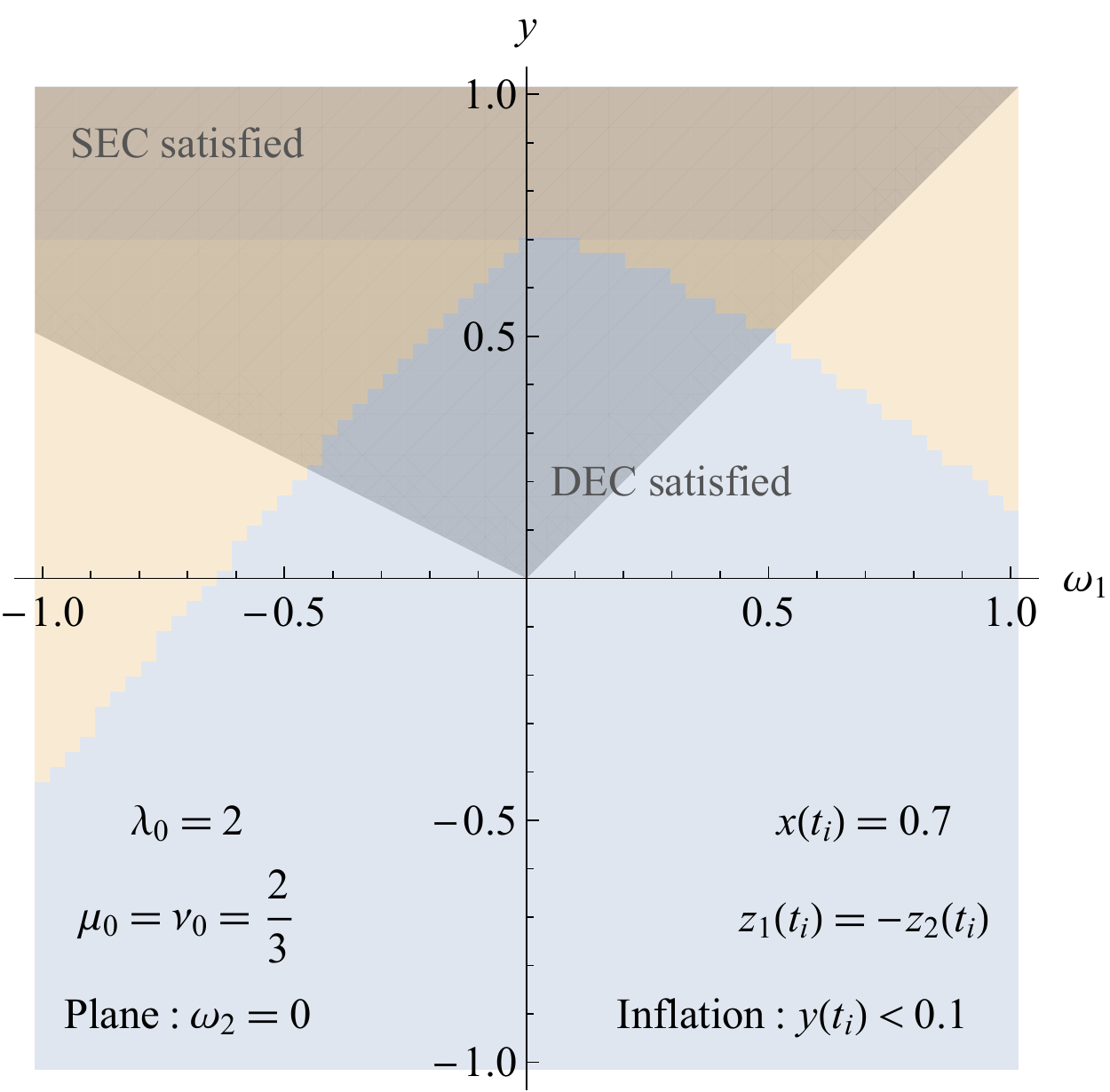}}
\end{minipage}
\caption{Slices (corresponding to $\omega_2=0$) through the internal space for six different starting points on the constraint surface. In all cases, $\lambda_0=2, \mu_0=\nu_0=2/3$. Each figure corresponds to one of the initial conditions identified by green dots in Fig.~\ref{FIG:ProjectedPoints}: (a) corresponds to point 1a; (b) corresponds to point 1b; (c) corresponds to point 2a; (d) corresponds to point 2b; (e) corresponds to point 3a; (f) corresponds to point 3b. Trajectories that begin within the light blue region flow to the de Sitter fixed point, {\bf dS}. Trajectories that begin in the orange region exit the EFT regime of validity with a background that is anisotropic. Fractions of the internal space that flow to various final states are displayed in Table~\ref{TAB:v11}.}
        \label{FIG:v11}
\end{figure*}

We first consider the selection $\lambda_0=2$, $\mu_0=\nu_0=2/3$. Slices through the internal space $\{ \omega_1, \omega_2, y \}$ for the six different starting points on the constraint surface are displayed in Fig.~\ref{FIG:v11}.
The fraction of the internal space that flows either to {\bf dS} or leaves the EFT regime of validity in an anisotropic state is presented in Table~\ref{TAB:v11}. Note that in this case, there are significant portions of the internal space (in light blue in Fig.~\ref{FIG:v11}) that both violate DEC and that flow to {\bf dS}. All of these initial conditions lie outside the domain to which Wald's theorem \cite{wald_83} applies.   
\begin{table*}
\begin{center}
 \begin{tabular}{ c | c | c | c | c | c | c} 
  		&& Initial $y$-values &&& \multicolumn{2}{c}{Fraction of the internal space that flows to \dots} \\ [1ex]
 {\it Fig.} & $x(t_i)$ & considered & $\lambda_0$ & $\mu_0=\nu_0$ & {\bf dS} & {\it anisotropic EXIT}\\ [1ex] 
 \Xhline{2pt}
\ref{FIG:v11}a&0.1& $|y(t_i)| \leq 1$ & 2& 2/3&0.2261&0.7739\Tstrut\Bstrut\\
\ref{FIG:v11}a&0.1& $0\leq y(t_i) \leq 1$ & 2& 2/3&0.0065&0.9935\Tstrut\Bstrut\\
 \Xhline{0.25pt}
\ref{FIG:v11}b&0.1& $|y(t_i)| \leq 1$ & 2&  2/3&0.2583&0.7417\Tstrut\Bstrut\\
\ref{FIG:v11}b&0.1& $0\leq y(t_i) \leq 1$ & 2&  2/3&0.0065&0.9935\Tstrut\Bstrut\\
 \Xhline{0.25pt}
 \ref{FIG:v11}c&0.5& $|y(t_i)| \leq 1$ & 2& 2/3&0.4303&0.5697\Tstrut\Bstrut\\
\ref{FIG:v11}c&0.5& $0\leq y(t_i) \leq 1$ & 2& 2/3&0.0996&0.9004\Tstrut\Bstrut\\
 \Xhline{0.25pt}
 \ref{FIG:v11}d&0.5& $|y(t_i)| \leq 1$ & 2& 2/3&0.4397&0.5603\Tstrut\Bstrut\\
\ref{FIG:v11}d&0.5& $0\leq y(t_i) \leq 1$ & 2& 2/3&0.1030&0.8970\Tstrut\Bstrut\\
 \Xhline{0.25pt}
\ref{FIG:v11}e&0.7& $|y(t_i)| \leq 1$ & 2& 2/3&0.5401&0.4599\Tstrut\Bstrut\\
\ref{FIG:v11}e&0.7& $0\leq y(t_i) \leq 1$ & 2& 2/3&0.1965&0.8035\Tstrut\Bstrut\\
 \Xhline{0.25pt}
\ref{FIG:v11}f&0.7& $|y(t_i)| \leq 1$ & 2& 2/3&0.5419&0.4581\Tstrut\Bstrut\\
\rowcolor{light-gray}\ref{FIG:v11}f&0.7& $0\leq y(t_i) \leq 1$ & 2& 2/3&0.2072&0.7928\Tstrut\Bstrut
 \end{tabular}
 \caption{The fraction of the internal space---coordinatized by $(\omega_1, \omega_2, y)$---that either flows to the isotropic de Sitter fixed point ({\bf dS}) or leaves the regime of validity of the EFT in an anisotropic state, denoted: `{\it anisotropic EXIT}\,'. The row that gives the highest fraction of the space that flows to {\bf dS}, for $y(t_i)\geq 0$, is highlighted in gray.}
 \label{TAB:v11}
\end{center}
\end{table*}

\begin{figure*}
\begin{minipage}{.3\linewidth}
\centering
\subfloat[]{\includegraphics[scale=0.425]{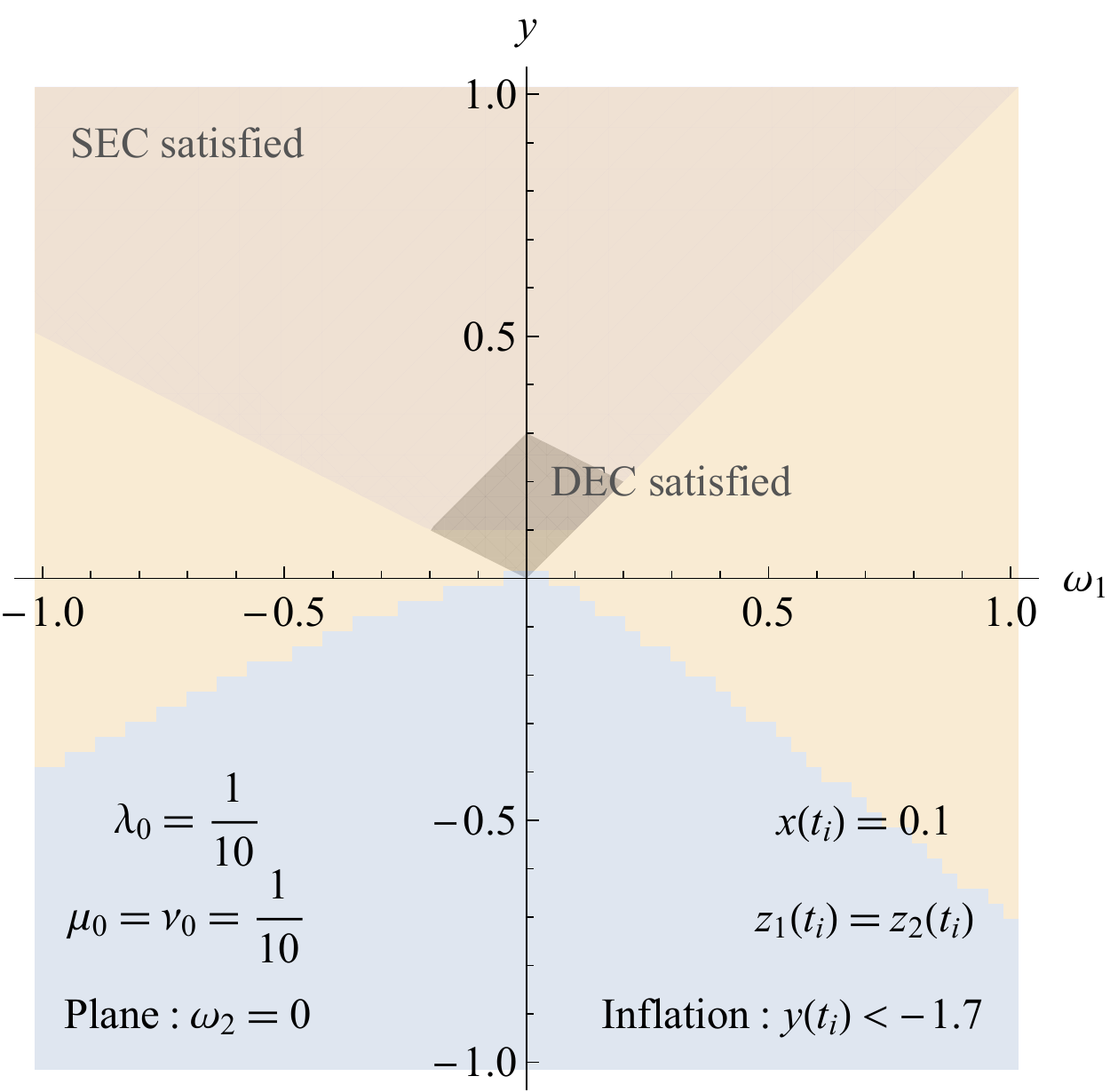}}
\end{minipage}
\begin{minipage}{.3\linewidth}
\centering
\subfloat[]{\includegraphics[scale=0.425]{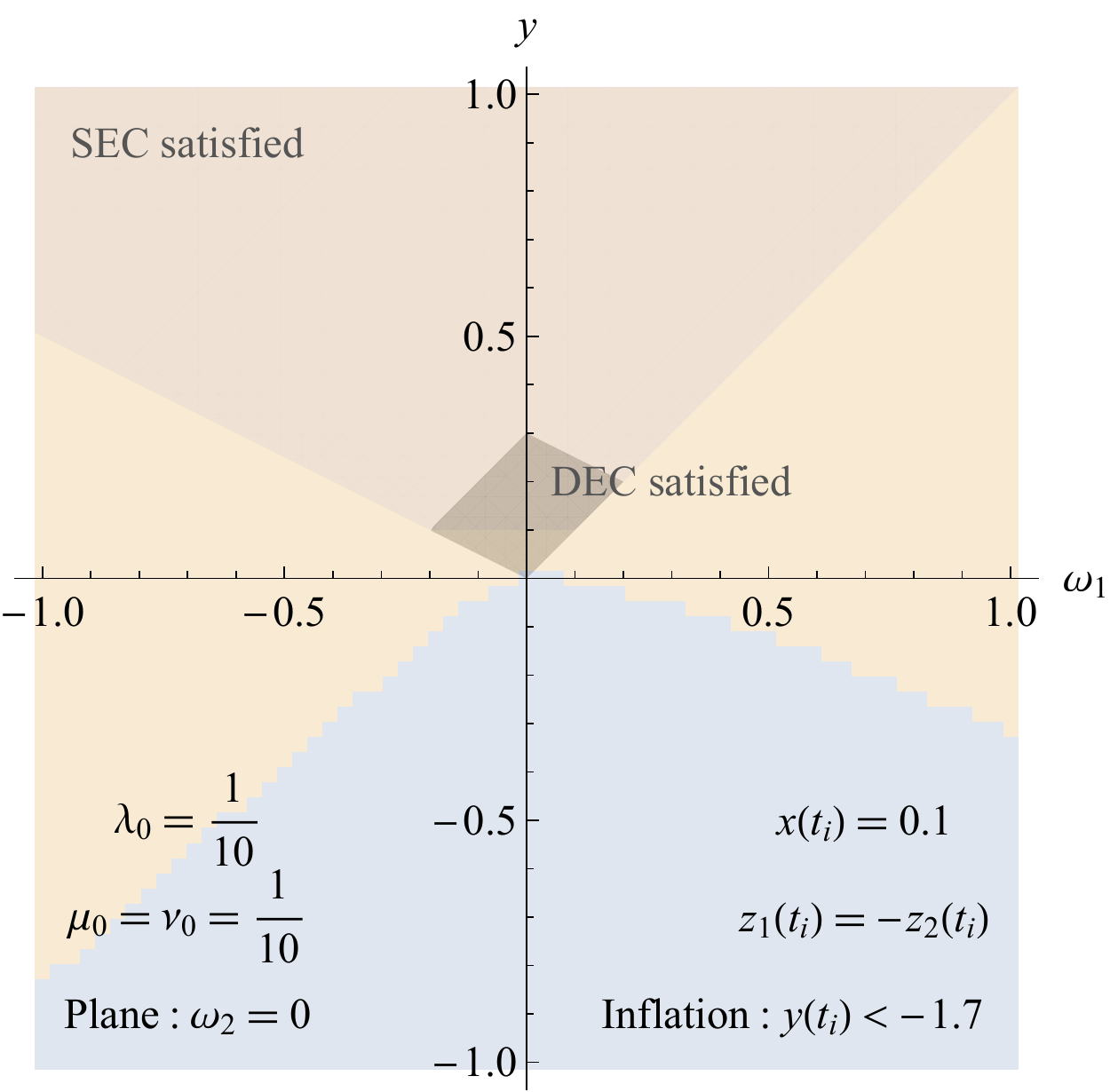}}
\end{minipage}
\begin{minipage}{.3\linewidth}
\centering
\subfloat[]{\includegraphics[scale=0.425]{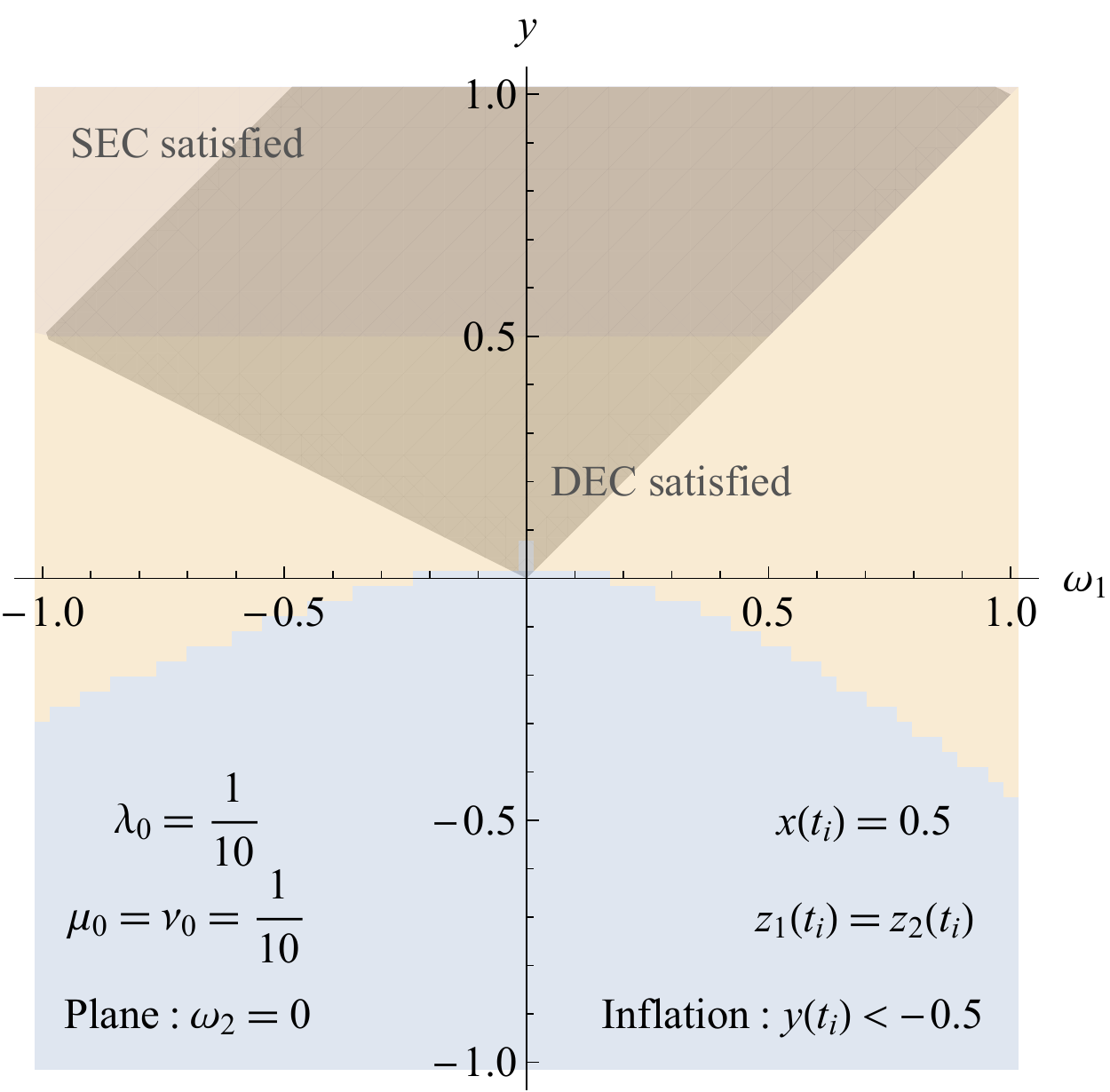}}
\end{minipage}
\begin{minipage}{.3\linewidth}
\centering
\subfloat[]{\includegraphics[scale=0.425]{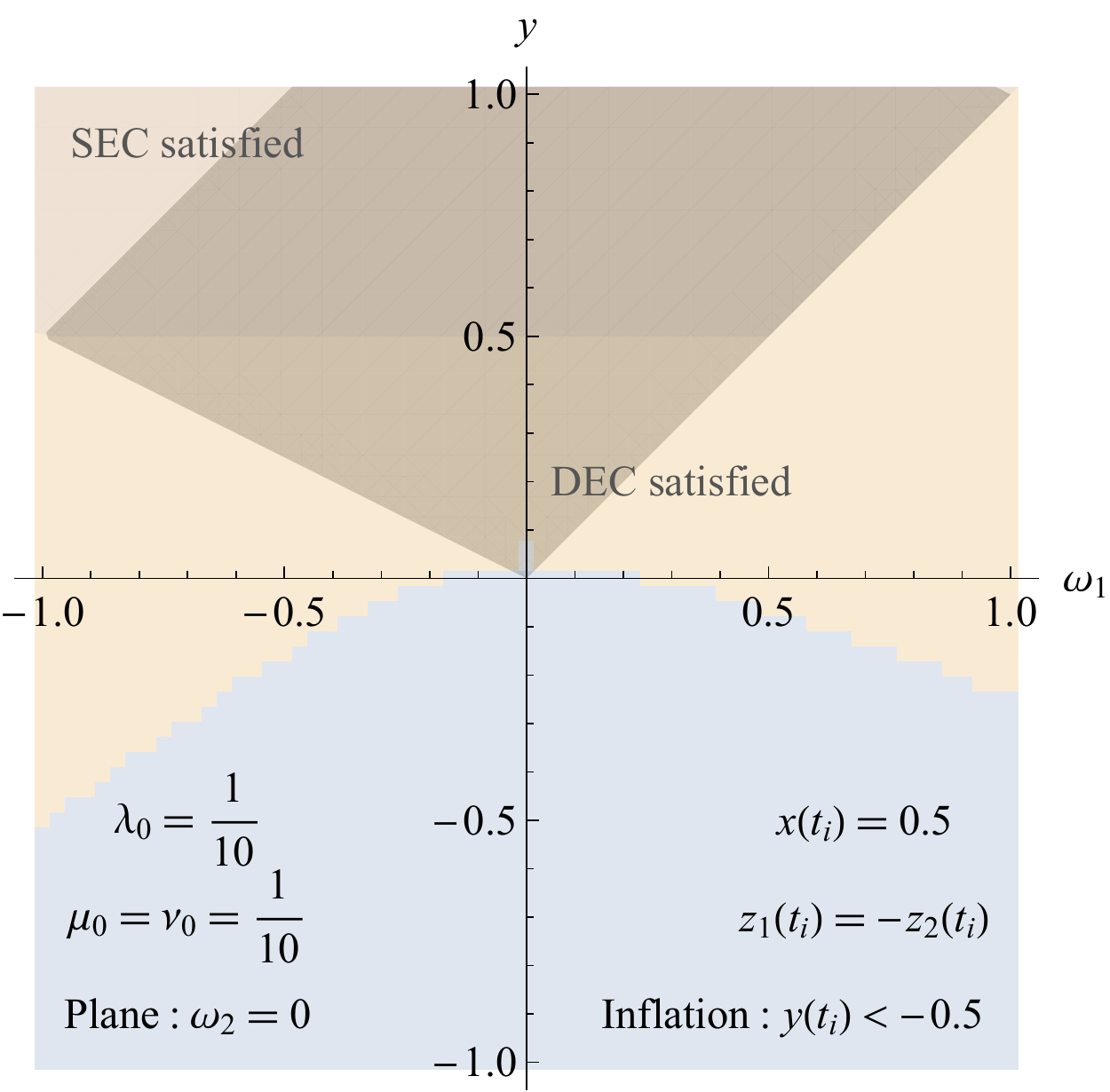}}
\end{minipage}
\begin{minipage}{.3\linewidth}
\centering
\subfloat[]{\includegraphics[scale=0.425]{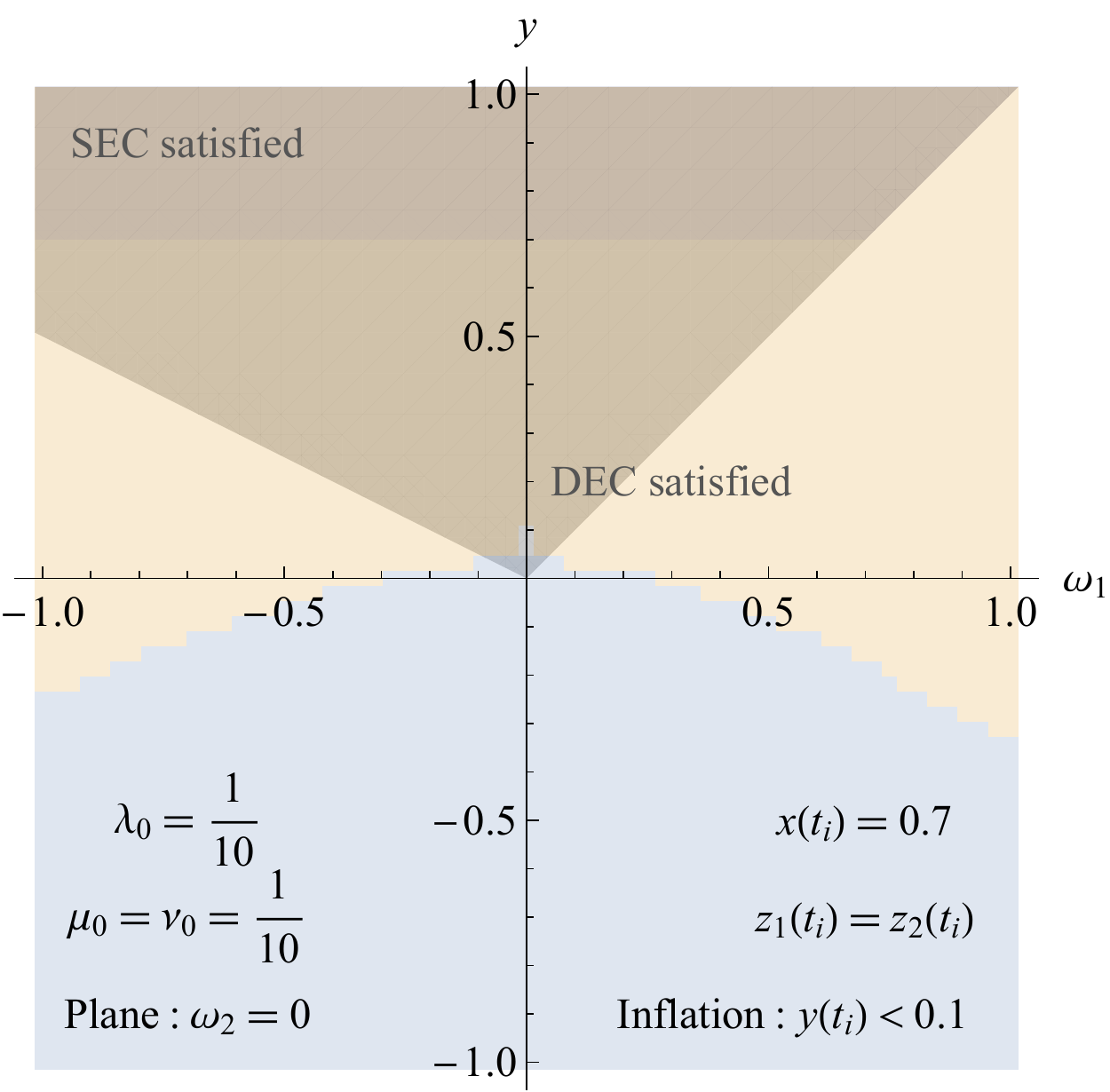}}
\end{minipage}
\begin{minipage}{.3\linewidth}
\centering
\subfloat[]{\includegraphics[scale=0.425]{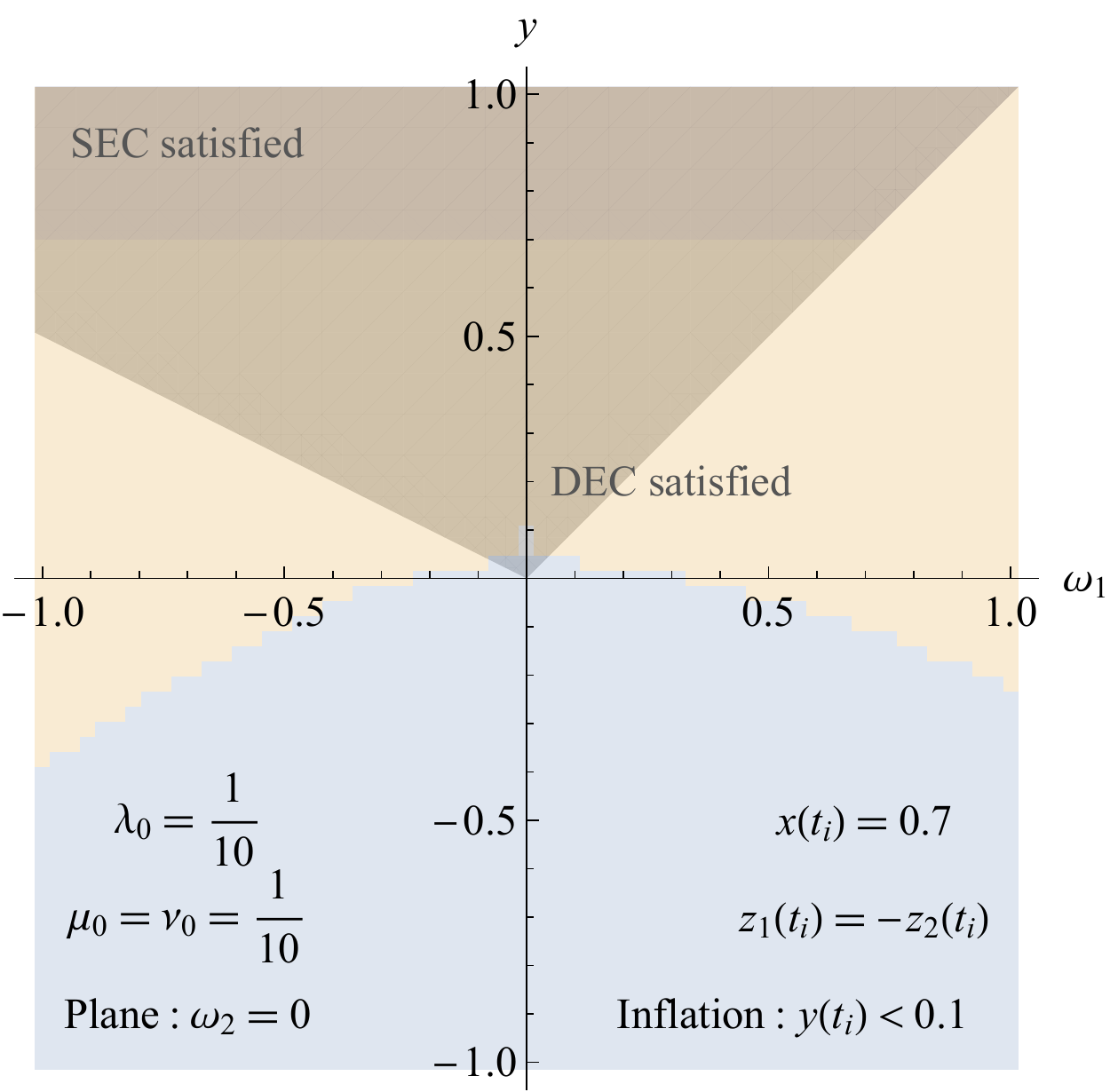}}
\end{minipage}
\caption{Slices (corresponding to $\omega_2=0$) through the internal space for six different starting points on the constraint surface. In all cases, $\lambda_0=1/10, \mu_0=\nu_0=1/10$. Each figure corresponds to one of the initial conditions identified by green dots in Fig.~\ref{FIG:ProjectedPoints}: (a) corresponds to point 1a; (b) corresponds to point 1b; (c) corresponds to point 2a; (d) corresponds to point 2b; (e) corresponds to point 3a; (f) corresponds to point 3b. Trajectories that begin within the light blue region flow to the de Sitter fixed point, {\bf dS}. Trajectories that begin in the orange region exit the EFT regime of validity with a background that is anisotropic. Fractions of the internal space that flow to various final states are displayed in Table~\ref{TAB:v16}.}
        \label{FIG:v16}
\end{figure*}

Next we consider $\lambda_0=1/10$, $\mu_0=\nu_0=1/10$. Compared to the scenarios featured in Fig.~\ref{FIG:v11} and Table \ref{TAB:v11}, for these selections of $\mu_0$ and $\nu_0$, the anisotropic pressures dissipate considerably more slowly. Slices through the internal space of $\{ \omega_1, \omega_2, y\}$ for the six different starting points on the constraint surface are displayed in Fig.~\ref{FIG:v16}.
The fraction of the internal space that flows either to {\bf dS} or leaves the EFT regime of validity in an anisotropic state is presented in Table~\ref{TAB:v16}. Compared to Fig.~\ref{FIG:v11}, there are fewer initial conditions that yield flows to {\bf dS} for non-negative $y (t_i)$, though trajectories with $y (t_i) < 0$ still generically flow to {\bf dS}. As we noted in Section \ref{SEC:ISOstability}, trajectories with $y (t_i) < 0$ begin with $P < - \rho$; hence they accelerate very efficiently at early times, diluting the anisotropic pressures and driving the system toward the {\bf dS} fixed point.
Again, all parts of the internal space that are shaded light blue correspond to initial conditions that lie outside the domain to which Wald's theorem \cite{wald_83} applies. 
\begin{table*}
\begin{center}
 \begin{tabular}{ c | c | c | c | c | c | c | c} 
  		&& Initial $y$-values &&& \multicolumn{3}{c}{Fraction of the internal space that flows to \dots} \\ [1ex]
 {\it Fig.} & $x(t_i)$ & considered & $\lambda_0$ & $\mu_0=\nu_0$ & {\bf dS} & {\it isotropic EXIT} & {\it anisotropic EXIT}\\ [1ex] 
 \Xhline{2pt}
\ref{FIG:v16}a&0.1& $|y(t_i)| \leq 1$ & 1/10& 1/10&0.2996&0&0.7004\Tstrut\Bstrut\\
\ref{FIG:v16}a&0.1& $0\leq y(t_i) \leq 1$ & 1/10& 1/10&0.0001&0&0.9999\Tstrut\Bstrut\\
 \Xhline{0.25pt}
\ref{FIG:v16}b&0.1& $|y(t_i)| \leq 1$ & 1/10&  1/10&0.2991&0&0.7009\Tstrut\Bstrut\\
\ref{FIG:v16}b&0.1& $0\leq y(t_i) \leq 1$ & 1/10&  1/10&0.0001&0&0.9999\Tstrut\Bstrut\\
 \Xhline{0.25pt}
 \ref{FIG:v16}c&0.5& $|y(t_i)| \leq 1$ & 1/10& 1/10&0.3776&--&0.6224\Tstrut\Bstrut\\
\ref{FIG:v16}c&0.5& $0\leq y(t_i) \leq 1$ & 1/10& 1/10&0.0012&--&0.9988\Tstrut\Bstrut\\
 \Xhline{0.25pt}
 \ref{FIG:v16}d&0.5& $|y(t_i)| \leq 1$ & 1/10& 1/10&0.3741&--&0.6259\Tstrut\Bstrut\\
\ref{FIG:v16}d&0.5& $0\leq y(t_i) \leq 1$ & 1/10& 1/10&0.0011&--&0.9988\Tstrut\Bstrut\\
 \Xhline{0.25pt}
\ref{FIG:v16}e&0.7& $|y(t_i)| \leq 1$ & 1/10& 1/10&0.4044&--&0.5956\Tstrut\Bstrut\\
\rowcolor{light-gray}\ref{FIG:v16}e&0.7& $0\leq y(t_i) \leq 1$ & 1/10& 1/10&0.0024&--&0.9976\Tstrut\Bstrut\\
 \Xhline{0.25pt}
\ref{FIG:v16}f&0.7& $|y(t_i)| \leq 1$ & 1/10& 1/10&0.4024&--&0.5975\Tstrut\Bstrut\\
\rowcolor{light-gray}\ref{FIG:v16}f&0.7& $0\leq y(t_i) \leq 1$ & 1/10& 1/10&0.0024&--&0.9976\Tstrut\Bstrut
 \end{tabular}
 \caption{The fraction of the internal space---coordinatized by $(\omega_1, \omega_2, y)$---that either flows to the isotropic de Sitter fixed point ({\bf dS}) or that leaves the regime of validity of the EFT in an (an)isotropic state, denoted: `{\it (an)isotropic EXIT}\,'. The rows that give the highest fraction of the space that flows to {\bf dS}, for $y(t_i)\geq 0$, are highlighted in gray. A dash `--' indicates a value $<0.0001$.}
\label{TAB:v16}
\end{center}
\end{table*}

We also studied scenarios intermediate between these two sets of cases, including scenarios in which $\lambda_0 \in \{ 1/10, 1, 2\}$ and $\mu_0=\nu_0 \in \{ 1/10, 2/3 \}$. See Table~\ref{TAB:IntScenarios} for a summary.
\begin{table}
\begin{center}
 \begin{tabular}{ c | c | c | c} 
	Initial $y$-values  & $\lambda_0$ & $\mu_0=\nu_0$ & Max.~fraction to {\bf dS} \\ [1ex] 
 \Xhline{2pt}
 $|y(t_i)| \leq 1$ & 2&  2/3&$\sim 54$\%\Tstrut\Bstrut\\
$0\leq y(t_i) \leq 1$ & 2&  2/3&$\sim 21$\%\Tstrut\Bstrut\\
 \Xhline{0.25pt}
$|y(t_i)| \leq 1$ & 2& 1/10&$\sim 11$\%\Tstrut\Bstrut\\
$0\leq y(t_i) \leq 1$ & 2& 1/10&$\sim 2$\%\Tstrut\Bstrut\\
 \Xhline{0.25pt}
$|y(t_i)| \leq 1$ & 1&  2/3&$\sim 51$\%\Tstrut\Bstrut\\
$0\leq y(t_i) \leq 1$ & 1&  2/3&$\sim 13$\%\Tstrut\Bstrut\\
 \Xhline{0.25pt}
 $|y(t_i)| \leq 1$ & 1& 1/10&$\sim 20$\%\Tstrut\Bstrut\\
$0\leq y(t_i) \leq 1$ & 1& 1/10&$\sim 1$\%\Tstrut\Bstrut\\
  \Xhline{0.25pt}
$|y(t_i)| \leq 1$ & 1/10&  2/3&$\sim 50$\%\Tstrut\Bstrut\\
$0\leq y(t_i) \leq 1$ & 1/10&  2/3&$\sim 7$\%\Tstrut\Bstrut\\
 \Xhline{0.25pt}
 $|y(t_i)| \leq 1$ & 1/10& 1/10&$\sim 40$\%\Tstrut\Bstrut\\
$0\leq y(t_i) \leq 1$ & 1/10& 1/10& $<1$\%\Tstrut\Bstrut
\end{tabular}
 \caption{A summary of results for the  maximum fraction of the internal space that flows to {\bf dS} for each of the values of $\lambda_0$ and $\mu_0=\nu_0$ considered. For each value of $\lambda_0$ and $\mu_0=\nu_0$, the points on the constraint surface with respect to which the maximum fraction has been computed are described in Fig.~\ref{FIG:ProjectedPoints}.}
\label{TAB:IntScenarios}
\end{center}
\end{table}
The results obtained collectively yield the following conclusions. 
\begin{itemize}
\item[--] An initial $x(t_i)=0.7$ always yields a larger fraction of the internal space that flows to {\bf dS} than scenarios with smaller $x (t_i)$, for all values of the parameters $\lambda_0$ and $\mu_0=\nu_0$ probed; this fraction gets progressively smaller as $x(t_i)$ decreases. This trend makes sense, since larger values of $x (t_i)$ correspond to initial states that are closer to initially being isotropic.  
\item[--] There were no flows to {\bf FPb} (the non-de Sitter isotropic fixed point): that fixed point is generally a saddle point (at least for values where $\lambda_0\neq\mu_0$ and $\lambda_0\neq\nu_0$), and one therefore expects that only trajectories with very specific initial conditions would reach it.
\item[--] For $0\leq y(t_i)\leq 1$ and for a fixed $\mu_0=\nu_0$, increasing $\lambda_0$ generally increases the fraction of the internal space that flows to {\bf dS}. This effect is far more appreciable for $\mu_0=\nu_0=2/3$ than for $\mu_0 = \nu_0 = 1/10$. 
\item[--] If we include trajectories with $y (t_i) < 0$, a larger fraction of the internal space generally flows to {\bf dS} for {\it smaller} values of $\lambda_0$ (holding $\mu_0=\nu_0$ fixed). The super-accelerated expansion driven by $y (t_i) < 0$ (corresponding to $P(t_i) < - \rho (t_i)$) lingers longer for smaller $\lambda_0$. For $\mu_0=\nu_0=2/3$, the maximum value of the fraction that flows to {\bf dS} is rather stable, at around 50\%.
\item[--] Much as we discussed at the beginning of Sec.~\ref{SEC:PPS}, we again find that larger values of $\lambda_0, \mu_0$, and $\nu_0$ generally create conditions that mimic evolution of a system with a cosmological constant. Correspondingly, a larger fraction of the internal space flows to {\bf dS}. 
\end{itemize}

\begin{figure*}
\begin{minipage}{.3\linewidth}
\centering
\subfloat[]{\includegraphics[scale=0.425]{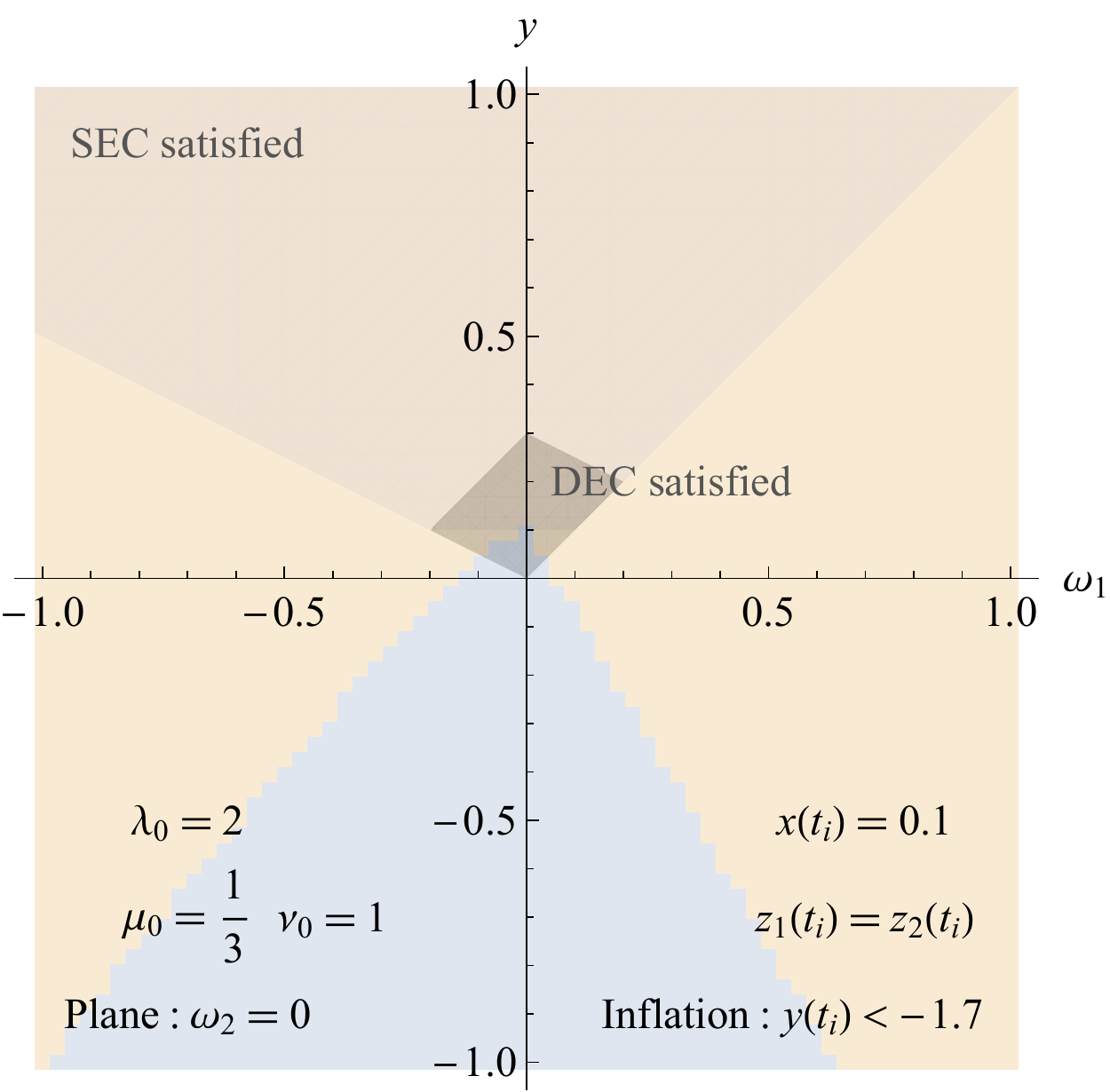}}
\end{minipage}
\begin{minipage}{.3\linewidth}
\centering
\subfloat[]{\includegraphics[scale=0.425]{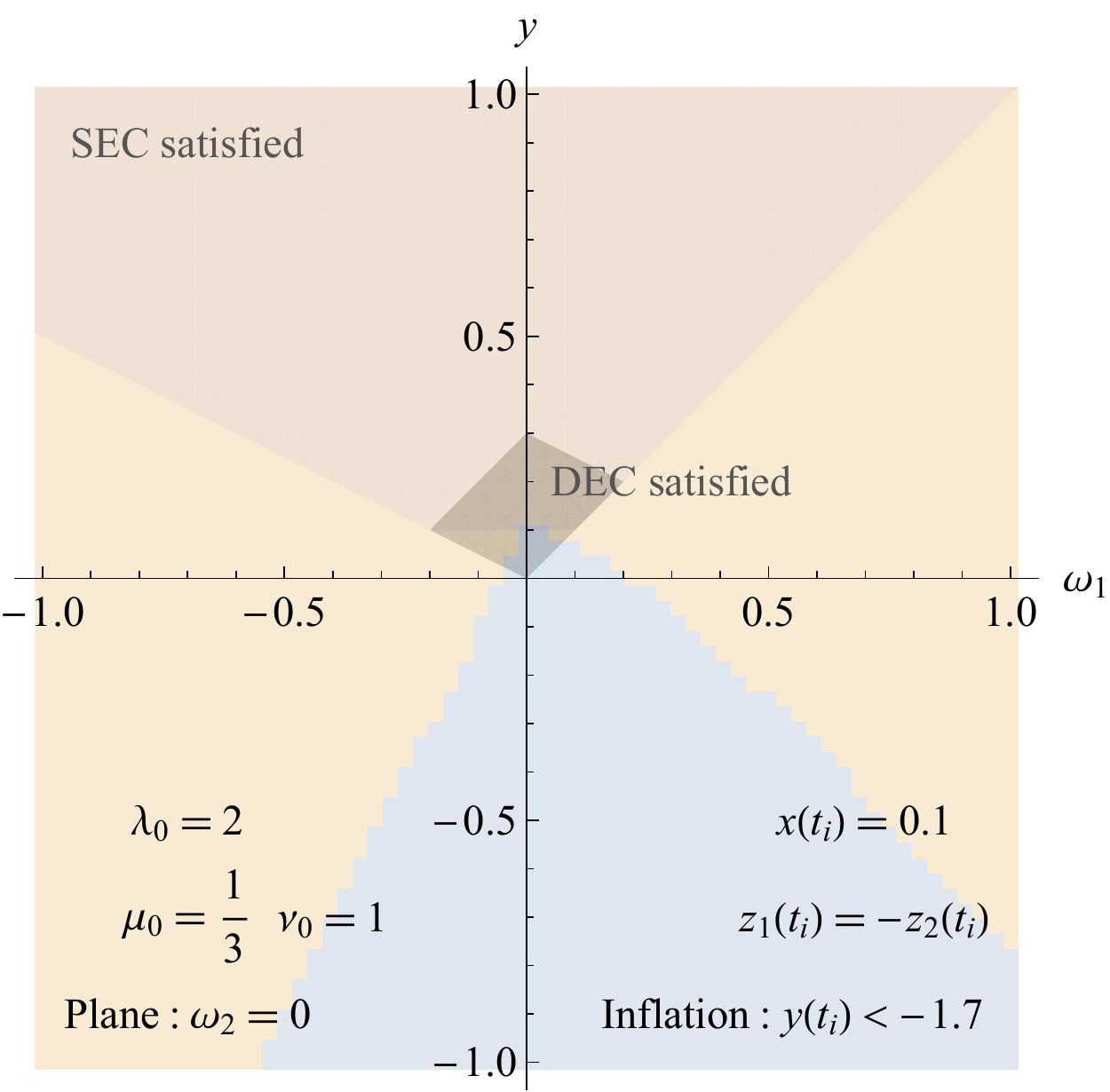}}
\end{minipage}
\begin{minipage}{.3\linewidth}
\centering
\subfloat[]{\includegraphics[scale=0.425]{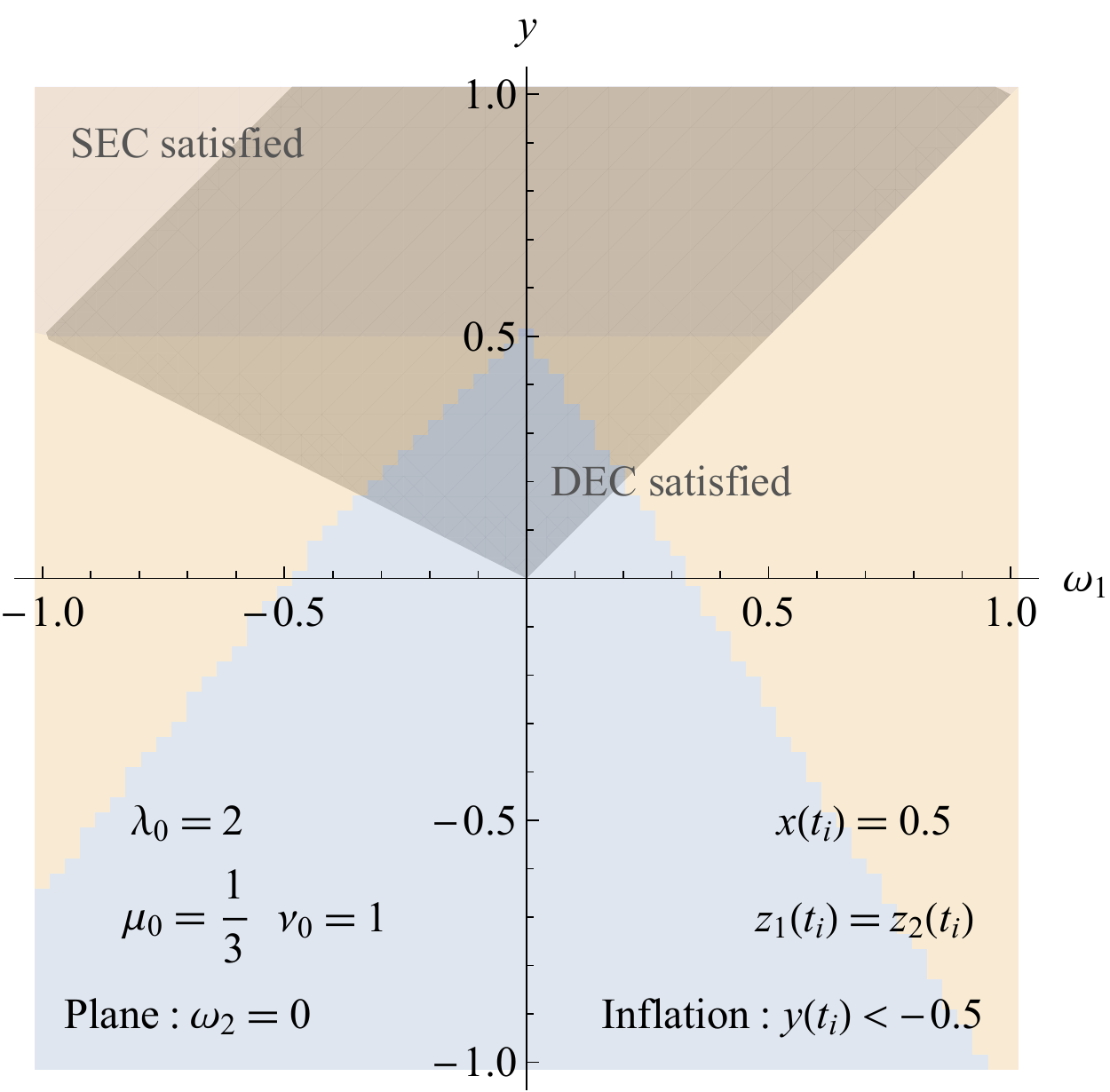}}
\end{minipage}
\begin{minipage}{.3\linewidth}
\centering
\subfloat[]{\includegraphics[scale=0.425]{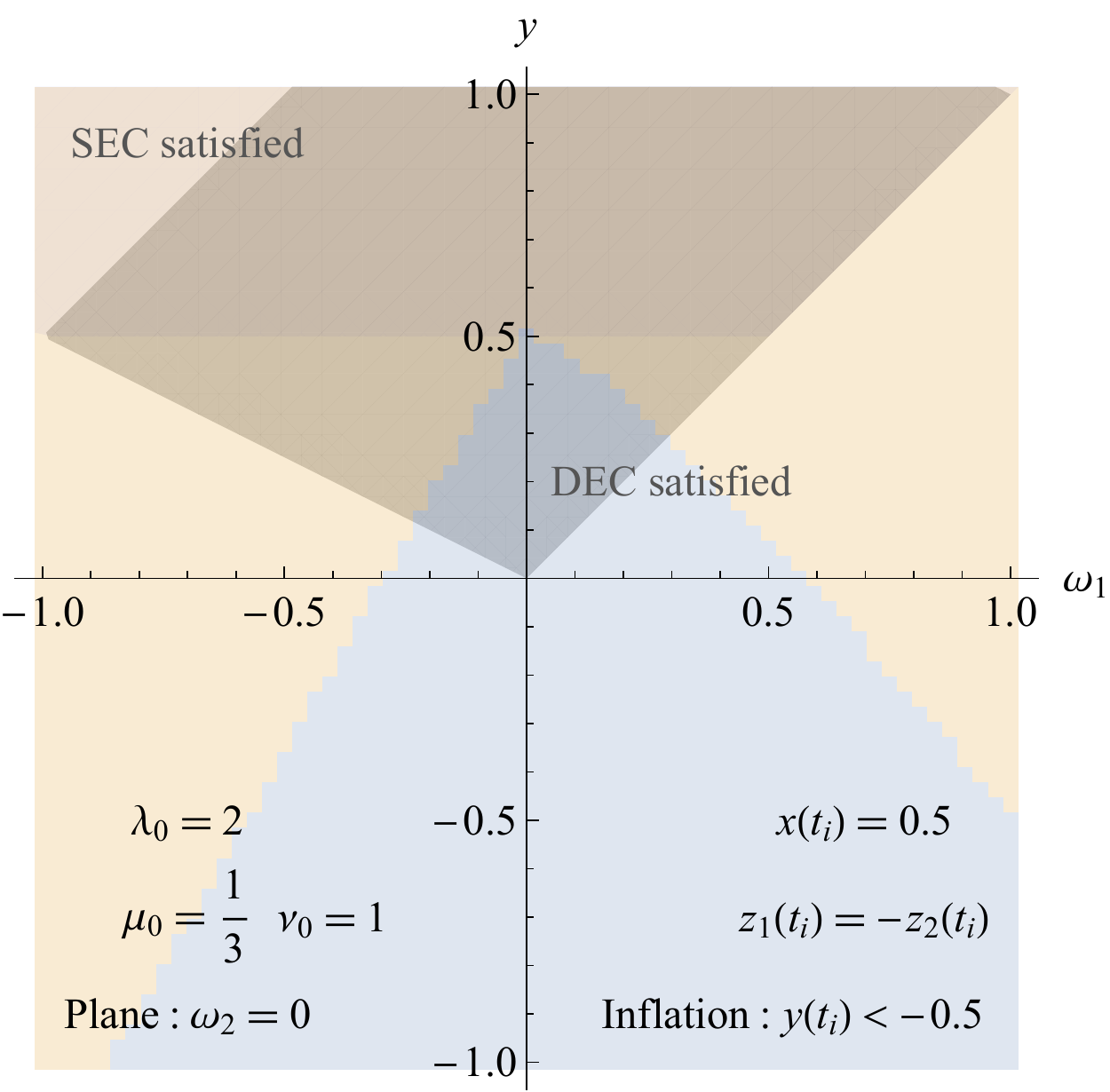}}
\end{minipage}
\begin{minipage}{.3\linewidth}
\centering
\subfloat[]{\includegraphics[scale=0.425]{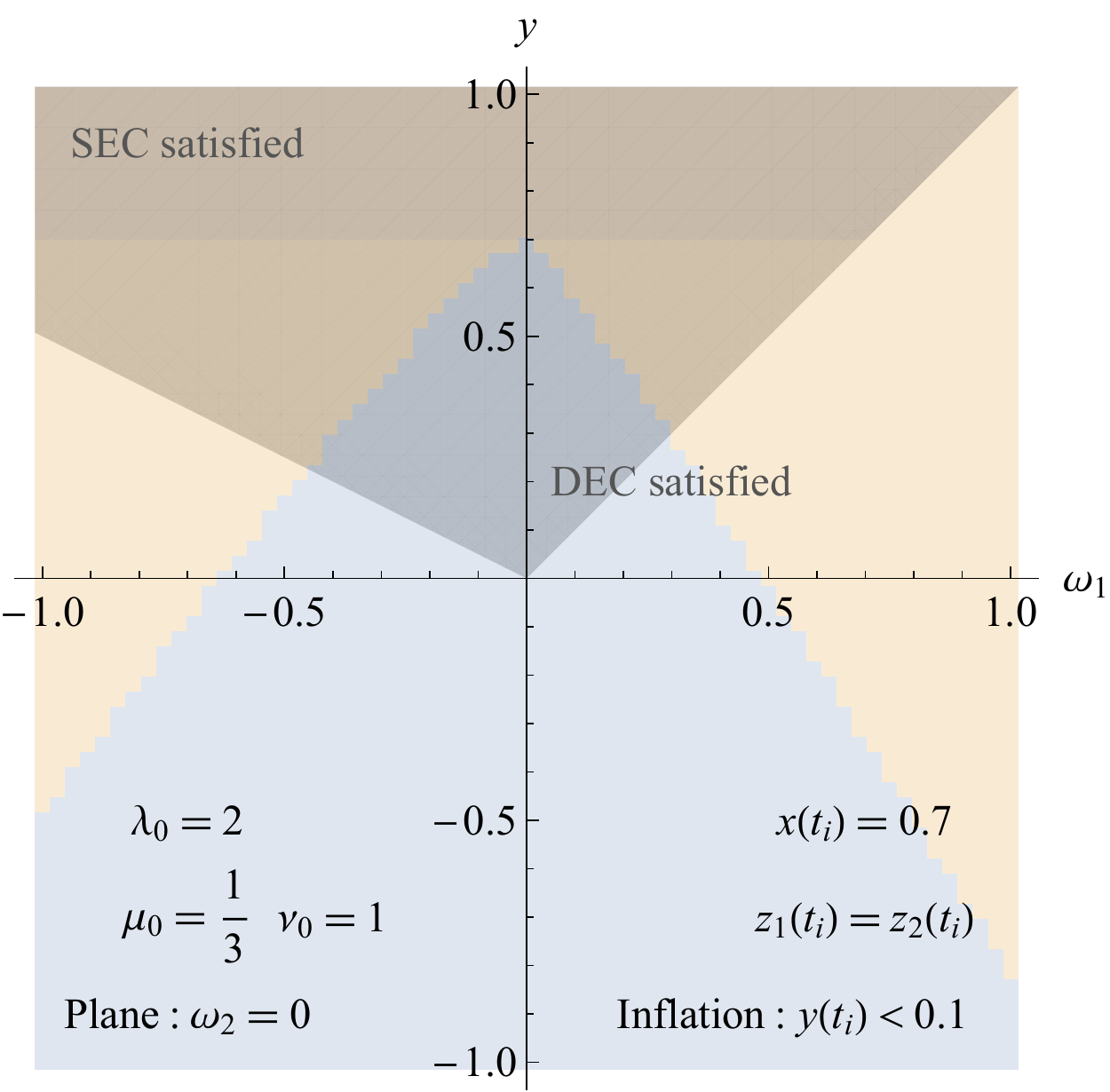}}
\end{minipage}
\begin{minipage}{.3\linewidth}
\centering
\subfloat[]{\includegraphics[scale=0.425]{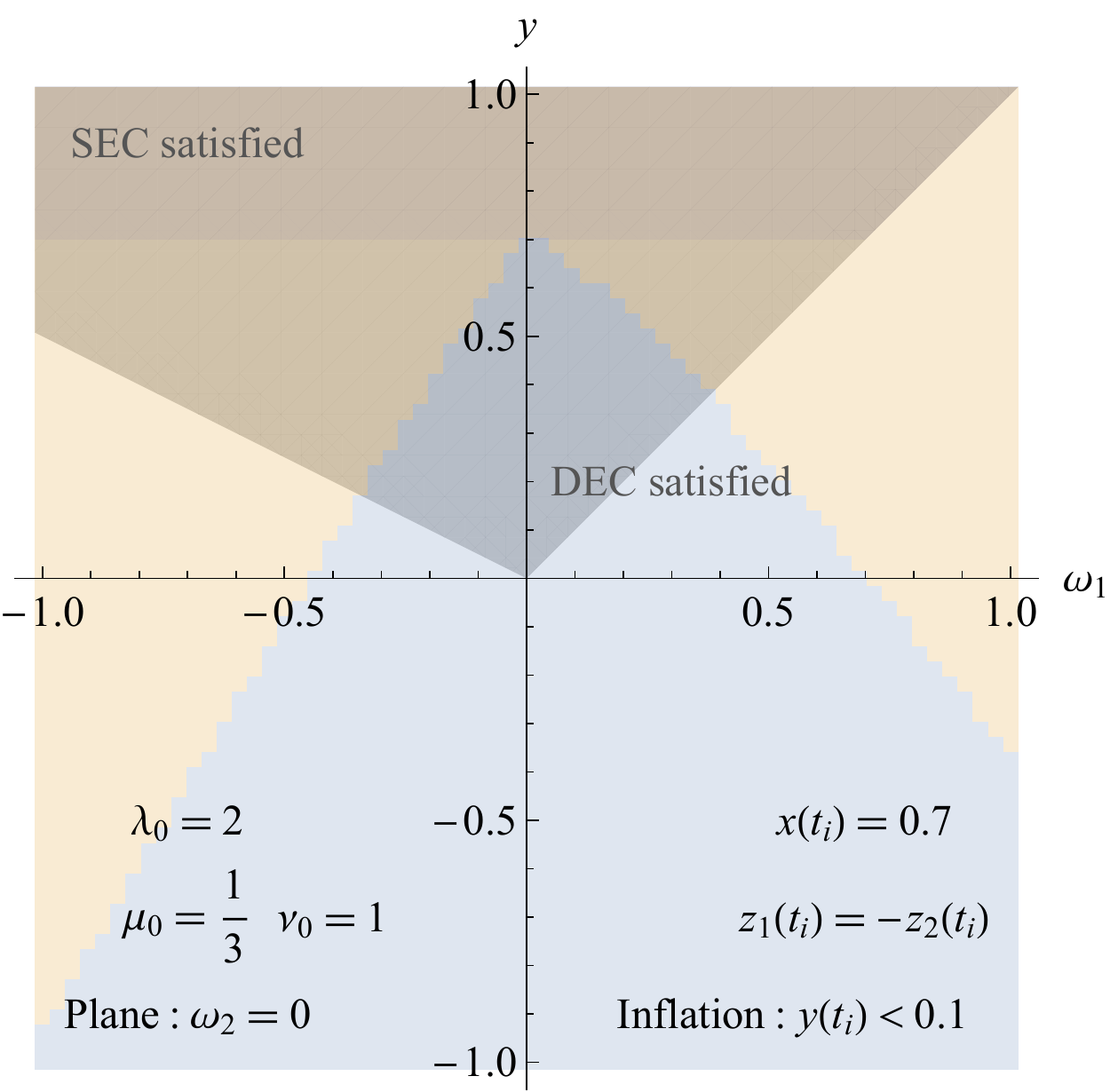}}
\end{minipage}
\caption{Slices (corresponding to $\omega_2=0$) through the internal space for six different starting points on the constraint surface. In all cases, $\lambda_0=2$, $\mu_0=1/3$, and $\nu_0 = 1$. Each figure corresponds to one of the initial conditions identified by green dots in Fig.~\ref{FIG:ProjectedPoints}: (a) corresponds to point 1a; (b) corresponds to point 1b; (c) corresponds to point 2a; (d) corresponds to point 2b; (e) corresponds to point 3a; (f) corresponds to point 3b. Trajectories that begin within the light blue region flow to the de Sitter fixed point, {\bf dS}. Trajectories that begin in the orange region exit the EFT regime of validity with a background that is anisotropic. Fractions of the internal space that flow to various final states are displayed in Table~\ref{TAB:v18}.}
        \label{FIG:v18}
\end{figure*}

Finally, we considered scenarios in which $\mu_0 \neq \nu_0$, such as the case $\mu_0 = 1/3$ and $\nu_0 = 1$, to keep their average the same as the cases with $\mu_0 = \nu_0 = 2/3$. In Fig.~\ref{FIG:v18} we display results for $\lambda_0 = 2$.
The fraction of the internal space that flows either to {\bf dS} or leaves the EFT regime of validity in an anisotropic state is presented in Table~\ref{TAB:v18}.
\begin{table*}
\begin{center}
 \begin{tabular}{ c | c | c | c | c | c | c | c} 
  		&& Initial $y$-values &&&& \multicolumn{2}{c}{Fraction of the internal space that flows to \dots} \\ [1ex]
 {\it Fig.} & $x(t_i)$ & considered & $\lambda_0$ & $\mu_0$ & $\nu_0$ & {\bf dS} & {\it anisotropic EXIT}\\ [1ex] 
 \Xhline{2pt}
\ref{FIG:v18}a&0.1& $|y(t_i)| \leq 1$ & 2 & 1/3 & 1 &0.1876&0.8124\Tstrut\Bstrut\\
\ref{FIG:v18}a&0.1& $0\leq y(t_i) \leq 1$ & 2 & 1/3 & 1 &0.0037&0.9963\Tstrut\Bstrut\\
 \Xhline{0.25pt}
\ref{FIG:v18}b&0.1& $|y(t_i)| \leq 1$ &  2 & 1/3 & 1&0.2110&0.7890\Tstrut\Bstrut\\
\ref{FIG:v18}b&0.1& $0\leq y(t_i) \leq 1$ & 2 & 1/3 & 1&0.0063&0.9937\Tstrut\Bstrut\\
 \Xhline{0.25pt}
 \ref{FIG:v18}c&0.5& $|y(t_i)| \leq 1$ & 2 & 1/3 & 1&0.3623&0.6377\Tstrut\Bstrut\\
\ref{FIG:v18}c&0.5& $0\leq y(t_i) \leq 1$ & 2 & 1/3 & 1&0.0711&0.9289\Tstrut\Bstrut\\
 \Xhline{0.25pt}
 \ref{FIG:v18}d&0.5& $|y(t_i)| \leq 1$ & 2 & 1/3 & 1&0.3759&0.6241\Tstrut\Bstrut\\
\ref{FIG:v18}d&0.5& $0\leq y(t_i) \leq 1$ & 2 & 1/3 & 1&0.0783&0.9217\Tstrut\Bstrut\\
 \Xhline{0.25pt}
\ref{FIG:v18}e&0.7& $|y(t_i)| \leq 1$ & 2 & 1/3 & 1&0.4668&0.5332\Tstrut\Bstrut\\
\ref{FIG:v18}e&0.7& $0\leq y(t_i) \leq 1$ & 2 & 1/3 & 1&0.1470&0.8530\Tstrut\Bstrut\\
 \Xhline{0.25pt}
\ref{FIG:v18}f&0.7& $|y(t_i)| \leq 1$ & 2 & 1/3 & 1&0.4716&0.5284\Tstrut\Bstrut\\
\rowcolor{light-gray}\ref{FIG:v18}f&0.7& $0\leq y(t_i) \leq 1$ & 2 & 1/3 & 1&0.1535&0.8465\Tstrut\Bstrut
 \end{tabular}
 \caption{Fraction of the internal space---coordinatized by $\{ \omega_1, \omega_2, y \}$---that either flows to the isotropic de Sitter fixed point ({\bf dS}) or that leaves the regime of validity of the EFT in an anisotropic, denoted: `{\it anisotropic EXIT}\,'. The row that gives the highest fraction of the space that flows to {\bf dS}, for $y(t_i)\geq 0$, is highlighted in gray.}
 \label{TAB:v18}
\end{center}
\end{table*}
Again we find regions (in light blue) where the DEC is not satisfied that nonetheless flow to the de Sitter fixed point. 
 
\section{Discussion}\label{SEC:Discussion}

By developing an effective field theory (EFT) treatment for model-independent, ``single-clock'' systems in homogeneous yet anisotropic spacetimes, we have analyzed dynamical flows of initially expanding spaces into isotropic states at late times---in particular, into de Sitter space. Our approach applies to all such single-clock scenarios and is not limited to realizations that are compatible with the usual single-scalar-field constructions. Our analysis can accommodate scenarios that fall outside the domain to which Wald's influential work \cite{wald_83} on the isotropization of homogeneous spacetimes applies.

Wald's cosmic no-hair theorem \cite{wald_83} applies to systems in which the energy-momentum tensor includes a bare cosmological constant plus additional matter degrees of freedom that satisfy the dominant and strong energy conditions. Our EFT treatment, on the other hand, does not include a bare cosmological constant. In addition, in light of work that describes how readily various energy conditions may be violated in otherwise well-behaved classical and semi-classical settings \cite{Kandrup:1992xw,Barcelo:1999hq,Visser:1999de,Barcelo:2000zf,Barcelo:2002bv,Bellucci:2001cc,Dubovsky:2005xd,Nicolis:2009qm,Rubakov:2014jja,Martin-Moruno:2017exc}, we drop the requirement that the effective matter degrees of freedom initially satisfy such energy conditions. 

The EFT we develop includes an action for matter degrees of freedom that is linear in perturbations about a 
suitably general anisotropic (Bianchi I) background. This action yields an energy-momentum tensor that may be reorganized into canonical form, including terms that describe anisotropic pressures. We identify salient variables that allow us to define expansion-normalized dimensionless dynamical variables, akin to the analysis in Refs.~\cite{vandenhoogen+coley_95, frusciante+al_14, Azhar:2018nol}. One may then construct a closed dynamical system in which one can uniquely identify isotropic states, including de Sitter evolution.

Our analysis of the ``principal phase space'' (the lowest-dimensional phase space) in this setting  reveals rich dynamical structure. We identify two fixed points that correspond to isotropic backgrounds, including a de Sitter attractor (labeled {\bf dS}) and a non-de Sitter fixed point (that is generally a saddle point). In addition to four further fixed points that correspond to anisotropic backgrounds, we identify a one-parameter set of (unstable) fixed points that are consistent with the {\it Kasner circle}---a one-parameter set of anisotropic vacuum solutions in which anisotropies are ``frozen-in'' (though they dissipate over time). 

Within this framework, we perform two different analyses. In Sec.~\ref{SEC:ISOstability} we investigate the stability of initially isotropic backgrounds in the presence of anisotropic pressures, identifying regions of phase space that flow to various final states, including flows to the isotropic de Sitter fixed point, {\bf dS}. (See Fig.~\ref{FIG:Stability_1} and Table~\ref{TAB:Stability_1}.) For certain reasonable choices of parameters, we find that as much as $70\%$ of the phase space can flow to {\bf dS}, including a significant portion that corresponds to initial conditions that do not satisfy the dominant or strong energy conditions.

In Sec.~\ref{SEC:anISOics} we investigate scenarios in which the background is initially anisotropic, and identify regions of phase space that flow to various final states, including {\bf dS}. (See Figs.~\ref{FIG:v11},~\ref{FIG:v16},~\ref{FIG:v18} and Tables~\ref{TAB:v11},~\ref{TAB:v16}, and~\ref{TAB:v18}.) Again, for reasonable choices of parameters that describe the dynamical systems, we find that as much as about $50\%$ of the phase space flows to {\bf dS}, including significant swaths of initial conditions that do not satisfy the dominant or strong energy conditions. (If we restrict attention to those regions of parameter space for which $P \geq - \rho$ at the initial time, then we find as much as $10\% - 20\%$ of the initial conditions flow to {\bf dS}.) Such flows lie outside the domain to which Wald’s no-hair theorem \cite{wald_83} applies.

Using our general EFT-based dynamical framework, we thus describe evolution of homogeneous spacetimes consistent with the emergence of an {\it effective cosmological constant}, for which the equation of state evolves toward $P / \rho \rightarrow -1$ while the anisotropic pressures redshift away. Our work reveals a broader context in which ``cosmic no-hair conjectures'' may apply, beyond the powerful results first identified by Wald \cite{wald_83}. Such generalized isotropization can further shed light on the range of conditions under which cosmic inflation may begin.

The present work can be extended in several interesting directions. One topic for future research is to identify fixed points and explore dynamical flows for these systems beyond the principal phase space, that is, upon setting at least one of $L, M, N > 0$. Another is to construct an EFT consistent with more general anisotropic backgrounds than the Bianchi I line element of Eq.~(\ref{ds}). Using tools like those developed here, such analyses would further help clarify just how general various cosmic no-hair conjectures may ultimately be.

Finally, an open question regarding any cosmic no-hair analysis is how to account for the second law of thermodynamics for gravitational systems. Penrose, for example, has suggested that entropy might scale with the Weyl curvature tensor \cite{PenroseWeyl1979,Penrose2008,Penrose:2018pyw}, in which case any system that flows from anisotropic to isotropic conditions would appear to violate the second law of thermodynamics. (For a recent review, see Ref.~\cite{Hu:2021pfh}.) Whether the Weyl tensor is an appropriate measure of entropy, or some distinct method of understanding the second law of thermodynamics for cosmological systems would be more appropriate, remains a fascinating question for further research.

\begin{acknowledgments}
We gratefully acknowledge helpful discussions with J\'{e}r\^{o}me Martin.
F.~A.~acknowledges support from the Faculty Research Support Program (FY2019) at the University of Notre Dame. Portions of this work were conducted in MIT's Center for Theoretical Physics and supported in part by the U.~S.~Department of Energy under Contract No.~DE-SC0012567.
\end{acknowledgments}

\appendix

\section{Anisotropic pressure in scalar field models}\label{APP:MultipleScalars}

We first consider a universe filled only with scalar fields, each with a canonical coupling to gravity, and further assume that the gravitational degrees of freedom are described by the canonical Einstein-Hilbert action. The action for such models can be written in the form
%%%%%%
\begin{widetext}
\beq
\begin{split}
S &= \int d^4 x \sqrt{-g} \left[ \frac{ M_{\rm pl}^2}{2} R + {\cal L}^{({ \cal M} )} \right] \\
&= \int d^4 x \sqrt{-g} \left[ \frac{ M_{\rm pl}^2}{2} R - \frac{1}{2} g^{\mu\nu} \, {\cal G}_{IJ} (\phi^K ) \partial_\mu \phi^I \partial_\nu \phi^J - V (\phi^K ) \right] \, .
\end{split}
\label{Smultifield}
\eeq
\end{widetext}
The energy-momentum tensor then takes the form (see, for example, Ref.~\cite{KMS})
%%%%%%%
\begin{widetext}
\beq
\begin{split}
T_{\mu\nu} = - \frac{2}{ \sqrt{-g} } \frac{ \delta S^{( {\cal M} )} }{\delta g^{\mu\nu} } &= g_{\mu\nu} {\cal L}^{( { \cal M} )} - 2 \left( \frac{ \delta {\cal L}^{( {\cal M} )} }{\delta g^{\mu\nu} } \right) \\
&= {\cal G}_{IJ} \partial_\mu \phi^I \partial_\nu \phi^J - g_{\mu\nu} \left[ \frac{1}{2} g^{\alpha \beta} \, {\cal G}_{IJ} \, \partial_\alpha \phi^I \partial_\beta \phi^J + V (\phi^K ) \right] \, ,
\end{split}
\label{Tmnscalars}
\eeq
\end{widetext}
where $S^{( {\cal M} )} \equiv \int d^4 x \sqrt{-g} \, {\cal L}^{( {\cal M} )}$. Even if the field-space manifold has nondiagonal terms, such that ${\cal G}_{IJ} (\phi^K) \neq 0$ for $I \neq J$, and each of the scalar fields couples directly to the others in the interaction potential $V (\phi^K)$, we still find that $T_{ij} \propto g_{ij}$ for a spatially homogeneous system, such that each $\phi^I \rightarrow \phi^I (t)$. In that case,
%%%%%%%
\beq
{\cal G}_{IJ} \, \partial_\mu \phi^I \partial_\nu \phi^J =  {\cal G}_{IJ} \, \dot{\phi}^I \dot{\phi}^J \delta^0_{\>\> \mu} \delta^0_{\>\> \nu} \, ,
\label{GIJkinetic}
\eeq
so that 
%%%%%%%
\beq
T_{ij} =  g_{ij} \left[ \frac{1}{2} {\cal G}_{IJ} \, \dot{\phi}^I \dot{\phi}^J - V (\phi^K ) \right] \, .
\label{Tijscalars}
\eeq
Given that $g_{0i} = 0$ for our background spacetime, we then readily find
%%%%%%
\beq
T^i_{\>\> j} = g^{ik} \, T_{kj} = g^{ik} \, g_{kj} \left[ \frac{1}{2} {\cal G}_{IJ} \dot{\phi}^I \dot{\phi}^J - V (\phi^K ) \right] \propto \delta^i_{\>\> j} \, .
\label{Tijscalars2}
\eeq
Comparing with Eq.~(\ref{Tmncomponents1}), we see that $T^i_{\>\> j} \propto \delta^i_{\>\> j}$ and hence $\pi_{ij} = 0$ for all such models, even those that include curved field-space manifolds characterized by nontrivial metrics ${\cal G}_{IJ} (\phi^K)$.

We may consider more general scenarios that move beyond the canonical Einstein-Hilbert gravitational action. In particular, we can examine the Horndeski action, which consists of the most general set of terms in $(3 + 1)$-dimensional spacetime that yields a second-order equation of motion for a scalar field $\phi (x^\mu)$ coupled to gravity. The action may be parameterized as \cite{HorndeskiRev,Rubakov:2014jja}
%%%%%%%
\begin{widetext}
\beq
\begin{split}
S = \int d^4 x \sqrt{-g} &\Bigg\{ G_2 (\phi, Z) - G_3 (\phi, Z) \, \Box \phi + G_4 (\phi, Z) R + G_{4, Z} \Big[ (\Box \phi )^2 - \nabla^\mu\nabla^\nu \phi \, \nabla_\mu \nabla_\nu \phi \Big] \\
&+ G_5 (\phi, Z) G^{\mu\nu} \, \nabla_\mu \nabla_\nu \phi - \frac{1}{6} G_{5, Z} \Big[ (\Box \phi )^3 - 3 \Box \phi \, \nabla^\mu \nabla^\nu \phi \, \nabla_\mu \nabla_\nu \phi  + 2 \nabla_\mu \nabla_\nu \phi \, \nabla^\nu \nabla^\lambda \phi \, \nabla^\mu \nabla_\lambda \phi \Big] \Bigg\} \, ,
\end{split}
\label{Horn1}
\eeq
\end{widetext}
with
%%%%%
\beq
Z \equiv  - \frac{1}{2} g^{\mu\nu} \partial_\mu \phi \, \partial_\nu \phi \, .
\label{YHorndef}
\eeq
(To avoid confusion with $X_{ij}$, we will use $Z$ to denote the kinetic term of the scalar field, as in Eq.~(\ref{YHorndef}); in the literature this term is usually denoted by $X$.) In Eq.~(\ref{Horn1}), each term $G_i (\phi, Z)$ is an arbitrary function of $\phi$ and $Z$, and $G_{i, Z} \equiv \partial G_i (\phi, Z) / \partial Z$. One candidate term of the form $g^{ij} \, X_{ij}$ appears within the term proportional to $G_5 (\phi, Z)$, which involves the Einstein tensor $G^{\mu\nu}$. In particular, we have, as usual,
%%%%%%
\beq
G^{\mu\nu} = R^{\mu\nu} - \frac{1}{2} g^{\mu\nu} R \, ,
\label{Gmndef}
\eeq
and hence the action in Eq.~(\ref{Horn1}) includes the term
%%%%%%%
\beq
g^{ij} X_{ij} = g^{ij} \left(- \frac{1}{2} G_5 (\phi, Z) \, R \,  \nabla_i \nabla_j \phi \right) \, ,
\label{Horn2}
\eeq
keeping in mind that even in a homogeneous background, for which $\phi \rightarrow \phi (t)$, $\nabla_i \nabla_j \phi = - \Gamma^0_{\>ij} \dot{\phi} \neq 0$.
Given that the dynamical coefficients $G_i (\phi, Z)$ are arbitrary, the most general statement we can make is that in a spatially homogeneous yet anisotropic spacetime, terms of the form $g^{ij} X_{ij}$ are {\it generic} in the action for single-field Horndeski models. 

\section{A unique isotropic state}\label{APP:uniqueISO}

Here we establish that in any (constrained) effective phase space described in Sec.~\ref{SEC:DynAnalysis}, the background space is isotropic if and only if $z_1=0=z_2$. This allows us to easily identify the isotropic background state in any phase space. 

For a finite (and nonzero) value of the Hubble expansion rate $H$, $z_1 = 0 = z_2 \iff \sigma_1 = 0 = \sigma_2\implies \sigma^2\equiv \frac{1}{3}\left(\sigma_1+\sigma_2\right)^2-\sigma_1\sigma_2=0$. Note however that it is also the case that the final implication sign in the previous sentence can be reversed: $\sigma^2 = 0 \implies \sigma_1 = 0 = \sigma_2$. This is because
\begin{align}
\sigma^2 = 0 &\equiv \frac{1}{3}\left(\sigma_1+\sigma_2\right)^2-\sigma_1\sigma_2=0\\
&\iff \sigma_2 = \left(\frac{1\pm\sqrt{3}i}{2}\right) \sigma_1,
\end{align}
so that $\sigma_1$ and $\sigma_2$ are both {\it real} if and only if $\sigma_1=0=\sigma_2$. Thus we have that 
\begin{equation}\label{EQN:zint}
z_1 = 0 = z_2 \iff \sigma_1 = 0 = \sigma_2\iff\sigma^2=0.
\end{equation}
Now, also by definition, we have that
\begin{align}
\sigma^2 & = \frac{1}{2}\left(\dot{\beta}_1^2+\dot{\beta}_2^2+\dot{\beta}_3^2\right)\\
&=\dot{\beta}_1^2+\dot{\beta}_2^2+\dot{\beta}_1\dot{\beta}_2,
\end{align}
where we have used the (derivative of the) constraint over the $\beta_i$'s of Eq.~(\ref{betaeom}) to establish the second line above. Thus we have
\begin{align}
\sigma^2 = 0 &\implies \dot{\beta}_2^2+\dot{\beta}_1\dot{\beta}_2+\dot{\beta}_1^2=0\\
&\iff \dot{\beta}_2 =  \left(\frac{-1\pm\sqrt{3}i}{2}\right) \dot{\beta}_1,
\end{align}
so that $\dot{\beta}_1$ and $\dot{\beta}_2$ are both {\it real} if and only if $\dot{\beta}_1=0=\dot{\beta}_2$. The (derivative of the) constraint on the $\beta_i$'s then ensures that $\dot{\beta}_3=0$, so that the $\beta_i$'s are all constant and the background space is  isotropic. Thus we have shown that 
\begin{equation}\label{EQN:sigint1}
\sigma^2=0\implies\textrm{Background is isotropic}.
\end{equation}
Note also that if the background is isotropic, then it has an FLRW form wherein the shear scalar also vanishes. Thus we have that
\begin{equation}\label{EQN:sigint2}
\textrm{Background is isotropic}\implies\sigma^2=0.
\end{equation}
Combining Eqs.~(\ref{EQN:zint}),~(\ref{EQN:sigint1}), and~(\ref{EQN:sigint2}) therefore gives us the required result: that (assuming $H$ is finite and nonzero)
\begin{align}
z_1 = 0 = z_2 & \iff \sigma_1 = 0 = \sigma_2\iff\sigma^2=0\nonumber\\
&\iff\textrm{Background is isotropic}.
\end{align}

\section{Spacetime structure associated with anisotropic fixed points {\bf FPc}--{\bf FPf}}\label{APP:dsAnisotropic}

Fixed points {\bf FPc}--{\bf FPf} in Table~\ref{TAB:FPs} are anisotropic for arbitrary (positive) values of $\mu_0$ and $\nu_0$. Here we describe the spacetime metric corresponding to each of these fixed points. Generally, we find that the metric has rotational symmetry about an axis (the $x^3$-axis for {\bf FPc} and {\bf FPd}, and the $x^2$-axis for {\bf FPe} and {\bf FPf}) and corresponds to a background that is expanding in each independent direction. 

Consider {\bf FPc}, for which $z_1=0$ and thus $\sigma_{1}=0$. (We will focus, in what follows, on expanding solutions where the Hubble expansion rate is finite, so that $0<H<\infty$.) This implies
\begin{align}\label{EQN:APbeta12dot}
\dot{\beta}_1 = \dot{\beta}_2.
\end{align}
This result, together with the derivative of the constraint on the $\beta_i$'s ($\sum_{i=1}^{3} \dot{\beta}_i=0$) yields
\begin{align}\label{EQN:APbeta31dot}
\dot{\beta}_3 = -2 \dot{\beta}_1. 
\end{align}
Furthermore, $z_2=-\sqrt{\nu_0/6}$ and thus from the definition of $z_2$ and $\sigma_2$:
\begin{align}\label{EQN:APbeta13dot}
\dot{\beta}_1 - \dot{\beta}_3 = -3H\sqrt{\frac{\nu_0}{6}}.
\end{align}
Substituting Eq.~(\ref{EQN:APbeta31dot}) into Eq.~(\ref{EQN:APbeta13dot}) we find
\begin{align}\label{EQN:APbeta1dot}
\dot{\beta}_1 = -\sqrt{\frac{\nu_0}{6}}\frac{\dot{a}}{a},
\end{align}
so that 
\begin{align}\label{EQN:APbeta1}
\beta_1 = -\sqrt{\frac{\nu_0}{6}}\ln a + C_1,
\end{align}
where $C_1$ is a constant of integration. Combining Eqs.~(\ref{EQN:APbeta12dot}) and~(\ref{EQN:APbeta1dot}) we find
\begin{align}\label{EQN:APbeta2}
\beta_2 = -\sqrt{\frac{\nu_0}{6}}\ln a + C_2,
\end{align}
where $C_2$ is a constant. Similarly, combining  Eqs.~(\ref{EQN:APbeta31dot}) and~(\ref{EQN:APbeta1dot}) we find
\begin{align}\label{EQN:APbeta3}
\beta_3 = \sqrt{\frac{2\nu_0}{3}}\ln a + C_3,
\end{align}
where $C_3$ is a constant. 

The scale factor in each direction, as described in Eq.~(\ref{EQN:Bprelim}), can be written as $a_i=a e^{\beta_i}$ so that, using Eqs.~(\ref{EQN:APbeta1}),~(\ref{EQN:APbeta2}), and~(\ref{EQN:APbeta3}), we find
\begin{align}
a_{1} &= A_1 a^{1-\sqrt{\nu_0/6}},\label{EQN:a1ap}\\
a_{2} &= A_2 a^{1-\sqrt{\nu_0/6}},\label{EQN:a2ap}\\
a_{3} &= A_3 a^{1+\sqrt{2\nu_0/3}},
\label{EQN:a3ap}
\end{align}
where $A_i \equiv e^{C_i}$ (for $i=1,2,3$) is a positive constant.  Substituting Eqs.~(\ref{EQN:a1ap})--(\ref{EQN:a3ap}) into the original form of the metric [Eq.~(\ref{EQN:Bprelim})] and implementing a (straightforward) rescaling of the spatial coordinates (which we now refer to using $\{x,y,z\}$), we find
\begin{align}
ds^2 = -dt^2
&+ a^{2\left(1-\sqrt{\nu_0/6}\right)}\left(dx^2+dy^2\right) \nonumber\\
&+ a^{2\left(1+\sqrt{2\nu_0/3}\right)}dz^2.
\end{align}
Note further that the Hubble slow-roll parameter is given by $\epsilon \equiv -\dot{H}/H^2= \nu_0/2$, so that 
\begin{align}\label{EQN:aFPc}
a \sim t^{2/\nu_0}. 
\end{align}
This metric has rotational symmetry about the $z$-axis. For $0<\nu_0<6$, scale factors in the $x$- and $y$-directions are growing with time though the expansion in these directions is slower than that in the $z$-direction.

A similar calculation for {\bf FPd} (where the only difference between {\bf FPd} and {\bf FPc}, for the purposes of computing the metric, is in the sign of $z_2$) yields
\begin{align}
ds^2 = -dt^2
&+ a^{2\left(1+\sqrt{\nu_0/6}\right)}\left(dx^2+dy^2\right) \nonumber\\
&+ a^{2\left(1-\sqrt{2\nu_0/3}\right)}dz^2,
\end{align}
where $a$ is given in Eq.~(\ref{EQN:aFPc}). This metric also has rotational symmetry about the $z$-axis. For $0<\nu_0<3/2$ the scale factor in the $z$-direction is growing with time though the expansion in this direction is slower than in the $x$- and $y$-directions.

Spacetime metrics for fixed points {\bf FPe} and {\bf FPf} can be obtained by symmetry from the metrics computed for {\bf FPc} and {\bf FPd} (respectively). For {\bf FPe} and {\bf FPf}, $z_2=0$, so that $\dot{\beta_1}=\dot{\beta_3}$. Thus the scale factor in the $x^2$-direction will differ from the scale factors in the other two directions (as opposed to {\bf FPc} and {\bf FPd}, where the scale factor in the $x^3$-direction was different from that in the other two directions). Note, in addition, that {\bf FPe} and {\bf FPf} are defined in terms of $\mu_0$ instead of $\nu_0$. 

The metric for {\bf FPe} can thus be obtained from that of {\bf FPc} by swapping $z\leftrightarrow y$ and setting $\nu_0\to\mu_0$. We find that
\begin{align}
ds^2 = -dt^2
&+ a^{2\left(1-\sqrt{\mu_0/6}\right)}\left(dx^2+dz^2\right) \nonumber\\
&+ a^{2\left(1+\sqrt{2\mu_0/3}\right)}dy^2,
\end{align}
with 
\begin{align}\label{EQN:aFPe}
a \sim t^{2/\mu_0}. 
\end{align}
This metric has rotational symmetry about the $y$-axis. For $0<\mu_0<6$, scale factors in the $x$- and $z$-directions are growing with time though the expansion in these directions is slower than that in the $y$-direction.

The metric for {\bf FPf} can similarly be obtained from that of {\bf FPd} by swapping $z\leftrightarrow y$ and setting $\nu_0\to\mu_0$. We find that
\begin{align}
ds^2 = -dt^2
&+ a^{2\left(1+\sqrt{\nu_0/6}\right)}\left(dx^2+dz^2\right) \nonumber\\
&+ a^{2\left(1-\sqrt{2\mu_0/3}\right)}dy^2,
\end{align}
where $a$ is given by Eq.~(\ref{EQN:aFPe}). This metric has rotational symmetry about the $y$-axis. For $0<\mu_0<3/2$ the scale factor in the $y$-direction is growing with time though the expansion in this direction is slower than in the $x$- and $z$-directions.

\section{A Kasner circle of fixed points}\label{APP:Kasner}

Here we demonstrate that fixed points in the principal phase space when $x=0$ correspond to the Kasner circle. Recall that those fixed points are given by
\begin{equation}
\label{EQN:FPs1}
x=y=\omega_1=\omega_2=0 \;\;\&\;\; 1 = (z_1+z_2)^2-3 z_1 z_2.
\end{equation}

Assume at the outset that $0<H<\infty$ (so that we only ever consider expanding solutions where the Hubble expansion rate is finite). Then, for these fixed points, the definitions of the dynamical variables yield:
\begin{align}
x = 0 & \iff c+L+\tilde{X}=0,\label{EQN:ximpli}\\
y = 0 & \iff 3c+\tilde{X}=0,\\
\omega_1 = 0 & \iff \pi_1=0\iff p_1=p_2,\\
\omega_2 = 0 & \iff \pi_2=0\iff p_1=p_3.\label{EQN:w2impli}
\end{align}
From the last two equations above, and the constraint that $p_1+p_2+p_3=0$ [Eq.~(\ref{psum})], we find
\begin{align}
p_1=p_2=p_3=0. \label{EQN:pzero}
\end{align}

Substituting Eqs.~(\ref{EQN:ximpli})--(\ref{EQN:w2impli}) and Eq.~(\ref{EQN:pzero}) into the first Friedmann equation [Eq.~(\ref{Friedmann})], the equation of evolution for the $\beta_i$'s [Eq.~(\ref{betaeom})], and the equation of evolution of the shear scalar yields:
\begin{align}
H^2&=\frac{1}{3}\sigma^2,\label{EQN:HKasner}\\
\ddot{\beta_i}+3H\dot{\beta_i} &=0,\label{EQN:SigKasner1}\\
 \frac{d}{dt}\sigma^2+6H\sigma^2 &= 0. \label{EQN:SigKasner2}
\end{align}
Note that the constraint on the $z_A$'s in Eq.~(\ref{EQN:FPs1}) (which we have not yet explicitly used) just yields the first Friedmann equation [viz.~Eq.~(\ref{EQN:HKasner})]. Integrating Eq.~(\ref{EQN:SigKasner2}), we find
\begin{align}\label{EQN:sigA}
\sigma^2 = e^{{C}}a^{-6},
\end{align}
where ${C}$ is a constant. Substituting this into Eq.~(\ref{EQN:HKasner}) and then integrating, we obtain
\begin{align}
\frac{a^3}{3}=\sqrt{\frac{e^{{C}}}{3}}t+{D},
\end{align}
where ${D}$ is a constant. Assuming the initial condition $a(t=0)=0$, yields
\begin{align}
a = \gamma t^{1/3},\;\;\textrm{where}\;\;\gamma \equiv \left(3 e^{{C}}\right)^{1/6}.\label{EQN:akas}
\end{align}
Note that therefore, we have
\begin{align}\label{EQN:HsolK}
H = \frac{1}{3t}. 
\end{align}

Integrating Eq.~(\ref{EQN:SigKasner1}), and using the  expression for $H$ in Eq.~(\ref{EQN:HsolK}), we find (for each $i=1,2,3$) that
\begin{align}
|\dot{\beta}_{i}|=\frac{e^{B_i}}{t},
\end{align}
where $B_i$ is a constant. Another way to write this, which will make things easier in what follows is:
\begin{align}\label{EQN:betadott}
\dot{\beta}_{i}=\pm\frac{e^{B_i}}{t}\equiv\frac{{A}_i}{t},
\end{align}
where ${A}_i\equiv\pm e^{B_i}$ is a constant that can take either sign. Constraints on these constants arise due to the derivative of the constraint on the sum of the $\beta_i$'s. In particular:
\begin{align}
\dot{\beta}_{1}+\dot{\beta}_{2}+\dot{\beta}_{3}=0\implies\frac{\left({A}_1+{A}_2+{A}_3\right)}{t}=0,
\end{align}
so that 
\begin{align}
{A}_1+{A}_2+{A}_3 = 0.\label{EQN:Bcons1}
\end{align}
Also, note that we have
\begin{align}
\sigma^2 &= \frac{1}{2}\left(\dot{\beta}_{1}^{\,2}+\dot{\beta}_{2}^{\,2}+\dot{\beta}_{3}^{\,2}\right)\\
&=\frac{1}{2t^2}\left({A}_1^2+{A}_2^2+{A}_3^2\right),\label{EQN:sigA1}
\end{align}
where the second line above uses Eq.~(\ref{EQN:betadott}). However, from Eqs.~(\ref{EQN:sigA}) and~(\ref{EQN:akas}), we also have
\begin{align}
\sigma^2 = \frac{e^{{C}}}{a^6}=\frac{e^{{C}}}{\gamma^6 t^2} = \frac{1}{3t^2}\label{EQN:sigA2}.
\end{align}
Combining Eqs.~(\ref{EQN:sigA1}) and~(\ref{EQN:sigA2}) gives
\begin{align}
{A}_1^2+{A}_2^2+{A}_3^2 = \frac{2}{3}.\label{EQN:Bcons2}
\end{align}

To complete the derivation of the metric, we may integrate Eq.~(\ref{EQN:betadott}) (for $i=1,2,3$), to find
\begin{align}
\beta_i = {A}_i \ln{t}+\mu_i,\label{EQN:finbeta}
\end{align}
where $\mu_i$ is a constant. Combining the expression for $a$ [in Eq.~(\ref{EQN:akas})] with the expressions for the $\beta_i$'s [in Eq.~(\ref{EQN:finbeta})] we can write the metric in the following way:
\begin{widetext}
\begin{align}
ds^2 &= -dt^2 + {  \gamma^2\,e^{2\mu_1}}\,t^{2/3}\,e^{2{A}_1\ln{t}}\,dx^2 + { \gamma^2\,e^{2\mu_2}}\,t^{2/3}\,e^{2{A}_2\ln{t}}\,dy^2 + { \gamma^2\,e^{2\mu_3}}\,t^{2/3}\,e^{2{A}_3\ln{t}}\,dz^2\\
&\equiv -dt^2 + { \gamma^2\,e^{2\mu_1}}\,t^{2q_1}\,dx^2+ { \gamma^2\,e^{2\mu_2}}\,t^{2q_2}\,dy^2+ { \gamma^2\,e^{2\mu_3}}\,t^{2q_3}\,dz^2,\label{EQN:kasner1}
\end{align}
\end{widetext}
where the final line uses the following definition (for each $i=1,2,3$):
\begin{align}
q_i& \equiv \frac{1}{3}+{A}_i.
\end{align}
A change of spatial coordinates allows us to bring the metric into a standard form (as quoted in, for example, Ref.~\cite[\S 30.2]{misner+al_73}):
\begin{align}
ds^2= -dt^2 + t^{2q_1}\,d\bar{x}^2+ t^{2q_2}\,d\bar{y}^2+ t^{2q_3}\,d\bar{z}^2.\label{EQN:kasner2}
\end{align}
Given the constraints on the ${A}_i$'s, this solution is {\it not} a function of three free parameters (the $q_i$'s)---there are (as we will describe) two constraints that these parameters must satisfy and hence we have a one-parameter family of solutions. From the definitions of the $q_i$'s and the constraints on the ${A}_i$'s, it is straightforward to show that 
\begin{align}
q_1+q_2+q_3 &= 1,\label{EQN:Kasnerplane}\\
q_1^2+q_2^2+q_3^2 &= 1.\label{EQN:Kasnersphere}
\end{align}
The metric is more precisely specified by those values of the $q_i$'s that are given by the intersection of the plane (in the first constraint above) and the sphere (in the second constraint). This circle (in general) is known as the {\it Kasner circle}. 

We have thus shown that the fixed points of interest [in Eq.~(\ref{EQN:FPs1})] are consistent with Einstein field equations whose solution is given by the {\it Kasner metric} in Eqs.~(\ref{EQN:kasner1}) or~(\ref{EQN:kasner2}). 

Note that our fixed-point analysis for the principal phase space yields a one-parameter family of solutions---as expressed by the constraint on the $z_A$'s, in Eq.~(\ref{EQN:FPs1}). One can expressly relate this one-parameter family of fixed points to the Kasner circle as follows.

The Kasner circle can be characterized by eliminating $q_1$ from Eq.~(\ref{EQN:Kasnersphere}), using Eq.~(\ref{EQN:Kasnerplane}). In particular, one obtains:
\begin{align}\label{EQN:Kasnerq}
q_2^2+q_3^2 +q_2 q_3- q_2-q_3=0.
\end{align}

Now, let the $z_A$'s satisfy Eq.~(\ref{EQN:FPs1}), and let us use the expressions for $H$ and $\dot{\beta}_i$ found earlier [in Eqs.~(\ref{EQN:HsolK}) and~(\ref{EQN:betadott})]. That is, we can write
\begin{align}
z_1\equiv\frac{\sigma_1}{3H}=\frac{\dot{\beta_1}-\dot{\beta_2}}{3H}
& = \frac{{{A}_1}/{t}-{{A}_2}/{t}}{{1}/{t}}\nonumber\\
&={A}_1-{A}_2
\equiv q_1 - q_2.\label{EQN:z1k}
\end{align}
Similarly, we can write
\begin{align}
z_2=q_1-q_3. \label{EQN:z2k}
\end{align}
It is now straightforward to show, using Eq.~(\ref{EQN:z1k}), Eq.~(\ref{EQN:z2k}), and the constraints on the $q_i$'s, that
\begin{align}
(z_1+z_2)^2-3 z_1 z_2-1 & = z_1^2-z_1 z_2 +z_2^2-1\\
& = q_2^2+q_3^2 +q_2 q_3- q_2-q_3.
\end{align}
Thus the set of points traced out by the constraint on the $z_A$'s is consistent with the set of points traced out by the constraints on the $q_i$'s. 

%\begin{thebibliography}{199}

%\bibliography{CC_EFT.bib}

\begin{thebibliography}{73}%
\makeatletter
\providecommand \@ifxundefined [1]{%
 \@ifx{#1\undefined}
}%
\providecommand \@ifnum [1]{%
 \ifnum #1\expandafter \@firstoftwo
 \else \expandafter \@secondoftwo
 \fi
}%
\providecommand \@ifx [1]{%
 \ifx #1\expandafter \@firstoftwo
 \else \expandafter \@secondoftwo
 \fi
}%
\providecommand \natexlab [1]{#1}%
\providecommand \enquote  [1]{``#1''}%
\providecommand \bibnamefont  [1]{#1}%
\providecommand \bibfnamefont [1]{#1}%
\providecommand \citenamefont [1]{#1}%
\providecommand \href@noop [0]{\@secondoftwo}%
\providecommand \href [0]{\begingroup \@sanitize@url \@href}%
\providecommand \@href[1]{\@@startlink{#1}\@@href}%
\providecommand \@@href[1]{\endgroup#1\@@endlink}%
\providecommand \@sanitize@url [0]{\catcode `\\12\catcode `\$12\catcode
  `\&12\catcode `\#12\catcode `\^12\catcode `\_12\catcode `\%12\relax}%
\providecommand \@@startlink[1]{}%
\providecommand \@@endlink[0]{}%
\providecommand \url  [0]{\begingroup\@sanitize@url \@url }%
\providecommand \@url [1]{\endgroup\@href {#1}{\urlprefix }}%
\providecommand \urlprefix  [0]{URL }%
\providecommand \Eprint [0]{\href }%
\providecommand \doibase [0]{http://dx.doi.org/}%
\providecommand \selectlanguage [0]{\@gobble}%
\providecommand \bibinfo  [0]{\@secondoftwo}%
\providecommand \bibfield  [0]{\@secondoftwo}%
\providecommand \translation [1]{[#1]}%
\providecommand \BibitemOpen [0]{}%
\providecommand \bibitemStop [0]{}%
\providecommand \bibitemNoStop [0]{.\EOS\space}%
\providecommand \EOS [0]{\spacefactor3000\relax}%
\providecommand \BibitemShut  [1]{\csname bibitem#1\endcsname}%
\let\auto@bib@innerbib\@empty
%</preamble>
\bibitem [{\citenamefont {Guth}\ and\ \citenamefont
  {Kaiser}(2005)}]{guth+kaiser_05}%
  \BibitemOpen
  \bibfield  {author} {\bibinfo {author} {\bibfnamefont {Alan~H.}\ \bibnamefont
  {Guth}}\ and\ \bibinfo {author} {\bibfnamefont {David~I.}\ \bibnamefont
  {Kaiser}},\ }\bibfield  {title} {\enquote {\bibinfo {title} {{Inflationary
  cosmology: Exploring the Universe from the smallest to the largest
  scales}},}\ }\href {\doibase 10.1126/science.1107483} {\bibfield  {journal}
  {\bibinfo  {journal} {Science}\ }\textbf {\bibinfo {volume} {307}},\ \bibinfo
  {pages} {884--890} (\bibinfo {year} {2005})},\ \Eprint
  {http://arxiv.org/abs/astro-ph/0502328} {arXiv:astro-ph/0502328} \BibitemShut
  {NoStop}%
\bibitem [{\citenamefont {Bassett}\ \emph {et~al.}(2006)\citenamefont
  {Bassett}, \citenamefont {Tsujikawa},\ and\ \citenamefont
  {Wands}}]{bassett+al_06}%
  \BibitemOpen
  \bibfield  {author} {\bibinfo {author} {\bibfnamefont {Bruce~A.}\
  \bibnamefont {Bassett}}, \bibinfo {author} {\bibfnamefont {Shinji}\
  \bibnamefont {Tsujikawa}}, \ and\ \bibinfo {author} {\bibfnamefont {David}\
  \bibnamefont {Wands}},\ }\bibfield  {title} {\enquote {\bibinfo {title}
  {{Inflation dynamics and reheating}},}\ }\href {\doibase
  10.1103/RevModPhys.78.537} {\bibfield  {journal} {\bibinfo  {journal} {Rev.
  Mod. Phys.}\ }\textbf {\bibinfo {volume} {78}},\ \bibinfo {pages} {537--589}
  (\bibinfo {year} {2006})},\ \Eprint {http://arxiv.org/abs/astro-ph/0507632}
  {arXiv:astro-ph/0507632} \BibitemShut {NoStop}%
\bibitem [{\citenamefont {Lyth}\ and\ \citenamefont
  {Liddle}(2009)}]{Lyth:2009zz}%
  \BibitemOpen
  \bibfield  {author} {\bibinfo {author} {\bibfnamefont {David~H.}\
  \bibnamefont {Lyth}}\ and\ \bibinfo {author} {\bibfnamefont {Andrew~R.}\
  \bibnamefont {Liddle}},\ }\href {\doibase
  https://doi.org/10.1017/CBO9780511819209} {\emph {\bibinfo {title} {{The
  Primordial Density Perturbation: Cosmology, Inflation and the Origin of
  Structure}}}}\ (\bibinfo  {publisher} {New York: Cambridge University
  Press},\ \bibinfo {year} {2009})\BibitemShut {NoStop}%
\bibitem [{\citenamefont {Martin}\ \emph {et~al.}(2014)\citenamefont {Martin},
  \citenamefont {Ringeval},\ and\ \citenamefont {Vennin}}]{martin+al_14}%
  \BibitemOpen
  \bibfield  {author} {\bibinfo {author} {\bibfnamefont {J\'{e}r\^{o}me}\ \bibnamefont
  {Martin}}, \bibinfo {author} {\bibfnamefont {Christophe}\ \bibnamefont
  {Ringeval}}, \ and\ \bibinfo {author} {\bibfnamefont {Vincent}\ \bibnamefont
  {Vennin}},\ }\bibfield  {title} {\enquote {\bibinfo {title}
  {{Encyclop\ae{}dia Inflationaris}},}\ }\href {\doibase
  10.1016/j.dark.2014.01.003} {\bibfield  {journal} {\bibinfo  {journal} {Phys.
  Dark Univ.}\ }\textbf {\bibinfo {volume} {5-6}},\ \bibinfo {pages} {75--235}
  (\bibinfo {year} {2014})},\ \Eprint {http://arxiv.org/abs/1303.3787}
  {arXiv:1303.3787 [astro-ph.CO]} \BibitemShut {NoStop}%
\bibitem [{\citenamefont {Guth}\ \emph {et~al.}(2014)\citenamefont {Guth},
  \citenamefont {Kaiser},\ and\ \citenamefont {Nomura}}]{Guth:2013sya}%
  \BibitemOpen
  \bibfield  {author} {\bibinfo {author} {\bibfnamefont {Alan~H.}\ \bibnamefont
  {Guth}}, \bibinfo {author} {\bibfnamefont {David~I.}\ \bibnamefont {Kaiser}},
  \ and\ \bibinfo {author} {\bibfnamefont {Yasunori}\ \bibnamefont {Nomura}},\
  }\bibfield  {title} {\enquote {\bibinfo {title} {{Inflationary paradigm after
  Planck 2013}},}\ }\href {\doibase 10.1016/j.physletb.2014.03.020} {\bibfield
  {journal} {\bibinfo  {journal} {Phys. Lett. B}\ }\textbf {\bibinfo {volume}
  {733}},\ \bibinfo {pages} {112--119} (\bibinfo {year} {2014})},\ \Eprint
  {http://arxiv.org/abs/1312.7619} {arXiv:1312.7619 [astro-ph.CO]} \BibitemShut
  {NoStop}%
\bibitem [{\citenamefont {Baumann}\ and\ \citenamefont
  {McAllister}(2015)}]{Baumann:2014nda}%
  \BibitemOpen
  \bibfield  {author} {\bibinfo {author} {\bibfnamefont {Daniel}\ \bibnamefont
  {Baumann}}\ and\ \bibinfo {author} {\bibfnamefont {Liam}\ \bibnamefont
  {McAllister}},\ }\href {\doibase 10.1017/CBO9781316105733} {\emph {\bibinfo
  {title} {{Inflation and String Theory}}}}\ (\bibinfo  {publisher} {New York:
  Cambridge University Press},\ \bibinfo {year} {2015})\ \Eprint
  {http://arxiv.org/abs/1404.2601} {arXiv:1404.2601 [hep-th]} \BibitemShut
  {NoStop}%
\bibitem [{\citenamefont {East}\ \emph {et~al.}(2016)\citenamefont {East},
  \citenamefont {Kleban}, \citenamefont {Linde},\ and\ \citenamefont
  {Senatore}}]{East:2015ggf}%
  \BibitemOpen
  \bibfield  {author} {\bibinfo {author} {\bibfnamefont {William~E.}\
  \bibnamefont {East}}, \bibinfo {author} {\bibfnamefont {Matthew}\
  \bibnamefont {Kleban}}, \bibinfo {author} {\bibfnamefont {Andrei}\
  \bibnamefont {Linde}}, \ and\ \bibinfo {author} {\bibfnamefont {Leonardo}\
  \bibnamefont {Senatore}},\ }\bibfield  {title} {\enquote {\bibinfo {title}
  {{Beginning inflation in an inhomogeneous universe}},}\ }\href {\doibase
  10.1088/1475-7516/2016/09/010} {\bibfield  {journal} {\bibinfo  {journal}
  {JCAP}\ }\textbf {\bibinfo {volume} {09}},\ \bibinfo {pages} {010} (\bibinfo
  {year} {2016})},\ \Eprint {http://arxiv.org/abs/1511.05143} {arXiv:1511.05143
  [hep-th]} \BibitemShut {NoStop}%
\bibitem [{\citenamefont {Kleban}\ and\ \citenamefont
  {Senatore}(2016)}]{kleban+senatore_16}%
  \BibitemOpen
  \bibfield  {author} {\bibinfo {author} {\bibfnamefont {Matthew}\ \bibnamefont
  {Kleban}}\ and\ \bibinfo {author} {\bibfnamefont {Leonardo}\ \bibnamefont
  {Senatore}},\ }\bibfield  {title} {\enquote {\bibinfo {title} {{Inhomogeneous
  Anisotropic Cosmology}},}\ }\href {\doibase 10.1088/1475-7516/2016/10/022}
  {\bibfield  {journal} {\bibinfo  {journal} {JCAP}\ }\textbf {\bibinfo
  {volume} {10}},\ \bibinfo {pages} {022} (\bibinfo {year} {2016})},\ \Eprint
  {http://arxiv.org/abs/1602.03520} {arXiv:1602.03520 [hep-th]} \BibitemShut
  {NoStop}%
\bibitem [{\citenamefont {Clough}\ \emph {et~al.}(2017)\citenamefont {Clough},
  \citenamefont {Lim}, \citenamefont {DiNunno}, \citenamefont {Fischler},
  \citenamefont {Flauger},\ and\ \citenamefont {Paban}}]{Clough:2016ymm}%
  \BibitemOpen
  \bibfield  {author} {\bibinfo {author} {\bibfnamefont {Katy}\ \bibnamefont
  {Clough}}, \bibinfo {author} {\bibfnamefont {Eugene~A.}\ \bibnamefont {Lim}},
  \bibinfo {author} {\bibfnamefont {Brandon~S.}\ \bibnamefont {DiNunno}},
  \bibinfo {author} {\bibfnamefont {Willy}\ \bibnamefont {Fischler}}, \bibinfo
  {author} {\bibfnamefont {Raphael}\ \bibnamefont {Flauger}}, \ and\ \bibinfo
  {author} {\bibfnamefont {Sonia}\ \bibnamefont {Paban}},\ }\bibfield  {title}
  {\enquote {\bibinfo {title} {{Robustness of Inflation to Inhomogeneous
  Initial Conditions}},}\ }\href {\doibase 10.1088/1475-7516/2017/09/025}
  {\bibfield  {journal} {\bibinfo  {journal} {JCAP}\ }\textbf {\bibinfo
  {volume} {09}},\ \bibinfo {pages} {025} (\bibinfo {year} {2017})},\ \Eprint
  {http://arxiv.org/abs/1608.04408} {arXiv:1608.04408 [hep-th]} \BibitemShut
  {NoStop}%
\bibitem [{\citenamefont {Clough}\ \emph {et~al.}(2018)\citenamefont {Clough},
  \citenamefont {Flauger},\ and\ \citenamefont {Lim}}]{Clough:2017efm}%
  \BibitemOpen
  \bibfield  {author} {\bibinfo {author} {\bibfnamefont {Katy}\ \bibnamefont
  {Clough}}, \bibinfo {author} {\bibfnamefont {Raphael}\ \bibnamefont
  {Flauger}}, \ and\ \bibinfo {author} {\bibfnamefont {Eugene~A.}\ \bibnamefont
  {Lim}},\ }\bibfield  {title} {\enquote {\bibinfo {title} {{Robustness of
  Inflation to Large Tensor Perturbations}},}\ }\href {\doibase
  10.1088/1475-7516/2018/05/065} {\bibfield  {journal} {\bibinfo  {journal}
  {JCAP}\ }\textbf {\bibinfo {volume} {05}},\ \bibinfo {pages} {065} (\bibinfo
  {year} {2018})},\ \Eprint {http://arxiv.org/abs/1712.07352} {arXiv:1712.07352
  [hep-th]} \BibitemShut {NoStop}%
\bibitem [{\citenamefont {Brandenberger}(2016)}]{brandenberger_17}%
  \BibitemOpen
  \bibfield  {author} {\bibinfo {author} {\bibfnamefont {Robert}\ \bibnamefont
  {Brandenberger}},\ }\bibfield  {title} {\enquote {\bibinfo {title} {{Initial
  conditions for inflation \textemdash{} A short review}},}\ }\href {\doibase
  10.1142/S0218271817400028} {\bibfield  {journal} {\bibinfo  {journal} {Int.
  J. Mod. Phys. D}\ }\textbf {\bibinfo {volume} {26}},\ \bibinfo {pages}
  {1740002} (\bibinfo {year} {2016})},\ \Eprint
  {http://arxiv.org/abs/1601.01918} {arXiv:1601.01918 [hep-th]} \BibitemShut
  {NoStop}%
\bibitem [{\citenamefont {Linde}(2018)}]{linde_18}%
  \BibitemOpen
  \bibfield  {author} {\bibinfo {author} {\bibfnamefont {Andrei}\ \bibnamefont
  {Linde}},\ }\bibfield  {title} {\enquote {\bibinfo {title} {{On the problem
  of initial conditions for inflation}},}\ }\href {\doibase
  10.1007/s10701-018-0177-9} {\bibfield  {journal} {\bibinfo  {journal} {Found.
  Phys.}\ }\textbf {\bibinfo {volume} {48}},\ \bibinfo {pages} {1246--1260}
  (\bibinfo {year} {2018})},\ \Eprint {http://arxiv.org/abs/1710.04278}
  {arXiv:1710.04278 [hep-th]} \BibitemShut {NoStop}%
\bibitem [{\citenamefont {Bloomfield}\ \emph {et~al.}(2019)\citenamefont
  {Bloomfield}, \citenamefont {Fitzpatrick}, \citenamefont {Hilbert},\ and\
  \citenamefont {Kaiser}}]{Bloomfield:2019rbs}%
  \BibitemOpen
  \bibfield  {author} {\bibinfo {author} {\bibfnamefont {Jolyon~K.}\
  \bibnamefont {Bloomfield}}, \bibinfo {author} {\bibfnamefont {Patrick}\
  \bibnamefont {Fitzpatrick}}, \bibinfo {author} {\bibfnamefont {Kiriakos}\
  \bibnamefont {Hilbert}}, \ and\ \bibinfo {author} {\bibfnamefont {David~I.}\
  \bibnamefont {Kaiser}},\ }\bibfield  {title} {\enquote {\bibinfo {title}
  {{Onset of inflation amid backreaction from inhomogeneities}},}\ }\href
  {\doibase 10.1103/PhysRevD.100.063512} {\bibfield  {journal} {\bibinfo
  {journal} {Phys. Rev. D}\ }\textbf {\bibinfo {volume} {100}},\ \bibinfo
  {pages} {063512} (\bibinfo {year} {2019})},\ \Eprint
  {http://arxiv.org/abs/1906.08651} {arXiv:1906.08651 [astro-ph.CO]}
  \BibitemShut {NoStop}%
\bibitem [{\citenamefont {Chowdhury}\ \emph {et~al.}(2019)\citenamefont
  {Chowdhury}, \citenamefont {Martin}, \citenamefont {Ringeval},\ and\
  \citenamefont {Vennin}}]{Chowdhury:2019otk}%
  \BibitemOpen
  \bibfield  {author} {\bibinfo {author} {\bibfnamefont {Debika}\ \bibnamefont
  {Chowdhury}}, \bibinfo {author} {\bibfnamefont {J\'er\^ome}\ \bibnamefont
  {Martin}}, \bibinfo {author} {\bibfnamefont {Christophe}\ \bibnamefont
  {Ringeval}}, \ and\ \bibinfo {author} {\bibfnamefont {Vincent}\ \bibnamefont
  {Vennin}},\ }\bibfield  {title} {\enquote {\bibinfo {title} {{Assessing the
  scientific status of inflation after {\it Planck}}},}\ }\href {\doibase
  10.1103/PhysRevD.100.083537} {\bibfield  {journal} {\bibinfo  {journal}
  {Phys. Rev. D}\ }\textbf {\bibinfo {volume} {100}},\ \bibinfo {pages}
  {083537} (\bibinfo {year} {2019})},\ \Eprint
  {http://arxiv.org/abs/1902.03951} {arXiv:1902.03951 [astro-ph.CO]}
  \BibitemShut {NoStop}%
\bibitem [{\citenamefont {Aurrekoetxea}\ \emph {et~al.}(2020)\citenamefont
  {Aurrekoetxea}, \citenamefont {Clough}, \citenamefont {Flauger},\ and\
  \citenamefont {Lim}}]{Aurrekoetxea:2019fhr}%
  \BibitemOpen
  \bibfield  {author} {\bibinfo {author} {\bibfnamefont {Josu~C.}\ \bibnamefont
  {Aurrekoetxea}}, \bibinfo {author} {\bibfnamefont {Katy}\ \bibnamefont
  {Clough}}, \bibinfo {author} {\bibfnamefont {Raphael}\ \bibnamefont
  {Flauger}}, \ and\ \bibinfo {author} {\bibfnamefont {Eugene~A.}\ \bibnamefont
  {Lim}},\ }\bibfield  {title} {\enquote {\bibinfo {title} {{The Effects of
  Potential Shape on Inhomogeneous Inflation}},}\ }\href {\doibase
  10.1088/1475-7516/2020/05/030} {\bibfield  {journal} {\bibinfo  {journal}
  {JCAP}\ }\textbf {\bibinfo {volume} {05}},\ \bibinfo {pages} {030} (\bibinfo
  {year} {2020})},\ \Eprint {http://arxiv.org/abs/1910.12547} {arXiv:1910.12547
  [astro-ph.CO]} \BibitemShut {NoStop}%
\bibitem [{\citenamefont {Creminelli}\ \emph
  {et~al.}(2020{\natexlab{a}})\citenamefont {Creminelli}, \citenamefont
  {Senatore},\ and\ \citenamefont {Vasy}}]{Creminelli:2019pdh}%
  \BibitemOpen
  \bibfield  {author} {\bibinfo {author} {\bibfnamefont {Paolo}\ \bibnamefont
  {Creminelli}}, \bibinfo {author} {\bibfnamefont {Leonardo}\ \bibnamefont
  {Senatore}}, \ and\ \bibinfo {author} {\bibfnamefont {Andr\'as}\ \bibnamefont
  {Vasy}},\ }\bibfield  {title} {\enquote {\bibinfo {title} {{Asymptotic
  Behavior of Cosmologies with $\Lambda >0$ in 2+1 Dimensions}},}\ }\href
  {\doibase 10.1007/s00220-020-03706-3} {\bibfield  {journal} {\bibinfo
  {journal} {Commun. Math. Phys.}\ }\textbf {\bibinfo {volume} {376}},\
  \bibinfo {pages} {1155--1170} (\bibinfo {year} {2020}{\natexlab{a}})},\
  \Eprint {http://arxiv.org/abs/1902.00519} {arXiv:1902.00519 [hep-th]}
  \BibitemShut {NoStop}%
\bibitem [{\citenamefont {Creminelli}\ \emph
  {et~al.}(2020{\natexlab{b}})\citenamefont {Creminelli}, \citenamefont
  {Hershkovits}, \citenamefont {Senatore},\ and\ \citenamefont
  {Vasy}}]{Creminelli:2020zvc}%
  \BibitemOpen
  \bibfield  {author} {\bibinfo {author} {\bibfnamefont {Paolo}\ \bibnamefont
  {Creminelli}}, \bibinfo {author} {\bibfnamefont {Or}~\bibnamefont
  {Hershkovits}}, \bibinfo {author} {\bibfnamefont {Leonardo}\ \bibnamefont
  {Senatore}}, \ and\ \bibinfo {author} {\bibfnamefont {Andr\'as}\ \bibnamefont
  {Vasy}},\ }\bibfield  {title} {\enquote {\bibinfo {title} {{A de Sitter
  no-hair theorem for 3+1d Cosmologies with isometry group forming
  2-dimensional orbits}},}\ }\href@noop {} {\  (\bibinfo {year}
  {2020}{\natexlab{b}})},\ \Eprint {http://arxiv.org/abs/2004.10754}
  {arXiv:2004.10754 [hep-th]} \BibitemShut {NoStop}%
\bibitem [{\citenamefont {{Azhar}}(2020)}]{azhar_20}%
  \BibitemOpen
  \bibfield  {author} {\bibinfo {author} {\bibfnamefont {Feraz}\ \bibnamefont
  {{Azhar}}},\ }\bibfield  {title} {\enquote {\bibinfo {title} {{Effective
  field theories as a novel probe of fine-tuning of cosmic inflation}},}\
  }\href {\doibase 10.1016/j.shpsb.2020.05.001} {\bibfield  {journal} {\bibinfo
   {journal} {Studies in the History and Philosophy of Modern Physics}\
  }\textbf {\bibinfo {volume} {71}},\ \bibinfo {pages} {87--100} (\bibinfo
  {year} {2020})},\ \Eprint {http://arxiv.org/abs/1911.05128} {arXiv:1911.05128
  [physics.hist-ph]} \BibitemShut {NoStop}%
\bibitem [{\citenamefont {Joana}\ and\ \citenamefont
  {Clesse}(2021)}]{Joana:2020rxm}%
  \BibitemOpen
  \bibfield  {author} {\bibinfo {author} {\bibfnamefont {Cristian}\
  \bibnamefont {Joana}}\ and\ \bibinfo {author} {\bibfnamefont {S\'ebastien}\
  \bibnamefont {Clesse}},\ }\bibfield  {title} {\enquote {\bibinfo {title}
  {{Inhomogeneous preinflation across Hubble scales in full general
  relativity}},}\ }\href {\doibase 10.1103/PhysRevD.103.083501} {\bibfield
  {journal} {\bibinfo  {journal} {Phys. Rev. D}\ }\textbf {\bibinfo {volume}
  {103}},\ \bibinfo {pages} {083501} (\bibinfo {year} {2021})},\ \Eprint
  {http://arxiv.org/abs/2011.12190} {arXiv:2011.12190 [astro-ph.CO]}
  \BibitemShut {NoStop}%
\bibitem [{\citenamefont {Tenkanen}\ and\ \citenamefont
  {Tomberg}(2020)}]{Tenkanen:2020cvw}%
  \BibitemOpen
  \bibfield  {author} {\bibinfo {author} {\bibfnamefont {Tommi}\ \bibnamefont
  {Tenkanen}}\ and\ \bibinfo {author} {\bibfnamefont {Eemeli}\ \bibnamefont
  {Tomberg}},\ }\bibfield  {title} {\enquote {\bibinfo {title} {{Initial
  conditions for plateau inflation: a case study}},}\ }\href {\doibase
  10.1088/1475-7516/2020/04/050} {\bibfield  {journal} {\bibinfo  {journal}
  {JCAP}\ }\textbf {\bibinfo {volume} {04}},\ \bibinfo {pages} {050} (\bibinfo
  {year} {2020})},\ \Eprint {http://arxiv.org/abs/2002.02420} {arXiv:2002.02420
  [astro-ph.CO]} \BibitemShut {NoStop}%
\bibitem [{\citenamefont {Wang}\ and\ \citenamefont
  {Senatore}(2021)}]{Wang:2021hzv}%
  \BibitemOpen
  \bibfield  {author} {\bibinfo {author} {\bibfnamefont {Jinhui}\ \bibnamefont
  {Wang}}\ and\ \bibinfo {author} {\bibfnamefont {Leonardo}\ \bibnamefont
  {Senatore}},\ }\bibfield  {title} {\enquote {\bibinfo {title} {{On the
  asymptotics of 3+1D cosmologies with bounded scalar potential and isometry
  group forming 2-dimensional orbits}},}\ }\href@noop {} {\  (\bibinfo {year}
  {2021})},\ \Eprint {http://arxiv.org/abs/2111.09257} {arXiv:2111.09257
  [hep-th]} \BibitemShut {NoStop}%
\bibitem [{\citenamefont {Corman}\ and\ \citenamefont
  {East}(2022)}]{Corman:2022alv}%
  \BibitemOpen
  \bibfield  {author} {\bibinfo {author} {\bibfnamefont {Maxence}\ \bibnamefont
  {Corman}}\ and\ \bibinfo {author} {\bibfnamefont {William~E.}\ \bibnamefont
  {East}},\ }\bibfield  {title} {\enquote {\bibinfo {title} {{Starting
  inflation from inhomogeneous initial conditions with momentum}},}\
  }\href@noop {} {\  (\bibinfo {year} {2022})},\ \Eprint
  {http://arxiv.org/abs/2212.04479} {arXiv:2212.04479 [gr-qc]} \BibitemShut
  {NoStop}%
\bibitem [{\citenamefont {Schmidt}(2004)}]{schmidt_04}%
  \BibitemOpen
  \bibfield  {author} {\bibinfo {author} {\bibfnamefont {Hans-Jurgen}\
  \bibnamefont {Schmidt}},\ }\bibfield  {title} {\enquote {\bibinfo {title}
  {{Lectures on mathematical cosmology}},}\ }\href@noop {} {\  (\bibinfo {year}
  {2004})},\ \Eprint {http://arxiv.org/abs/gr-qc/0407095} {arXiv:gr-qc/0407095}
  \BibitemShut {NoStop}%
\bibitem [{\citenamefont {Wald}(1983)}]{wald_83}%
  \BibitemOpen
  \bibfield  {author} {\bibinfo {author} {\bibfnamefont {Robert~M.}\
  \bibnamefont {Wald}},\ }\bibfield  {title} {\enquote {\bibinfo {title}
  {{Asymptotic behavior of homogeneous cosmological models in the presence of a
  positive cosmological constant}},}\ }\href {\doibase
  10.1103/PhysRevD.28.2118} {\bibfield  {journal} {\bibinfo  {journal} {Phys.
  Rev. D}\ }\textbf {\bibinfo {volume} {28}},\ \bibinfo {pages} {2118--2120}
  (\bibinfo {year} {1983})}\BibitemShut {NoStop}%
\bibitem [{\citenamefont {Watanabe}\ \emph {et~al.}(2009)\citenamefont
  {Watanabe}, \citenamefont {Kanno},\ and\ \citenamefont
  {Soda}}]{watanabe+al_09}%
  \BibitemOpen
  \bibfield  {author} {\bibinfo {author} {\bibfnamefont {Masa-aki}\
  \bibnamefont {Watanabe}}, \bibinfo {author} {\bibfnamefont {Sugumi}\
  \bibnamefont {Kanno}}, \ and\ \bibinfo {author} {\bibfnamefont {Jiro}\
  \bibnamefont {Soda}},\ }\bibfield  {title} {\enquote {\bibinfo {title}
  {Inflationary universe with anisotropic hair},}\ }\href {\doibase
  10.1103/PhysRevLett.102.191302} {\bibfield  {journal} {\bibinfo  {journal}
  {Phys. Rev. Lett.}\ }\textbf {\bibinfo {volume} {102}},\ \bibinfo {pages}
  {191302} (\bibinfo {year} {2009})},\ \Eprint {http://arxiv.org/abs/0902.2833}
  {arXiv:0902.2833 [hep-th]} \BibitemShut {NoStop}%
\bibitem [{\citenamefont {Maleknejad}\ and\ \citenamefont
  {Sheikh-Jabbari}(2012)}]{maleknejad+sheikh-jabbari_12}%
  \BibitemOpen
  \bibfield  {author} {\bibinfo {author} {\bibfnamefont {A.}~\bibnamefont
  {Maleknejad}}\ and\ \bibinfo {author} {\bibfnamefont {M.~M.}\ \bibnamefont
  {Sheikh-Jabbari}},\ }\bibfield  {title} {\enquote {\bibinfo {title}
  {Revisiting cosmic no-hair theorem for inflationary settings},}\ }\href
  {\doibase 10.1103/PhysRevD.85.123508} {\bibfield  {journal} {\bibinfo
  {journal} {Phys. Rev. D}\ }\textbf {\bibinfo {volume} {85}},\ \bibinfo
  {pages} {123508} (\bibinfo {year} {2012})},\ \Eprint
  {http://arxiv.org/abs/1203.0219} {arXiv:1203.0219 [hep-th]} \BibitemShut
  {NoStop}%
\bibitem [{\citenamefont {Cannone}\ \emph {et~al.}(2015)\citenamefont
  {Cannone}, \citenamefont {Gong},\ and\ \citenamefont
  {Tasinato}}]{Cannone:2015rra}%
  \BibitemOpen
  \bibfield  {author} {\bibinfo {author} {\bibfnamefont {Dario}\ \bibnamefont
  {Cannone}}, \bibinfo {author} {\bibfnamefont {Jinn-Ouk}\ \bibnamefont
  {Gong}}, \ and\ \bibinfo {author} {\bibfnamefont {Gianmassimo}\ \bibnamefont
  {Tasinato}},\ }\bibfield  {title} {\enquote {\bibinfo {title} {{Breaking
  discrete symmetries in the effective field theory of inflation}},}\ }\href
  {\doibase 10.1088/1475-7516/2015/08/003} {\bibfield  {journal} {\bibinfo
  {journal} {JCAP}\ }\textbf {\bibinfo {volume} {08}},\ \bibinfo {pages} {003}
  (\bibinfo {year} {2015})},\ \Eprint {http://arxiv.org/abs/1505.05773}
  {arXiv:1505.05773 [hep-th]} \BibitemShut {NoStop}%
\bibitem [{\citenamefont {Andr\'{e}asson}\ and\ \citenamefont
  {Ringstr\"{o}m}(2016)}]{andreasson+ringstrom_16}%
  \BibitemOpen
  \bibfield  {author} {\bibinfo {author} {\bibfnamefont {H\r{a}kan}\
  \bibnamefont {Andr\'{e}asson}}\ and\ \bibinfo {author} {\bibfnamefont {Hans}\
  \bibnamefont {Ringstr\"{o}m}},\ }\bibfield  {title} {\enquote {\bibinfo
  {title} {Proof of the cosmic no-hair conjecture in the $\mathbb{T}^3$-{G}owdy
  symmetric {Einstein---Vlasov} setting},}\ }\href {\doibase 10.4171/JEMS/623}
  {\bibfield  {journal} {\bibinfo  {journal} {J. Eur. Math. Soc.}\ }\textbf
  {\bibinfo {volume} {18}},\ \bibinfo {pages} {1565--1650} (\bibinfo {year}
  {2016})},\ \Eprint {http://arxiv.org/abs/1306.6223} {arXiv:1306.6223 [gr-qc]}
  \BibitemShut {NoStop}%
\bibitem [{\citenamefont {Carroll}\ and\ \citenamefont
  {Chatwin-Davies}(2018)}]{carroll+chatwin-davies_18}%
  \BibitemOpen
  \bibfield  {author} {\bibinfo {author} {\bibfnamefont {Sean~M.}\ \bibnamefont
  {Carroll}}\ and\ \bibinfo {author} {\bibfnamefont {Aidan}\ \bibnamefont
  {Chatwin-Davies}},\ }\bibfield  {title} {\enquote {\bibinfo {title} {Cosmic
  equilibration: A holographic no-hair theorem from the generalized second
  law},}\ }\href {\doibase 10.1103/PhysRevD.97.046012} {\bibfield  {journal}
  {\bibinfo  {journal} {Phys. Rev. D}\ }\textbf {\bibinfo {volume} {97}},\
  \bibinfo {pages} {046012} (\bibinfo {year} {2018})},\ \Eprint
  {http://arxiv.org/abs/1703.09241} {arXiv:1703.09241 [hep-th]} \BibitemShut
  {NoStop}%
\bibitem [{\citenamefont {Moss}\ and\ \citenamefont
  {Sahni}(1986)}]{moss+sahni_86}%
  \BibitemOpen
  \bibfield  {author} {\bibinfo {author} {\bibfnamefont {Ian}\ \bibnamefont
  {Moss}}\ and\ \bibinfo {author} {\bibfnamefont {Varon}\ \bibnamefont
  {Sahni}},\ }\bibfield  {title} {\enquote {\bibinfo {title} {Anisotropy in the
  chaotic inflationary universe},}\ }\href {\doibase
  https://doi.org/10.1016/0370-2693(86)91488-7} {\bibfield  {journal} {\bibinfo
   {journal} {Phys. Lett. B}\ }\textbf {\bibinfo {volume} {178}},\ \bibinfo
  {pages} {159--162} (\bibinfo {year} {1986})}\BibitemShut {NoStop}%
\bibitem [{\citenamefont {Kitada}\ and\ \citenamefont
  {Maeda}(1993)}]{kitada+maeda_93}%
  \BibitemOpen
  \bibfield  {author} {\bibinfo {author} {\bibfnamefont {Yuichi}\ \bibnamefont
  {Kitada}}\ and\ \bibinfo {author} {\bibfnamefont {Kei-ichi}\ \bibnamefont
  {Maeda}},\ }\bibfield  {title} {\enquote {\bibinfo {title} {{Cosmic no hair
  theorem in homogeneous space-times. 1. Bianchi models}},}\ }\href {\doibase
  10.1088/0264-9381/10/4/008} {\bibfield  {journal} {\bibinfo  {journal}
  {Class. Quant. Grav.}\ }\textbf {\bibinfo {volume} {10}},\ \bibinfo {pages}
  {703--734} (\bibinfo {year} {1993})}\BibitemShut {NoStop}%
\bibitem [{\citenamefont {Rendall}(2004)}]{rendall_04}%
  \BibitemOpen
  \bibfield  {author} {\bibinfo {author} {\bibfnamefont {Alan~D.}\ \bibnamefont
  {Rendall}},\ }\bibfield  {title} {\enquote {\bibinfo {title} {{Accelerated
  cosmological expansion due to a scalar field whose potential has a positive
  lower bound}},}\ }\href {\doibase 10.1088/0264-9381/21/9/018} {\bibfield
  {journal} {\bibinfo  {journal} {Class. Quant. Grav.}\ }\textbf {\bibinfo
  {volume} {21}},\ \bibinfo {pages} {2445--2454} (\bibinfo {year} {2004})},\
  \Eprint {http://arxiv.org/abs/gr-qc/0403070} {arXiv:gr-qc/0403070}
  \BibitemShut {NoStop}%
\bibitem [{\citenamefont {Kandrup}(1992)}]{Kandrup:1992xw}%
  \BibitemOpen
  \bibfield  {author} {\bibinfo {author} {\bibfnamefont {H.~E.}\ \bibnamefont
  {Kandrup}},\ }\bibfield  {title} {\enquote {\bibinfo {title} {{Violations of
  the strong energy condition for interacting systems of particles}},}\ }\href
  {\doibase 10.1103/PhysRevD.46.5360} {\bibfield  {journal} {\bibinfo
  {journal} {Phys. Rev. D}\ }\textbf {\bibinfo {volume} {46}},\ \bibinfo
  {pages} {5360--5366} (\bibinfo {year} {1992})}\BibitemShut {NoStop}%
\bibitem [{\citenamefont {Barcelo}\ and\ \citenamefont
  {Visser}(1999)}]{Barcelo:1999hq}%
  \BibitemOpen
  \bibfield  {author} {\bibinfo {author} {\bibfnamefont {Carlos}\ \bibnamefont
  {Barcel\'{o}}}\ and\ \bibinfo {author} {\bibfnamefont {Matt}\ \bibnamefont
  {Visser}},\ }\bibfield  {title} {\enquote {\bibinfo {title} {{Traversable
  wormholes from massless conformally coupled scalar fields}},}\ }\href
  {\doibase 10.1016/S0370-2693(99)01117-X} {\bibfield  {journal} {\bibinfo
  {journal} {Phys. Lett. B}\ }\textbf {\bibinfo {volume} {466}},\ \bibinfo
  {pages} {127--134} (\bibinfo {year} {1999})},\ \Eprint
  {http://arxiv.org/abs/gr-qc/9908029} {arXiv:gr-qc/9908029} \BibitemShut
  {NoStop}%
\bibitem [{\citenamefont {Visser}\ and\ \citenamefont
  {Barcelo}(2000)}]{Visser:1999de}%
  \BibitemOpen
  \bibfield  {author} {\bibinfo {author} {\bibfnamefont {Matt}\ \bibnamefont
  {Visser}}\ and\ \bibinfo {author} {\bibfnamefont {Carlos}\ \bibnamefont
  {Barcel\'{o}}},\ }\bibfield  {title} {\enquote {\bibinfo {title} {{Energy
  conditions and their cosmological implications}},}\ }in\ \href {\doibase
  10.1142/9789812792129_0014} {\emph {\bibinfo {booktitle} {{Proceedings of the Third International Workshop on Particle Physics and the Early Universe (COSMO-99)}}}},\ \bibinfo
  {editor} {edited by\ \bibinfo {editor} {\bibfnamefont {U.}\ \bibnamefont
  {{Cotti}}},\ \bibinfo {editor} {\bibfnamefont {R.}\ \bibnamefont
  {{Jeannerot}}},\ \bibinfo {editor} {\bibfnamefont {G.}\ \bibnamefont
  {{Senjanovi\'{c}}}},\ and\ \bibinfo {editor} {\bibfnamefont {A.}~\bibnamefont
  {{Smirnov}}}}\ (\bibinfo {year} {{Singapore: World Scientific,
  2000}})\ pp.\ \bibinfo {pages} {98--112},\ \Eprint
  {http://arxiv.org/abs/gr-qc/0001099} {arXiv:gr-qc/0001099}
   \BibitemShut {NoStop}%
\bibitem [{\citenamefont {Barcelo}\ and\ \citenamefont
  {Visser}(2000)}]{Barcelo:2000zf}%
  \BibitemOpen
  \bibfield  {author} {\bibinfo {author} {\bibfnamefont {Carlos}\ \bibnamefont
  {Barcel\'{o}}}\ and\ \bibinfo {author} {\bibfnamefont {Matt}\ \bibnamefont
  {Visser}},\ }\bibfield  {title} {\enquote {\bibinfo {title} {{Scalar fields,
  energy conditions, and traversable wormholes}},}\ }\href {\doibase
  10.1088/0264-9381/17/18/318} {\bibfield  {journal} {\bibinfo  {journal}
  {Class. Quant. Grav.}\ }\textbf {\bibinfo {volume} {17}},\ \bibinfo {pages}
  {3843--3864} (\bibinfo {year} {2000})},\ \Eprint
  {http://arxiv.org/abs/gr-qc/0003025} {arXiv:gr-qc/0003025} \BibitemShut
  {NoStop}%
\bibitem [{\citenamefont {Bellucci}\ and\ \citenamefont
  {Faraoni}(2002)}]{Bellucci:2001cc}%
  \BibitemOpen
  \bibfield  {author} {\bibinfo {author} {\bibfnamefont {S.}~\bibnamefont
  {Bellucci}}\ and\ \bibinfo {author} {\bibfnamefont {V.}~\bibnamefont
  {Faraoni}},\ }\bibfield  {title} {\enquote {\bibinfo {title} {{Energy
  conditions and classical scalar fields}},}\ }\href {\doibase
  10.1016/S0550-3213(02)00437-6} {\bibfield  {journal} {\bibinfo  {journal}
  {Nucl. Phys. B}\ }\textbf {\bibinfo {volume} {640}},\ \bibinfo {pages}
  {453--468} (\bibinfo {year} {2002})},\ \Eprint
  {http://arxiv.org/abs/hep-th/0106168} {arXiv:hep-th/0106168} \BibitemShut
  {NoStop}%
\bibitem [{\citenamefont {Barcelo}\ and\ \citenamefont
  {Visser}(2002)}]{Barcelo:2002bv}%
  \BibitemOpen
  \bibfield  {author} {\bibinfo {author} {\bibfnamefont {Carlos}\ \bibnamefont
  {Barcel\'{o}}}\ and\ \bibinfo {author} {\bibfnamefont {Matt}\ \bibnamefont
  {Visser}},\ }\bibfield  {title} {\enquote {\bibinfo {title} {{Twilight for
  the energy conditions?}}}\ }\href {\doibase 10.1142/S0218271802002888}
  {\bibfield  {journal} {\bibinfo  {journal} {Int. J. Mod. Phys. D}\ }\textbf
  {\bibinfo {volume} {11}},\ \bibinfo {pages} {1553--1560} (\bibinfo {year}
  {2002})},\ \Eprint {http://arxiv.org/abs/gr-qc/0205066} {arXiv:gr-qc/0205066}
  \BibitemShut {NoStop}%
\bibitem [{\citenamefont {Dubovsky}\ \emph {et~al.}(2006)\citenamefont
  {Dubovsky}, \citenamefont {Gregoire}, \citenamefont {Nicolis},\ and\
  \citenamefont {Rattazzi}}]{Dubovsky:2005xd}%
  \BibitemOpen
  \bibfield  {author} {\bibinfo {author} {\bibfnamefont {S.}~\bibnamefont
  {Dubovsky}}, \bibinfo {author} {\bibfnamefont {T.}~\bibnamefont {Gr\'{e}goire}},
  \bibinfo {author} {\bibfnamefont {A.}~\bibnamefont {Nicolis}}, \ and\
  \bibinfo {author} {\bibfnamefont {R.}~\bibnamefont {Rattazzi}},\ }\bibfield
  {title} {\enquote {\bibinfo {title} {{Null energy condition and superluminal
  propagation}},}\ }\href {\doibase 10.1088/1126-6708/2006/03/025} {\bibfield
  {journal} {\bibinfo  {journal} {JHEP}\ }\textbf {\bibinfo {volume} {03}},\
  \bibinfo {pages} {025} (\bibinfo {year} {2006})},\ \Eprint
  {http://arxiv.org/abs/hep-th/0512260} {arXiv:hep-th/0512260} \BibitemShut
  {NoStop}%
\bibitem [{\citenamefont {Nicolis}\ \emph {et~al.}(2010)\citenamefont
  {Nicolis}, \citenamefont {Rattazzi},\ and\ \citenamefont
  {Trincherini}}]{Nicolis:2009qm}%
  \BibitemOpen
  \bibfield  {author} {\bibinfo {author} {\bibfnamefont {Alberto}\ \bibnamefont
  {Nicolis}}, \bibinfo {author} {\bibfnamefont {Riccardo}\ \bibnamefont
  {Rattazzi}}, \ and\ \bibinfo {author} {\bibfnamefont {Enrico}\ \bibnamefont
  {Trincherini}},\ }\bibfield  {title} {\enquote {\bibinfo {title} {{Energy's
  and amplitudes' positivity}},}\ }\href {\doibase 10.1007/JHEP05(2010)095}
  {\bibfield  {journal} {\bibinfo  {journal} {JHEP}\ }\textbf {\bibinfo
  {volume} {05}},\ \bibinfo {pages} {095} (\bibinfo {year} {2010})},\ \bibinfo
  {note} {[Erratum: JHEP 11, 128 (2011)]},\ \Eprint
  {http://arxiv.org/abs/0912.4258} {arXiv:0912.4258 [hep-th]} \BibitemShut
  {NoStop}%
\bibitem [{\citenamefont {Rubakov}(2014)}]{Rubakov:2014jja}%
  \BibitemOpen
  \bibfield  {author} {\bibinfo {author} {\bibfnamefont {V.~A.}\ \bibnamefont
  {Rubakov}},\ }\bibfield  {title} {\enquote {\bibinfo {title} {{The Null
  Energy Condition and its violation}},}\ }\href {\doibase
  10.3367/UFNe.0184.201402b.0137} {\bibfield  {journal} {\bibinfo  {journal}
  {Phys. Usp.}\ }\textbf {\bibinfo {volume} {57}},\ \bibinfo {pages} {128--142}
  (\bibinfo {year} {2014})},\ \Eprint {http://arxiv.org/abs/1401.4024}
  {arXiv:1401.4024 [hep-th]} \BibitemShut {NoStop}%
\bibitem [{\citenamefont {Martin-Moruno}\ and\ \citenamefont
  {Visser}(2017)}]{Martin-Moruno:2017exc}%
  \BibitemOpen
  \bibfield  {author} {\bibinfo {author} {\bibfnamefont {Prado}\ \bibnamefont
  {Martin-Moruno}}\ and\ \bibinfo {author} {\bibfnamefont {Matt}\ \bibnamefont
  {Visser}},\ }\bibfield  {title} {\enquote {\bibinfo {title} {{Classical and
  semi-classical energy conditions}},}\ }\href {\doibase
  10.1007/978-3-319-55182-1_9} {\bibfield  {journal} {\bibinfo  {journal}
  {Fundam. Theor. Phys.}\ }\textbf {\bibinfo {volume} {189}},\ \bibinfo {pages}
  {193--213} (\bibinfo {year} {2017})},\ \Eprint
  {http://arxiv.org/abs/1702.05915} {arXiv:1702.05915 [gr-qc]} \BibitemShut
  {NoStop}%
\bibitem [{\citenamefont {Cheung}\ \emph {et~al.}(2008)\citenamefont {Cheung},
  \citenamefont {Creminelli}, \citenamefont {Fitzpatrick}, \citenamefont
  {Kaplan},\ and\ \citenamefont {Senatore}}]{cheung+al_08}%
  \BibitemOpen
  \bibfield  {author} {\bibinfo {author} {\bibfnamefont {Clifford}\
  \bibnamefont {Cheung}}, \bibinfo {author} {\bibfnamefont {Paolo}\
  \bibnamefont {Creminelli}}, \bibinfo {author} {\bibfnamefont {A.~Liam}\
  \bibnamefont {Fitzpatrick}}, \bibinfo {author} {\bibfnamefont {Jared}\
  \bibnamefont {Kaplan}}, \ and\ \bibinfo {author} {\bibfnamefont {Leonardo}\
  \bibnamefont {Senatore}},\ }\bibfield  {title} {\enquote {\bibinfo {title}
  {{The Effective Field Theory of Inflation}},}\ }\href {\doibase
  10.1088/1126-6708/2008/03/014} {\bibfield  {journal} {\bibinfo  {journal}
  {JHEP}\ }\textbf {\bibinfo {volume} {03}},\ \bibinfo {pages} {014} (\bibinfo
  {year} {2008})},\ \Eprint {http://arxiv.org/abs/0709.0293} {arXiv:0709.0293
  [hep-th]} \BibitemShut {NoStop}%
\bibitem [{\citenamefont {Weinberg}(2008)}]{Weinberg:2008hq}%
  \BibitemOpen
  \bibfield  {author} {\bibinfo {author} {\bibfnamefont {Steven}\ \bibnamefont
  {Weinberg}},\ }\bibfield  {title} {\enquote {\bibinfo {title} {{Effective
  Field Theory for Inflation}},}\ }\href {\doibase 10.1103/PhysRevD.77.123541}
  {\bibfield  {journal} {\bibinfo  {journal} {Phys. Rev. D}\ }\textbf {\bibinfo
  {volume} {77}},\ \bibinfo {pages} {123541} (\bibinfo {year} {2008})},\
  \Eprint {http://arxiv.org/abs/0804.4291} {arXiv:0804.4291 [hep-th]}
  \BibitemShut {NoStop}%
\bibitem [{\citenamefont {Azhar}\ and\ \citenamefont
  {Kaiser}(2018)}]{Azhar:2018nol}%
  \BibitemOpen
  \bibfield  {author} {\bibinfo {author} {\bibfnamefont {Feraz}\ \bibnamefont
  {Azhar}}\ and\ \bibinfo {author} {\bibfnamefont {David~I.}\ \bibnamefont
  {Kaiser}},\ }\bibfield  {title} {\enquote {\bibinfo {title} {{Flows into
  inflation: An effective field theory approach}},}\ }\href {\doibase
  10.1103/PhysRevD.98.063515} {\bibfield  {journal} {\bibinfo  {journal} {Phys.
  Rev. D}\ }\textbf {\bibinfo {volume} {98}},\ \bibinfo {pages} {063515}
  (\bibinfo {year} {2018})},\ \Eprint {http://arxiv.org/abs/1807.02088}
  {arXiv:1807.02088 [astro-ph.CO]} \BibitemShut {NoStop}%
\bibitem [{\citenamefont {Burgess}\ \emph {et~al.}(2016)\citenamefont
  {Burgess}, \citenamefont {Holman},\ and\ \citenamefont
  {Tasinato}}]{burgess+al_16}%
  \BibitemOpen
  \bibfield  {author} {\bibinfo {author} {\bibfnamefont {C.~P.}\ \bibnamefont
  {Burgess}}, \bibinfo {author} {\bibfnamefont {R.}~\bibnamefont {Holman}}, \
  and\ \bibinfo {author} {\bibfnamefont {G.}~\bibnamefont {Tasinato}},\
  }\bibfield  {title} {\enquote {\bibinfo {title} {{Open EFTs, IR effects \&
  late-time resummations: Systematic corrections in stochastic inflation}},}\
  }\href {\doibase 10.1007/JHEP01(2016)153} {\bibfield  {journal} {\bibinfo
  {journal} {JHEP}\ }\textbf {\bibinfo {volume} {01}},\ \bibinfo {pages} {153}
  (\bibinfo {year} {2016})},\ \Eprint {http://arxiv.org/abs/1512.00169}
  {arXiv:1512.00169 [gr-qc]} \BibitemShut {NoStop}%
\bibitem [{\citenamefont {Gorbenko}\ and\ \citenamefont
  {Senatore}(2019)}]{gorbenko+senatore_19}%
  \BibitemOpen
  \bibfield  {author} {\bibinfo {author} {\bibfnamefont {Victor}\ \bibnamefont
  {Gorbenko}}\ and\ \bibinfo {author} {\bibfnamefont {Leonardo}\ \bibnamefont
  {Senatore}},\ }\bibfield  {title} {\enquote {\bibinfo {title} {{$\lambda
  \phi^4$ in dS}},}\ }\href@noop {} {\  (\bibinfo {year} {2019})},\ \Eprint
  {http://arxiv.org/abs/1911.00022} {arXiv:1911.00022 [hep-th]} \BibitemShut
  {NoStop}%
\bibitem [{\citenamefont {Baumgart}\ and\ \citenamefont
  {Sundrum}(2020)}]{baumgart+sundrum_20}%
  \BibitemOpen
  \bibfield  {author} {\bibinfo {author} {\bibfnamefont {Matthew}\ \bibnamefont
  {Baumgart}}\ and\ \bibinfo {author} {\bibfnamefont {Raman}\ \bibnamefont
  {Sundrum}},\ }\bibfield  {title} {\enquote {\bibinfo {title} {{De Sitter
  Diagrammar and the Resummation of Time}},}\ }\href {\doibase
  10.1007/JHEP07(2020)119} {\bibfield  {journal} {\bibinfo  {journal} {JHEP}\
  }\textbf {\bibinfo {volume} {07}},\ \bibinfo {pages} {119} (\bibinfo {year}
  {2020})},\ \Eprint {http://arxiv.org/abs/1912.09502} {arXiv:1912.09502
  [hep-th]} \BibitemShut {NoStop}%
\bibitem [{\citenamefont {Frusciante}\ \emph {et~al.}(2014)\citenamefont
  {Frusciante}, \citenamefont {Raveri},\ and\ \citenamefont
  {Silvestri}}]{frusciante+al_14}%
  \BibitemOpen
  \bibfield  {author} {\bibinfo {author} {\bibfnamefont {Noemi}\ \bibnamefont
  {Frusciante}}, \bibinfo {author} {\bibfnamefont {Marco}\ \bibnamefont
  {Raveri}}, \ and\ \bibinfo {author} {\bibfnamefont {Alessandra}\ \bibnamefont
  {Silvestri}},\ }\bibfield  {title} {\enquote {\bibinfo {title} {{Effective
  Field Theory of Dark Energy: a Dynamical Analysis}},}\ }\href {\doibase
  10.1088/1475-7516/2014/02/026} {\bibfield  {journal} {\bibinfo  {journal}
  {JCAP}\ }\textbf {\bibinfo {volume} {02}},\ \bibinfo {pages} {026} (\bibinfo
  {year} {2014})},\ \Eprint {http://arxiv.org/abs/1310.6026} {arXiv:1310.6026
  [astro-ph.CO]} \BibitemShut {NoStop}%
\bibitem [{\citenamefont {van~den Hoogen}\ and\ \citenamefont
  {Coley}(1995)}]{vandenhoogen+coley_95}%
  \BibitemOpen
  \bibfield  {author} {\bibinfo {author} {\bibfnamefont {R.~J.}\ \bibnamefont
  {van~den Hoogen}}\ and\ \bibinfo {author} {\bibfnamefont {A.~A.}\
  \bibnamefont {Coley}},\ }\bibfield  {title} {\enquote {\bibinfo {title}
  {{Qualitative analysis of causal anisotropic viscous fluid cosmological
  models}},}\ }\href {\doibase 10.1088/0264-9381/12/9/019} {\bibfield
  {journal} {\bibinfo  {journal} {Class. Quant. Grav.}\ }\textbf {\bibinfo
  {volume} {12}},\ \bibinfo {pages} {2335--2354} (\bibinfo {year} {1995})},\
  \Eprint {http://arxiv.org/abs/gr-qc/9605062} {arXiv:gr-qc/9605062}
  \BibitemShut {NoStop}%
\bibitem [{\citenamefont {Hawking}\ and\ \citenamefont
  {Ellis}(1973)}]{Hawking:1973uf}%
  \BibitemOpen
  \bibfield  {author} {\bibinfo {author} {\bibfnamefont {S.~W.}\ \bibnamefont
  {Hawking}}\ and\ \bibinfo {author} {\bibfnamefont {G.~F.~R.}\ \bibnamefont
  {Ellis}},\ }\href {\doibase 10.1017/CBO9780511524646} {\emph {\bibinfo
  {title} {{The Large Scale Structure of Space-Time}}}}\ (\bibinfo  {publisher}
  {New York: Cambridge University Press},\ \bibinfo {year} {1973})\BibitemShut
  {NoStop}%
\bibitem [{\citenamefont {Wald}(1984)}]{wald_83text}%
  \BibitemOpen
  \bibfield  {author} {\bibinfo {author} {\bibfnamefont {Robert~M.}\
  \bibnamefont {Wald}},\ }\href {\doibase
  10.7208/chicago/9780226870373.001.0001} {\emph {\bibinfo {title} {{General
  Relativity}}}}\ (\bibinfo  {publisher} {Chicago: University of Chicago
  Press},\ \bibinfo {year} {1984})\BibitemShut {NoStop}%
\bibitem [{\citenamefont {Visser}(1995)}]{visser_95}%
  \BibitemOpen
  \bibfield  {author} {\bibinfo {author} {\bibfnamefont {Matt}\ \bibnamefont
  {Visser}},\ }\href@noop {} {\emph {\bibinfo {title} {{Lorentzian wormholes:
  From Einstein to Hawking}}}}\ (\bibinfo  {publisher} {{Woodbury, New York:
  American Institute of Physics Press}},\ \bibinfo {year} {1995})\BibitemShut
  {NoStop}%
\bibitem [{\citenamefont {Cropp}\ and\ \citenamefont
  {Visser}(2011)}]{Cropp:2010yj}%
  \BibitemOpen
  \bibfield  {author} {\bibinfo {author} {\bibfnamefont {Bethan}\ \bibnamefont
  {Cropp}}\ and\ \bibinfo {author} {\bibfnamefont {Matt}\ \bibnamefont
  {Visser}},\ }\bibfield  {title} {\enquote {\bibinfo {title} {{Any spacetime
  has a Bianchi type I spacetime as a limit}},}\ }\href {\doibase
  10.1088/0264-9381/28/5/055007} {\bibfield  {journal} {\bibinfo  {journal}
  {Class. Quant. Grav.}\ }\textbf {\bibinfo {volume} {28}},\ \bibinfo {pages}
  {055007} (\bibinfo {year} {2011})},\ \Eprint {http://arxiv.org/abs/1008.4639}
  {arXiv:1008.4639 [gr-qc]} \BibitemShut {NoStop}%
\bibitem [{\citenamefont {Pereira}\ \emph {et~al.}(2007)\citenamefont
  {Pereira}, \citenamefont {Pitrou},\ and\ \citenamefont
  {Uzan}}]{pereira+al_07}%
  \BibitemOpen
  \bibfield  {author} {\bibinfo {author} {\bibfnamefont {Thiago~S.}\
  \bibnamefont {Pereira}}, \bibinfo {author} {\bibfnamefont {Cyril}\
  \bibnamefont {Pitrou}}, \ and\ \bibinfo {author} {\bibfnamefont
  {Jean-Philippe}\ \bibnamefont {Uzan}},\ }\bibfield  {title} {\enquote
  {\bibinfo {title} {{Theory of cosmological perturbations in an anisotropic
  universe}},}\ }\href {\doibase 10.1088/1475-7516/2007/09/006} {\bibfield
  {journal} {\bibinfo  {journal} {JCAP}\ }\textbf {\bibinfo {volume} {09}},\
  \bibinfo {pages} {006} (\bibinfo {year} {2007})},\ \Eprint
  {http://arxiv.org/abs/0707.0736} {arXiv:0707.0736 [astro-ph]} \BibitemShut
  {NoStop}%
\bibitem [{\citenamefont {Starobinsky}(2000)}]{starobinsky_00}%
  \BibitemOpen
  \bibfield  {author} {\bibinfo {author} {\bibfnamefont {Alexei~A.}\
  \bibnamefont {Starobinsky}},\ }\bibfield  {title} {\enquote {\bibinfo {title}
  {{Future and origin of our universe: Modern view}},}\ }\href@noop {}
  {\bibfield  {journal} {\bibinfo  {journal} {Grav. Cosmol.}\ }\textbf
  {\bibinfo {volume} {6}},\ \bibinfo {pages} {157--163} (\bibinfo {year}
  {2000})},\ \Eprint {http://arxiv.org/abs/astro-ph/9912054}
  {arXiv:astro-ph/9912054} \BibitemShut {NoStop}%
\bibitem [{\citenamefont {Caldwell}(2002)}]{caldwell_02}%
  \BibitemOpen
  \bibfield  {author} {\bibinfo {author} {\bibfnamefont {R.~R.}\ \bibnamefont
  {Caldwell}},\ }\bibfield  {title} {\enquote {\bibinfo {title} {{A phantom
  menace? Cosmological consequences of a dark energy component with super-negative equation of state}},}\ }\href {\doibase 10.1016/S0370-2693(02)02589-3} {\bibfield
  {journal} {\bibinfo  {journal} {Phys. Lett. B}\ }\textbf {\bibinfo {volume}
  {545}},\ \bibinfo {pages} {23--29} (\bibinfo {year} {2002})},\ \Eprint
  {http://arxiv.org/abs/astro-ph/9908168} {arXiv:astro-ph/9908168} \BibitemShut
  {NoStop}%
\bibitem [{\citenamefont {Caldwell}\ \emph {et~al.}(2003)\citenamefont
  {Caldwell}, \citenamefont {Kamionkowski},\ and\ \citenamefont
  {Weinberg}}]{caldwell+al_03}%
  \BibitemOpen
  \bibfield  {author} {\bibinfo {author} {\bibfnamefont {Robert~R.}\
  \bibnamefont {Caldwell}}, \bibinfo {author} {\bibfnamefont {Marc}\
  \bibnamefont {Kamionkowski}}, \ and\ \bibinfo {author} {\bibfnamefont
  {Nevin~N.}\ \bibnamefont {Weinberg}},\ }\bibfield  {title} {\enquote
  {\bibinfo {title} {{Phantom energy: Dark energy with $w < -1$ causes a cosmic doomsday}},}\ }\href {\doibase
  10.1103/PhysRevLett.91.071301} {\bibfield  {journal} {\bibinfo  {journal}
  {Phys. Rev. Lett.}\ }\textbf {\bibinfo {volume} {91}},\ \bibinfo {pages}
  {071301} (\bibinfo {year} {2003})},\ \Eprint
  {http://arxiv.org/abs/astro-ph/0302506} {arXiv:astro-ph/0302506} \BibitemShut
  {NoStop}%
\bibitem [{\citenamefont {Kasner}(1921)}]{kasner_21}%
  \BibitemOpen
  \bibfield  {author} {\bibinfo {author} {\bibfnamefont {Edward}\ \bibnamefont
  {Kasner}},\ }\bibfield  {title} {\enquote {\bibinfo {title} {{Geometrical
  theorems on Einstein's cosmological equations}},}\ }\href {\doibase
  10.2307/2370192} {\bibfield  {journal} {\bibinfo  {journal} {Am. J. Math.}\
  }\textbf {\bibinfo {volume} {43}},\ \bibinfo {pages} {217--221} (\bibinfo
  {year} {1921})}\BibitemShut {NoStop}%
\bibitem [{\citenamefont {Misner}\ \emph {et~al.}(1973)\citenamefont {Misner},
  \citenamefont {Thorne},\ and\ \citenamefont {Wheeler}}]{misner+al_73}%
  \BibitemOpen
  \bibfield  {author} {\bibinfo {author} {\bibfnamefont {Charles~W.}\
  \bibnamefont {Misner}}, \bibinfo {author} {\bibfnamefont {K.~S.}\
  \bibnamefont {Thorne}}, \ and\ \bibinfo {author} {\bibfnamefont {J.~A.}\
  \bibnamefont {Wheeler}},\ }\href@noop {} {\emph {\bibinfo {title}
  {{Gravitation}}}}\ (\bibinfo  {publisher} {W. H. Freeman},\ \bibinfo
  {address} {San Francisco},\ \bibinfo {year} {1973})\BibitemShut {NoStop}%
\bibitem [{\citenamefont {Coley}(1999)}]{coley_99}%
  \BibitemOpen
  \bibfield  {author} {\bibinfo {author} {\bibfnamefont {Alan~A.}\ \bibnamefont
  {Coley}},\ }\bibfield  {title} {\enquote {\bibinfo {title} {{Dynamical
  systems in cosmology}},}\ }in\ \href@noop {} {\emph {\bibinfo {booktitle}
  {{Spanish Relativity Meeting (ERE 99)}}}}\ (\bibinfo {year} {1999})\ \Eprint
  {http://arxiv.org/abs/gr-qc/9910074} {arXiv:gr-qc/9910074} \BibitemShut
  {NoStop}%
\bibitem [{\citenamefont {Frampton}\ \emph
  {et~al.}(2012{\natexlab{a}})\citenamefont {Frampton}, \citenamefont
  {Ludwick},\ and\ \citenamefont {Scherrer}}]{Frampton:2011aa}%
  \BibitemOpen
  \bibfield  {author} {\bibinfo {author} {\bibfnamefont {Paul~H.}\ \bibnamefont
  {Frampton}}, \bibinfo {author} {\bibfnamefont {Kevin~J.}\ \bibnamefont
  {Ludwick}}, \ and\ \bibinfo {author} {\bibfnamefont {Robert~J.}\ \bibnamefont
  {Scherrer}},\ }\bibfield  {title} {\enquote {\bibinfo {title} {{Pseudo-rip:
  Cosmological models intermediate between the cosmological constant and the
  little rip}},}\ }\href {\doibase 10.1103/PhysRevD.85.083001} {\bibfield
  {journal} {\bibinfo  {journal} {Phys. Rev. D}\ }\textbf {\bibinfo {volume}
  {85}},\ \bibinfo {pages} {083001} (\bibinfo {year} {2012}{\natexlab{a}})},\
  \Eprint {http://arxiv.org/abs/1112.2964} {arXiv:1112.2964 [astro-ph.CO]}
  \BibitemShut {NoStop}%
\bibitem [{\citenamefont {Stefancic}(2005)}]{Stefancic:2004kb}%
  \BibitemOpen
  \bibfield  {author} {\bibinfo {author} {\bibfnamefont {Hrvoje}\ \bibnamefont
  {\v{S}tefan\v{c}i\'{c}}},\ }\bibfield  {title} {\enquote {\bibinfo {title} {{Expansion
  around the vacuum equation of state: Sudden future singularities and
  asymptotic behavior}},}\ }\href {\doibase 10.1103/PhysRevD.71.084024}
  {\bibfield  {journal} {\bibinfo  {journal} {Phys. Rev. D}\ }\textbf {\bibinfo
  {volume} {71}},\ \bibinfo {pages} {084024} (\bibinfo {year} {2005})},\
  \Eprint {http://arxiv.org/abs/astro-ph/0411630} {arXiv:astro-ph/0411630}
  \BibitemShut {NoStop}%
\bibitem [{\citenamefont {Nojiri}\ \emph {et~al.}(2005)\citenamefont {Nojiri},
  \citenamefont {Odintsov},\ and\ \citenamefont {Tsujikawa}}]{Nojiri:2005sx}%
  \BibitemOpen
  \bibfield  {author} {\bibinfo {author} {\bibfnamefont {Shin'ichi}\
  \bibnamefont {Nojiri}}, \bibinfo {author} {\bibfnamefont {Sergei~D.}\
  \bibnamefont {Odintsov}}, \ and\ \bibinfo {author} {\bibfnamefont {Shinji}\
  \bibnamefont {Tsujikawa}},\ }\bibfield  {title} {\enquote {\bibinfo {title}
  {{Properties of singularities in (phantom) dark energy universe}},}\ }\href
  {\doibase 10.1103/PhysRevD.71.063004} {\bibfield  {journal} {\bibinfo
  {journal} {Phys. Rev. D}\ }\textbf {\bibinfo {volume} {71}},\ \bibinfo
  {pages} {063004} (\bibinfo {year} {2005})},\ \Eprint
  {http://arxiv.org/abs/hep-th/0501025} {arXiv:hep-th/0501025} \BibitemShut
  {NoStop}%
\bibitem [{\citenamefont {Nojiri}\ and\ \citenamefont
  {Odintsov}(2005)}]{Nojiri:2005sr}%
  \BibitemOpen
  \bibfield  {author} {\bibinfo {author} {\bibfnamefont {Shin'ichi}\
  \bibnamefont {Nojiri}}\ and\ \bibinfo {author} {\bibfnamefont {Sergei~D.}\
  \bibnamefont {Odintsov}},\ }\bibfield  {title} {\enquote {\bibinfo {title}
  {{Inhomogeneous equation of state of the universe: Phantom era, future
  singularity and crossing the phantom barrier}},}\ }\href {\doibase
  10.1103/PhysRevD.72.023003} {\bibfield  {journal} {\bibinfo  {journal} {Phys.
  Rev. D}\ }\textbf {\bibinfo {volume} {72}},\ \bibinfo {pages} {023003}
  (\bibinfo {year} {2005})},\ \Eprint {http://arxiv.org/abs/hep-th/0505215}
  {arXiv:hep-th/0505215} \BibitemShut {NoStop}%
\bibitem [{\citenamefont {Frampton}\ \emph {et~al.}(2011)\citenamefont
  {Frampton}, \citenamefont {Ludwick},\ and\ \citenamefont
  {Scherrer}}]{Frampton:2011sp}%
  \BibitemOpen
  \bibfield  {author} {\bibinfo {author} {\bibfnamefont {Paul~H.}\ \bibnamefont
  {Frampton}}, \bibinfo {author} {\bibfnamefont {Kevin~J.}\ \bibnamefont
  {Ludwick}}, \ and\ \bibinfo {author} {\bibfnamefont {Robert~J.}\ \bibnamefont
  {Scherrer}},\ }\bibfield  {title} {\enquote {\bibinfo {title} {{The Little
  Rip}},}\ }\href {\doibase 10.1103/PhysRevD.84.063003} {\bibfield  {journal}
  {\bibinfo  {journal} {Phys. Rev. D}\ }\textbf {\bibinfo {volume} {84}},\
  \bibinfo {pages} {063003} (\bibinfo {year} {2011})},\ \Eprint
  {http://arxiv.org/abs/1106.4996} {arXiv:1106.4996 [astro-ph.CO]} \BibitemShut
  {NoStop}%
\bibitem [{\citenamefont {Frampton}\ \emph
  {et~al.}(2012{\natexlab{b}})\citenamefont {Frampton}, \citenamefont
  {Ludwick}, \citenamefont {Nojiri}, \citenamefont {Odintsov},\ and\
  \citenamefont {Scherrer}}]{Frampton:2011rh}%
  \BibitemOpen
  \bibfield  {author} {\bibinfo {author} {\bibfnamefont {Paul~H.}\ \bibnamefont
  {Frampton}}, \bibinfo {author} {\bibfnamefont {Kevin~J.}\ \bibnamefont
  {Ludwick}}, \bibinfo {author} {\bibfnamefont {Shin'ichi}\ \bibnamefont
  {Nojiri}}, \bibinfo {author} {\bibfnamefont {Sergei~D.}\ \bibnamefont
  {Odintsov}}, \ and\ \bibinfo {author} {\bibfnamefont {Robert~J.}\
  \bibnamefont {Scherrer}},\ }\bibfield  {title} {\enquote {\bibinfo {title}
  {{Models for Little Rip Dark Energy}},}\ }\href {\doibase
  10.1016/j.physletb.2012.01.048} {\bibfield  {journal} {\bibinfo  {journal}
  {Phys. Lett. B}\ }\textbf {\bibinfo {volume} {708}},\ \bibinfo {pages}
  {204--211} (\bibinfo {year} {2012}{\natexlab{b}})},\ \Eprint
  {http://arxiv.org/abs/1108.0067} {arXiv:1108.0067 [hep-th]} \BibitemShut
  {NoStop}%
\bibitem [{\citenamefont {{Penrose}}({New York: Cambridge University Press,
  1979})}]{PenroseWeyl1979}%
  \BibitemOpen
  \bibfield  {author} {\bibinfo {author} {\bibfnamefont {R.}~\bibnamefont
  {{Penrose}}},\ }\bibfield  {title} {\enquote {\bibinfo {title}
  {{Singularities and time-asymmetry}},}\ }in\ \href@noop {} {\emph {\bibinfo
  {booktitle} {General Relativity: An Einstein Centenary Survey}}},\ \bibinfo
  {editor} {edited by\ \bibinfo {editor} {\bibfnamefont {S.~W.}\ \bibnamefont
  {{Hawking}}}\ and\ \bibinfo {editor} {\bibfnamefont {W.}~\bibnamefont
  {{Israel}}}}\ (\bibinfo {year} {{New York: Cambridge University Press,
  1979}})\ pp.\ \bibinfo {pages} {581--638}\BibitemShut {NoStop}%
\bibitem [{\citenamefont {{Penrose}}({New York: Cambridge University Press,
  2008})}]{Penrose2008}%
  \BibitemOpen
  \bibfield  {author} {\bibinfo {author} {\bibfnamefont {Roger}\ \bibnamefont
  {{Penrose}}},\ }\bibfield  {title} {\enquote {\bibinfo {title} {{Causality,
  quantum theory, and cosmology}},}\ }in\ \href@noop {} {\emph {\bibinfo
  {booktitle} {On Space and Time}}},\ \bibinfo {editor} {edited by\ \bibinfo
  {editor} {\bibfnamefont {S.}~\bibnamefont {{Majid}}}}\ (\bibinfo {year} {{New
  York: Cambridge University Press, 2008}})\ pp.\ \bibinfo {pages}
  {141--195}\BibitemShut {NoStop}%
\bibitem [{\citenamefont {Penrose}(2018)}]{Penrose:2018pyw}%
  \BibitemOpen
  \bibfield  {author} {\bibinfo {author} {\bibfnamefont {Roger}\ \bibnamefont
  {Penrose}},\ }\bibfield  {title} {\enquote {\bibinfo {title} {{The Big Bang
  and its Dark-Matter Content: Whence, Whither, and Wherefore}},}\ }\href
  {\doibase 10.1007/s10701-018-0162-3} {\bibfield  {journal} {\bibinfo
  {journal} {Found. Phys.}\ }\textbf {\bibinfo {volume} {48}},\ \bibinfo
  {pages} {1177--1190} (\bibinfo {year} {2018})}\BibitemShut {NoStop}%
\bibitem [{\citenamefont {Hu}(2021)}]{Hu:2021pfh}%
  \BibitemOpen
  \bibfield  {author} {\bibinfo {author} {\bibfnamefont {Bei-Lok}\ \bibnamefont
  {Hu}},\ }\bibfield  {title} {\enquote {\bibinfo {title} {{Weyl Curvature
  Hypothesis in Light of Quantum Backreaction at Cosmological Singularities or
  Bounces}},}\ }\href {\doibase 10.3390/universe7110424} {\bibfield  {journal}
  {\bibinfo  {journal} {Universe}\ }\textbf {\bibinfo {volume} {7}},\ \bibinfo
  {pages} {424} (\bibinfo {year} {2021})},\ \Eprint
  {http://arxiv.org/abs/2110.01104} {arXiv:2110.01104 [gr-qc]} \BibitemShut
  {NoStop}%
\bibitem [{\citenamefont {Kaiser}\ \emph {et~al.}(2013)\citenamefont {Kaiser},
  \citenamefont {Mazenc},\ and\ \citenamefont {Sfakianakis}}]{KMS}%
  \BibitemOpen
  \bibfield  {author} {\bibinfo {author} {\bibfnamefont {David~I.}\
  \bibnamefont {Kaiser}}, \bibinfo {author} {\bibfnamefont {Edward~A.}\
  \bibnamefont {Mazenc}}, \ and\ \bibinfo {author} {\bibfnamefont
  {Evangelos~I.}\ \bibnamefont {Sfakianakis}},\ }\bibfield  {title} {\enquote
  {\bibinfo {title} {{Primordial Bispectrum from Multifield Inflation with
  Nonminimal Couplings}},}\ }\href {\doibase 10.1103/PhysRevD.87.064004}
  {\bibfield  {journal} {\bibinfo  {journal} {Phys. Rev. D}\ }\textbf {\bibinfo
  {volume} {87}},\ \bibinfo {pages} {064004} (\bibinfo {year} {2013})},\
  \Eprint {http://arxiv.org/abs/1210.7487} {arXiv:1210.7487 [astro-ph.CO]}
  \BibitemShut {NoStop}%
\bibitem [{\citenamefont {Kobayashi}(2019)}]{HorndeskiRev}%
  \BibitemOpen
  \bibfield  {author} {\bibinfo {author} {\bibfnamefont {Tsutomu}\ \bibnamefont
  {Kobayashi}},\ }\bibfield  {title} {\enquote {\bibinfo {title} {{Horndeski
  theory and beyond: a review}},}\ }\href {\doibase 10.1088/1361-6633/ab2429}
  {\bibfield  {journal} {\bibinfo  {journal} {Rept. Prog. Phys.}\ }\textbf
  {\bibinfo {volume} {82}},\ \bibinfo {pages} {086901} (\bibinfo {year}
  {2019})},\ \Eprint {http://arxiv.org/abs/1901.07183} {arXiv:1901.07183
  [gr-qc]} \BibitemShut {NoStop}%
\end{thebibliography}


\end{document}